\documentclass[aps,prb,a4paper,groupedaddress,showpacs,twocolumn,floatfix]{revtex4}

\pdfoutput=1

\usepackage{amssymb,amsbsy,amsmath}

\usepackage{graphicx}
\usepackage{dcolumn}
\usepackage{bm}
\usepackage{subfigure}

\usepackage{multirow}
\usepackage{array}
\usepackage{rotating}

\usepackage[latin1, ansinew]{inputenc}

\usepackage{color}

\newcommand{\bra}[1]{\ensuremath{\langle #1|}}
\newcommand{\ket}[1]{\ensuremath{|#1\rangle}}
\newcommand{\braket}[2]{\ensuremath{\langle #1|#2\rangle}}

\newcommand{\op}[1]{%
    \fontdimen12\textfont3=2pt\fontdimen12\scriptfont3=1.4pt%
    \!\null\mathop{\vphantom{#1}\smash{#1}}\limits_{\sim}\null\!}
\newcommand{\vecop}[1]{%
    \fontdimen12\textfont3=2pt\fontdimen12\scriptfont3=1.4pt%
    \!\null\mathop{\textbf{\vphantom{#1}\smash{#1}}}\limits_{\sim}\null\!}

\newcommand{\mat}[1]{%
    \fontdimen12\textfont3=2pt\fontdimen12\scriptfont3=1.4pt%
    \!\null\mathop{\textbf{\vphantom{#1}\smash{#1}}}\null\!}

\newcommand{\fmref}[1]{(\protect\ref{#1})}
\newcommand{\xref}[1]{\protect\ref{#1}}
\newcommand{\figref}[1]{Fig.~\protect\ref{#1}}

\newcommand{\EinsOp}
   {\;\smash{\raisebox{-1.1ex}{$\!\!\stackrel{\!\mbox{1}
   \hspace{-0.4ex}\rule[0.0ex]{0.06ex}{1.60ex}}{\sim}$}}}

\newcommand{\reduced}[3]{\ensuremath{\langle #1||#2 || #3 \rangle}}

\newcommand{\com}[2]{\ensuremath{\left[ #1,\,#2 \right]}}

\newcommand{\threej}[6]{\ensuremath{\begin{pmatrix} #1 & #2 & #3 \\ #4 & #5 & #6 \end{pmatrix}}}
\newcommand{\sixj}[6]{\ensuremath{\begin{Bmatrix} #1& #2 & #3 \\ #4 & #5 & #6 \end{Bmatrix}}}
\newcommand{\ninej}[9]{\ensuremath{\begin{pmatrix} #1& #2 & #3 \\ #4 & #5 & #6 \\ #7 & #8 & #9 \end{pmatrix}}}

\newcommand{\ito}[3]{\ensuremath{\op{#1}^{(#2)}_{#3}}}
\newcommand{\vecito}[2]{\ensuremath{\vecop{#1}^{(#2)}}}

\newcommand {\mofe}[1]{\{$\textrm{Mo}_{72}\textrm{Fe}_{30}$\}}

\begin{document}

\title{Calculating the energy spectra of magnetic molecules:
  application of real- and spin-space symmetries}

\author{Roman Schnalle}
\email{rschnall@uni-bielefeld.de}
\author{J{\"u}rgen Schnack}
\email{jschnack@uni-bielefeld.de}
\affiliation{Universit{\"a}t Bielefeld, Fakult{\"a}t f{\"u}r Physik, Postfach 100131, D-33501 Bielefeld, Germany}

\date{\today}

\begin{abstract}
The determination of the energy spectra of small spin systems as
for instance given by magnetic molecules is a demanding
numerical problem. In this work we review numerical approaches
to diagonalize the Heisenberg Hamiltonian that employ
symmetries; in particular we focus on the spin-rotational
symmetry $SU(2)$ in combination with point-group symmetries.
With these methods one is able to block-diagonalize the
Hamiltonian and thus to treat spin systems of unprecedented
size. In addition it provides a spectroscopic labeling by
irreducible representations that is helpful when interpreting
transitions induced by Electron Paramagnetic Resonance (EPR),
Nuclear Magnetic Resonance (NMR) or Inelastic Neutron Scattering
(INS). It is our aim to provide the reader with detailed
knowledge on how to set up such a diagonalization scheme.
\end{abstract}


\maketitle

%

\section{Introduction} \label{sec-1}

Magnetism is a research field that is almost as old as
human writing. It took several thousand years until its nature,
which is quantum, could be determined. In 1928 Werner Heisenberg
published his work on the theory of ferromagnetism \emph{Zur
Theorie des Ferromagnetismus}, in which he introduced what is
today called the Heisenberg model.\cite{Heisenberg:ZP28} That
this spin-only model is successfully applicable to magnetism
rests on the property of many iron group elements to possess a
quenched angular momentum in chemical compounds.\cite{Kittel}
Therefore, for many magnetic substances the Heisenberg
Hamiltonian provides the dominant term whereas effects connected
to spin-orbit interaction are treated perturbatively in these
systems. For theoretical work on non-Heisenberg systems see
e.g. Refs.~\onlinecite{BCC:JPCA98,CCH:CR00,PTC:JCP03,MCC:JACS03,PTC:IRPC10}.

The Heisenberg Hamiltonian 
\begin{equation}
   \op{H}_\text{Heisenberg} = - \sum_{i,j} J_{ij} \vecop{s}(i)
   \cdot \vecop{s}(j)  \label{eq:heisenberg}
\end{equation}
models the magnetic system by a sum of pairwise interactions
between spins. The interaction strength (exchange parameter)
between spins at sites $i$ and $j$ is given by a number $J_{ij}$
with $J_{ij}<0$ referring to an antiferromagnetic and $J_{ij}>0$
to a ferromagnetic coupling. The spins are described by vector
operators. 

In order to understand magnetic observables such as
magnetization, susceptibility, heat capacity or EPR, NMR and INS
spectra the knowledge of the full energy spectrum of the
investigated small magnetic system as for instance a magnetic
molecule is often indispensable. Although the Heisenberg
Hamiltonian, Eq.~\fmref{eq:heisenberg}, appears to be not too
complicated, analytical solutions are known only for very small
numbers of spins\cite{Kou:JMMM97,Kou:JMMM98,BSS:JMMM00} or for
instance for the spin-$1/2$ chain via the Bethe
ansatz.\cite{Bet:ZP31} The attempt to diagonalize the
Hamilton matrix numerically is very often severely restricted
due to the huge dimension of the underlying Hilbert space. For a
magnetic system of $N$ spins of spin quantum number $s$ the
dimension is $(2s+1)^N$ which grows exponentially with $N$.

Group theoretical methods can help to ease this numerical
problem. A further benefit is given by the characterization of
the obtained energy levels by quantum numbers and the
classification according to irreducible representations. This
review intends to provide an overview of the latest developments
in efficient numerical diagonalization techniques of the
Heisenberg model using symmetries. In particular we focus on the
spin-rotational symmetry $SU(2)$ in combination with point-group
symmetries.

The full rotational symmetry of angular momenta has been
employed for quite a while. In quantum chemistry the method of
irreducible tensor operators was adapted to few spin systems
along with the upcoming field of molecular
magnetism.\cite{GaP:GCI93,BCC:IC99,BeG:EPR,Tsu:group_theory,Tsu:ICA08,BBO:PRB07}
Nowadays the computer program \emph{MAGPACK}, that completely
diagonalizes the Heisenberg Hamiltonian using $SU(2)$ symmetry,
is freely available.\cite{BCC:JCC99} Also for the approximate
determination of energy eigenvalues by means of Density Matrix
Renormalization Group (DMRG) methods,\cite{Whi:PRB93,WhD:PRB93B}
that can for instance treat chains of a few hundreds of spins
with high accuracy, spin-rotational symmetry was
employed.\cite{MCG:EPL02} In other fields such as nuclear
physics this method was also adapted to model finite Fermi
systems such as nuclei employing $SU(2)$
symmetry.\cite{DuP:RPP04} Early applications are also known for
Hubbard models, where one can actually exploit two $SU(2)$
symmetries.\cite{AZH:PRB88,Zha:IJMPB91,LNN:PRB98,Sch:AP02}

Besides spin-rotational symmetry many magnetic molecules or spin
lattices possess spatial symmetries that can be expressed as
point-group symmetries. Nevertheless, a combination of $SU(2)$
with point-group symmetries is not very common. The reason, as
will become more apparent later, is that a rearrangement of
spins due to point-group operations easily leads to complicated
basis transformations between different coupling schemes. A
possible compromise is to use only part of the spin-rotational
symmetry (namely rotations about the $z$--axis) together with
point-group
symmetries\cite{SZP:JPI96,RLM:PRB08,RSH:PRB04,RiS:EPJB10} or to
expand all basis states in terms of simpler product
states.\cite{PhysRevB.64.064419,PhysRevB.68.029902,PhysRevB.64.014408}
During the past years a few attempts have been undertaken to
combine the full spin-rotational symmetry with point-group
symmetries. Oliver Waldmann combines the full spin-rotational
symmetry with those point-group symmetries that are compatible
with the spin coupling scheme, i.e. avoid complicated basis
transformations between different coupling
schemes.\cite{Wal:PRB00} Along the same lines, especially
low-symmetry groups such as $D_2$ are often applicable since the
coupling scheme can be organized accordingly, compare
Ref.~\onlinecite{DGP:IC93} for an early investigation.
Sinitsyn, Bostrem, and Ovchinnikov follow a similar route for
the square lattice antiferromagnet by employing $D_4$
point-group symmetry.\cite{BOS:TMP06,SBO:JPA07}

Very recently a general scheme was developed that allows to
combine spin-rotational symmetry with general point-group
symmetries.\cite{Schnalle:Diss09,ScS:PRB09,ScS:P09} The key
problem, that the application of point-group operations leads to
states belonging to a basis characterized by a different
coupling scheme whose representation in the original basis is
not (easily) known, can be solved by means of graph theoretical
methods that have been developed in another
context.\cite{FPV:CPC97,FPV:CPC95} We discuss in detail how this
method can be implemented and present results for numerical
exact diagonalizations of Heisenberg spin systems of
unprecedented size.\cite{ScS:PRB09,ScS:P09,SLS:CMP09,Sch:DT10}
Our aim is to provide the reader with sufficient material to be
able to employ these powerful group theoretical methods. They
can for instance also be applied to calculate higher order
Wigner-$nJ$ symbols that appear when the double exchange is
modeled in mixed-valent spin
systems.\cite{BCC:JCP96,CBC:IC09,BCC:JCC10}

The article is organized as follows. Section~\ref{sec-2}
introduces the basic concepts, i.e. the Hamiltonian and its
properties, the irreducible tensor operators, point group
operations, and the construction of basis states for irreducible
representations.  Section~\ref{sec-3} demonstrates with the help
of three examples that the Hamiltonian of spin systems of
unprecedented size can be diagonalized completely. The outlook
in Section~\ref{sec-4} shortly summarizes and shows
perspectives. The main part of this article is contained in an
extended appendix that explains all technical details to set up
the discussed diagonalization scheme.

%

\section{Conceptual ideas} \label{sec-2}

\subsection{Spin Hamiltonian of magnetic molecules} \label{sec-2-1}
\begin{figure}[ht!]
\centering
\includegraphics[width=40mm]{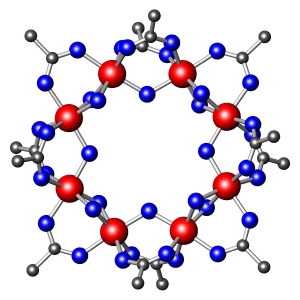} \quad
\includegraphics[width=35mm]{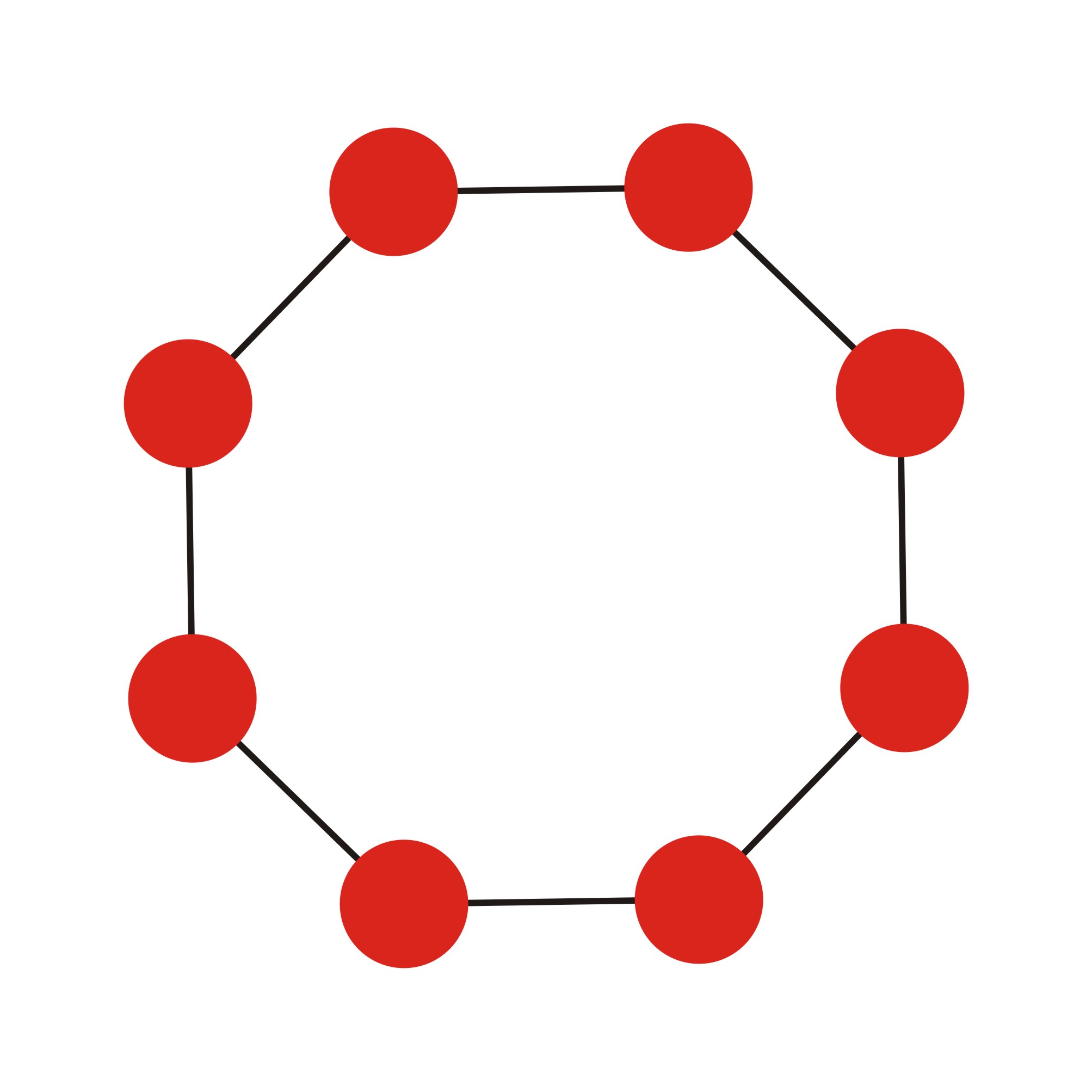}
\caption{Simplification of the chemical structure of a
$\text{Cr}_{8}$ compound (l.h.s.) to the corresponding spin
system for the same molecule (r.h.s.). Dots represent spin
sites, lines exchange interactions.}
\label{F-1}
\end{figure}

The research field of molecular magnetism deals with the
investigation of the magnetic properties of chemical compounds
composed of a number of ions that reaches from only a few up to
dozens of it. For the magnetic modeling of the molecule only
those ions are taken into account which possess unpaired
electrons and thus a non-vanishing magnetic moment. Since the
molecules, which are prepared in the form of a crystal or powder
sample, are often quite well separated from each other by their
ligands, inter-molecular interactions can be neglected in most
cases. Additionally, the electrons can very often be treated as
localized so that both features lead to a simplified sketch of
the chemical compound, namely a spin system. The interactions
between different spins of the system then depend of course on
the chemical surrounding and stem from direct exchange or
super-exchange\cite{And:PR59} via chemical bridges.  Figure
\ref{F-1} shows such a simplification for a $\text{Cr}_{8}$
compound which can easily be modeled by a ring-like system of
interacting spins.\cite{SSG:CAEJ02,CVG:PRB03,AGC:PRB03}

A general Hamiltonian that models magnetic molecules could be
written as
\begin{equation} \label{eq:Hamiltonian_gen}
   \op{H}_\text{general}= \op{H}_\text{exchange} + \op{H}_\text{Zeeman}
\ ,
\end{equation}
where, to be more specific, in a system of $N$ spins,
i.e. $i,j=1,\dots,N$, the two terms assume the form
\begin{eqnarray}
   \op{H}_\text{exchange} &=& \sum_{i,j} \vecop{s}(i) \cdot \mathbf{J}_{ij} \cdot \vecop{s}(j) \ , \label{eq:exchange} \\
   \op{H}_\text{Zeeman} &=& \mu_B \sum_{i} \vecop{s}(i) \cdot \mathbf{g}_{i} \cdot \vec{B} \ .
\end{eqnarray}
$\op{H}_\text{exchange}$ describes in a compact way the
(isotropic and anisotropic) exchange interaction between two
single-spin vector operators $\vecop{s}(i)$ and $\vecop{s}(j)$
as well as the single-ion anisotropy.\cite{BeG:EPR} The quantity
$\mathbf{J}_{ij}$ is a second rank Cartesian tensor containing
the corresponding parameters. $\op{H}_\text{Zeeman}$ couples the
spins to an external magnetic field $\vec{B}$. In general, the
coupling to an external field can be anisotropic and is thus
described by local tensors $\mathbf{g}_{i}$.

It turns out that for many magnetic molecules the isotropic
Heisenberg Hamilton operator provides a very good model. In
addition, we assume that for the highly symmetric spin systems to
be treated in this article the $g$-tensors are scalars, and the
same for all ions. Then the resulting Hamiltonian that models
the system simplifies to
\begin{equation}
   \op{H} = \op{H}_\text{Heisenberg} + \op{H}_\text{Zeeman}
\end{equation}
with 
\begin{eqnarray}
   \op{H}_\text{Heisenberg} &=& - \sum_{i,j} J_{ij} \vecop{s}(i)
   \cdot \vecop{s}(j) 
\\
 \label{eq:zeeman}
 \op{H}_\text{Zeeman} &=& g \mu_B \vecop{S} \cdot \vec{B}
\ .
\end{eqnarray}
$\vecop{S}=\sum_i \vecop{s}(i)$ is the total spin. As already
mentioned, $J_{ij}<0$ refers to an antiferromagnetic and
$J_{ij}>0$ to a ferromagnetic coupling.

The Heisenberg Hamiltonian is completely isotropic in spin space
($SU(2)$ symmetry), i.e. the commutators of the square of the
total spin $\vecop{S}$ and its $z$-component $\op{S}^z$ with
$\op{H}_\text{Heisenberg}$ vanish
\begin{equation} \label{eq:commutator}
  \com{\op{H}_\text{Heisenberg}}{\op{S}^z}=0 \ , 
\quad \com{\op{H}_\text{Heisenberg}}{\vecop{S}^2}=0
  \ .
\end{equation}
Since $\com{\vecop{S}^2}{\op{S}^z}=0$ the total magnetic quantum
number $M$ and the quantum number of the total spin $S$ serve as
good quantum numbers and a simultaneous eigenbasis of
$\op{S}^z$, $\vecop{S}^2$ and $\op{H}_\text{Heisenberg}$ can be
found.

A well adapted basis is then given by states of the form
$\ket{\alpha \, S \, M}$ which can be constructed according to a
vector-coupling scheme (see App. \ref{sec-A-2}). These states
are already eigenstates of $\vecop{S}^2$ and $\op{S}^z$ and
$\alpha$ denotes a set of additional quantum numbers resulting
from the coupling of the single spins $\vecop{s}(i)$ to the
total spin $\vecop{S}$. Due to Eqs. $\fmref{eq:commutator}$ the
matrix elements of the Heisenberg Hamiltonian $\bra{\alpha' \,
S' \, M'}\op{H}_\text{Heisenberg}\ket{\alpha \, S \, M}$ between
states with different $S$ and $M$ vanish, leading to a
block-factorized Hamilton matrix in which each block can be
diagonalized separately. In this case the field dependence of
the energies induced by the Zeeman term in Eq.~\fmref{eq:zeeman}
can easily be added without further complicated
calculations. This is because the $z$-direction can be chosen to
point along the external field, so that the Zeeman term commutes
with $\op{H}_\text{Heisenberg}$, $\vecop{S}^2$ and
$\op{S}^z$. Then $M$ still serves as a good quantum number, and
the effect of the external field $\vec{B}=B\cdot \vec{e}^z$ on
eigenstates of $\op{H}_\text{Heisenberg}$ results in a simple
field dependence of the energy eigenvalues $E_i$ according to
\begin{equation} \label{eq:energies}
   E_i(B)=E_i + g \mu_B B M_i
\ .
\end{equation}
This way thermodynamic properties depending on the temperature
$T$ and the external magnetic field $\vec{B}$ can easily be
calculated from the energy spectrum of the investigated magnetic
molecule once the energies $E_i$ are known.

\subsection{Irreducible tensor operator method} \label{sec-2-2}
The determination of the matrix elements of the Heisenberg
Hamiltonian can elegantly be achieved with the help of
irreducible tensor
operators.\cite{GaP:GCI93,BCC:IC99,BeG:EPR,Tsu:group_theory,Tsu:ICA08,BBO:PRB07}
To this end, it is necessary to reformulate the spin
vector-operators in terms of irreducible tensor operators and to
subsequently use tensorial algebra. In this regard the
underlying theory is clearly based on group as well as
representation theory. At this point it would probably not make
sense to introduce all group-theoretical tools which lead to a
complete understanding of the technical implementations used in
this work. Several textbooks provide deep knowledge about these
topics and the authors would like to refer to
those.\cite{Sil:ito,Wig:group,FaR:ito,Edm:angularmomentum,CoS:atomicspectra,Tin:Group_theory}
Nevertheless, at least the origin of appearing concepts and
formulations shall be explained. Some understanding of abstract
group and representation theory is assumed.

\subsubsection{Irreducible tensor operators} \label{sec-2-2-1}
An irreducible tensor operator $\vecito{T}{k}$ of rank $k$ is
defined by the transformation properties of its components $q$
under a general coordinate rotation $R$ according to
\begin{equation} \label{eq:ito}
\op{D}(R) \ito{T}{k}{q} \op{D}^{-1}(R) = \sum_{q'} \ito{T}{k}{q'} D_{qq'}^{(k)}(R)
\ .
\end{equation}
Here $\op{D}(R)$ denotes the operator associated with the
coordinate rotation $R$. The subscripts $q$ as well as $q'$ take
the values $-k,-k+1,\dots,k$. By $D_{qq'}^{(k)}(R)$ the matrix
elements of the so-called \textit{Wigner rotation matrices}
$\mat{D}^{(k)}(R)$ are denoted (cf. App. \ref{sec-A-1}).

The case $k=0$ in Eq.~\fmref{eq:ito} directly leads to what is
called a \textit{scalar operator}. As can easily be seen, a
scalar operator is invariant under coordinate rotation,
i.e. with $D_{00}^{(0)}(R)=1$
\begin{equation} \begin{split}
\op{D}(R) \ito{T}{0}{} \op{D}^{-1}(R) &= \ito{T}{0}{} D_{00}^{(0)}(R) \\
&= \ito{T}{0}{} \ .
\end{split}\end{equation}
In analogy to the states which span the irreducible
representation $D^{(1)}$ of $R_3$ and which are said to behave
under coordinate rotations like the components of a vector the
irreducible tensor operator $\vecito{T}{1}$ is called a
\textit{vector operator}. For example, the components of a
first-rank irreducible tensor operator $\vecito{s}{1}$ derived
from the Cartesian components of the spin vector operator are
given by
\begin{equation} \begin{split} \label{eq:spintensors}
   \ito{s}{1}{0} &= \op{s}^z \ , \\
   \ito{s}{1}{\pm 1} &= \mp \sqrt{\frac{1}{2}} \left( \op{s}^x \pm i \op{s}^y \right)
   \ .
\end{split} \end{equation}

Stressing the analogy between the behavior of states and
irreducible tensor operators under coordinate rotations, the
role of the components $\ito{T}{k}{q}$ in Eq.~\fmref{eq:ito} has
to be specified. The components of the irreducible tensor
operator $\vecito{T}{k}$ of rank $k$ serve as a basis and
therefore span the $(2k+1)$-dimensional irreducible
representation $D^{(k)}$ of the rotation group $R_3$.

For comparison, consider a group $\mathcal{G}$ and its
irreducible representations $\Gamma(\mathcal{G})$. The direct
product of the irreducible representations $\Gamma^{(i)}$ and
$\Gamma^{(j)}$ separately spanned by two sets of basis vectors
is given by $\Gamma^{(i)} \otimes \Gamma^{(j)}$. It is reducible
(cf. Eq.~\fmref{eq:coup}) if linear combinations of the product
functions can be found which transform as basis functions for an
irreducible representation. This concept can -- in a one-to-one
correspondence -- be extended to tensor operators since they
behave like the above mentioned functions. As a result, the
direct product of two irreducible tensor operators spanning
$D^{(k_1)}$ and $D^{(k_2)}$ can be decomposed into irreducible
representations spanned by linear combinations of the products
$\ito{T}{k_1}{q_1}\ito{T}{k_2}{q_2}$. The coefficients of these
linear combinations are the Clebsch-Gordan coefficients of
Eq.~\fmref{eq:clebsch}.

Formally, the direct product of two irreducible tensor operators is given by
\begin{equation} \label{eq:prod_ito}
\left\{\vecito{T}{k_1} \otimes \vecito{T}{k_2} \right\}^{(k)}_q
= \sum_{q_1,q_2} C_{q_1 \, q_2 \, q}^{k_1 \, k_2 \, k} \,
\ito{T}{k_1}{q_1} \ito{T}{k_2}{q_2} \ ,
\end{equation}
where possible values of the resulting rank $k$ can be
determined in analogy to the vectorial coupling of spins and are
given by $k=|k_1-k_2|,|k_1-k_2|+1,\dots,k_1+k_2$. Equation
\fmref{eq:prod_ito} is a fundamental expression for the
application of the irreducible tensor operator method within a
numerical exact diagonalization routine. It leads to the desired
formulation of the spin Hamiltonian in terms of irreducible
tensor operators.

As an example, the coupling of the first rank irreducible tensor
operators $\vecito{U}{1}$ and $\vecito{V}{1}$ according to
Eq.~\fmref{eq:prod_ito} shall be presented here. Considering a
compound irreducible tensor operator with $k=0$, the coupling
results in
\begin{equation} \begin{split}
   & \left\{\vecito{U}{1} \otimes \vecito{V}{1} \right\}^{(0)}
   =\\ & \qquad \frac{1}{\sqrt{3}} \left( \ito{U}{1}{1}
   \ito{V}{1}{-1} - \ito{U}{1}{0} \ito{V}{1}{0} + \ito{U}{1}{-1}
   \ito{V}{1}{1} \right) \ , \end{split}
\end{equation}
where for the Clebsch-Gordan coefficients the equation $C_{q_1
\, q_2 \, 0}^{1 \, 1 \, 0} = 1/\sqrt{3} \cdot (-1)^{1-q_1}
\delta_{q_1,q_2}$ was used.\cite{VMK:quantum_theory} Expressing
the spherical components of $\vecito{V}{1}$ and $\vecito{U}{1}$
in terms of the Cartesian components in analogy to
Eq.~\fmref{eq:spintensors} yields
\begin{equation} \begin{split} \label{eq:scalar_prod_tensor}
   & \left\{\vecito{U}{1} \otimes \vecito{V}{1} \right\}^{(0)} =\\
   & \qquad - \frac{1}{\sqrt{3}} \left( \op{U}^x \op{V}^x + \op{U}^y \op{V}^y + \op{U}^z \op{V}^z \right)
   \ , \end{split}
\end{equation}
which is, apart from the prefactor, the scalar product of the
Cartesian vector operators $\vecop{U}$ and $\vecop{V}$.

Finally, the problem of coupling irreducible tensor operators is
identical to the coupling of angular momenta
(cf. App. \ref{sec-A-1}). Thus, from a mathematical point of
view an advantage when using irreducible tensor operators is
that one can adapt the mathematical approaches for coupling
angular momenta and that one can refer to them.

\subsubsection{Matrix elements of irreducible tensor operators} \label{sec-2-2-2}
In the case of an irreducible tensor operator $\vecito{T}{k}$
the matrix elements of this operator can be calculated according
to the \textit{Wigner-Eckart theorem}. It states for matrix
elements with respect to spin states of the form $\ket{\alpha \,
S \, M}$ that
\begin{equation} \begin{split} \label{eq:wigner-eckart}
  &\bra{\alpha \, S \, M} \ito{T}{k}{q} \ket{\alpha' \, S' \, M'} = \\
  & \quad (-1)^{S-M} \reduced{\alpha \, S}{\vecito{T}{k}}{\alpha' \, S'} \threej{S}{k}{S'}{-M}{q}{M'}
  \ .
\end{split} \end{equation}
The matrix element is apart from a phase factor decoupled into a
Wigner-3J symbol and a quantity $\reduced{\alpha \,
S}{\vecito{T}{k}}{\alpha' \, S'}$ called the \textit{reduced
matrix element} of the irreducible tensor operator
$\vecito{T}{k}$. The proof of Eq.~\fmref{eq:wigner-eckart} is
given in standard textbooks about group theory and quantum
mechanics.\cite{Tin:Group_theory,Sil:ito} However, the physical
meaning of this theorem and the consequences for the use within
the irreducible tensor operator method shall be briefly
discussed here.

First of all, it should be mentioned that the Wigner-Eckart
theorem relies on the transformation properties of the appearing
wave functions and operators. Furthermore, since the reduced
matrix element is completely independent of any magnetic quantum
number, the Wigner-Eckart theorem separates the physical part of
the matrix element -- the reduced matrix element -- from the
purely geometrical part reflected by the Wigner-3J symbol. The
value of the reduced matrix element depends on the particular
form of the tensor operator and the states
(cf. App. \ref{sec-A-4}) whereas the Wigner-3J symbol only
depends on rotational symmetry properties. If a zero-rank
($k=0$) irreducible tensor operator is assumed, the Wigner-3J
symbol in Eq.~\fmref{eq:wigner-eckart} directly reflects that
there is no transition between states $\ket{\alpha \, S \, M}$
with different $S$ or $M$. The matrix of the Heisenberg
Hamiltonian from Eq.~\fmref{eq:Heisenberg_WW}, which can be
written as a zero-rank irreducible tensor operator (see
App. \ref{sec-A-3}), therefore takes block-diagonal form without
further calculations.

According to Eq.~\fmref{eq:wigner-eckart} the calculation of the
matrix element of an irreducible tensor operator is directly
related to the calculation of the reduced matrix element of this
operator. The reduced matrix element of an irreducible tensor
operator $\vecito{s}{k}$ with $k=0,1$, acting on a basis
function of a single spin, can be derived from the evaluation of
the Wigner-Eckart theorem. It yields the expressions
\begin{eqnarray} \label{eq:red_spin1}
   \reduced{s}{\vecito{s}{0}}{s} &=& \reduced{s}{\EinsOp}{s} = (2s +1)^\frac{1}{2} \ , \\
   \label{eq:red_spin2}
   \reduced{s}{\vecito{s}{1}}{s} &=& \left[ s \left(s + 1 \right) \left( 2s + 1 \right)\right]^\frac{1}{2}
   \ ,
\end{eqnarray}
where the zero-rank irreducible tensor operator of a single spin
$\vecito{s}{0}$ is given by the unity operator $\EinsOp$ and the
components of $\vecito{s}{1}$ are given by
Eq.~\fmref{eq:spintensors}.

Using Eq.~\fmref{eq:prod_ito} in combination with the
Wigner-Eckart theorem in Eq.~\fmref{eq:wigner-eckart} and a
decomposition of states $\ket{\alpha \, S \, M}$ into product
states according to Eq.~\fmref{eq:clebsch}, one obtains the
expression\cite{Lue:Magnetochemie}
\begin{equation} \begin{split} \label{eq:red_compound}
   & \reduced{\alpha_1 \, s_1 \, \alpha_2 \, s_2 \, S}{\left\{ \vecito{T}{k_1} \otimes \vecito{T}{k_2} \right\}^{(k)}_{q} }{\alpha_1' \, s_1' \, \alpha_2' \, s_2' \, S'} = \\
   & \quad \left[ \left( 2S + 1 \right) \left( 2S' + 1 \right) \left( 2k + 1 \right) \right]^\frac{1}{2} \ninej{s_1}{s_1'}{k_1}{s_2}{s_2'}{k_2}{S}{S'}{k} \\
   & \quad \times \reduced{\alpha_1 \, s_1}{\vecito{T}{k_1}}{\alpha_1' \, s_1'} \reduced{\alpha_2 \, s_2}{\vecito{T}{k_2}}{\alpha_2' \, s_2'}
   \ .
\end{split} \end{equation}
This is the reduced matrix element of a compound irreducible
tensor operator of rank $k$ which consists of the direct product
of two irreducible tensor operators of general ranks $k_1$ and
$k_2$.

Equation \fmref{eq:red_compound} is the basic formula for
calculating reduced matrix elements of irreducible tensor
operators composed of single-spin tensor operators as they
appear in the Heisenberg Hamiltonian. By a successive
application every irreducible tensor operator of that kind can
be decoupled into a series of phase factors, Wigner-9J symbols
and the reduced matrix elements of single-spin tensor operators
(cf. Eq.~\fmref{eq:spintensors}). The successive application of
Eq.~\fmref{eq:red_compound} is often called \textit{decoupling
procedure} since the compound tensor operator that describes the
system under consideration is decoupled so that its reduced
matrix element can be calculated (see App. \ref{sec-A-4}).

\subsection{Point-group symmetries in Heisenberg spin systems} \label{sec-2-3}

\begin{figure}[ht!]
\centering
\includegraphics[width=40mm]{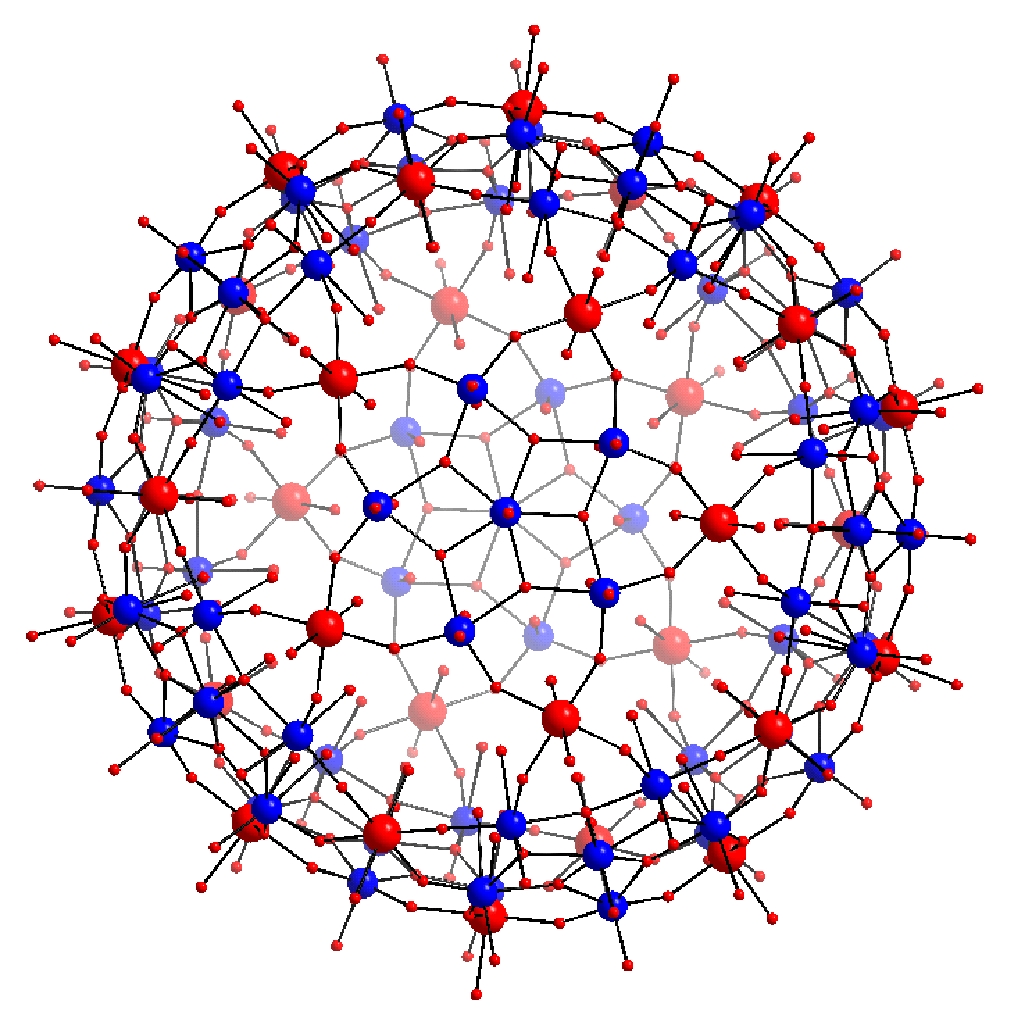}\quad
\includegraphics[width=40mm]{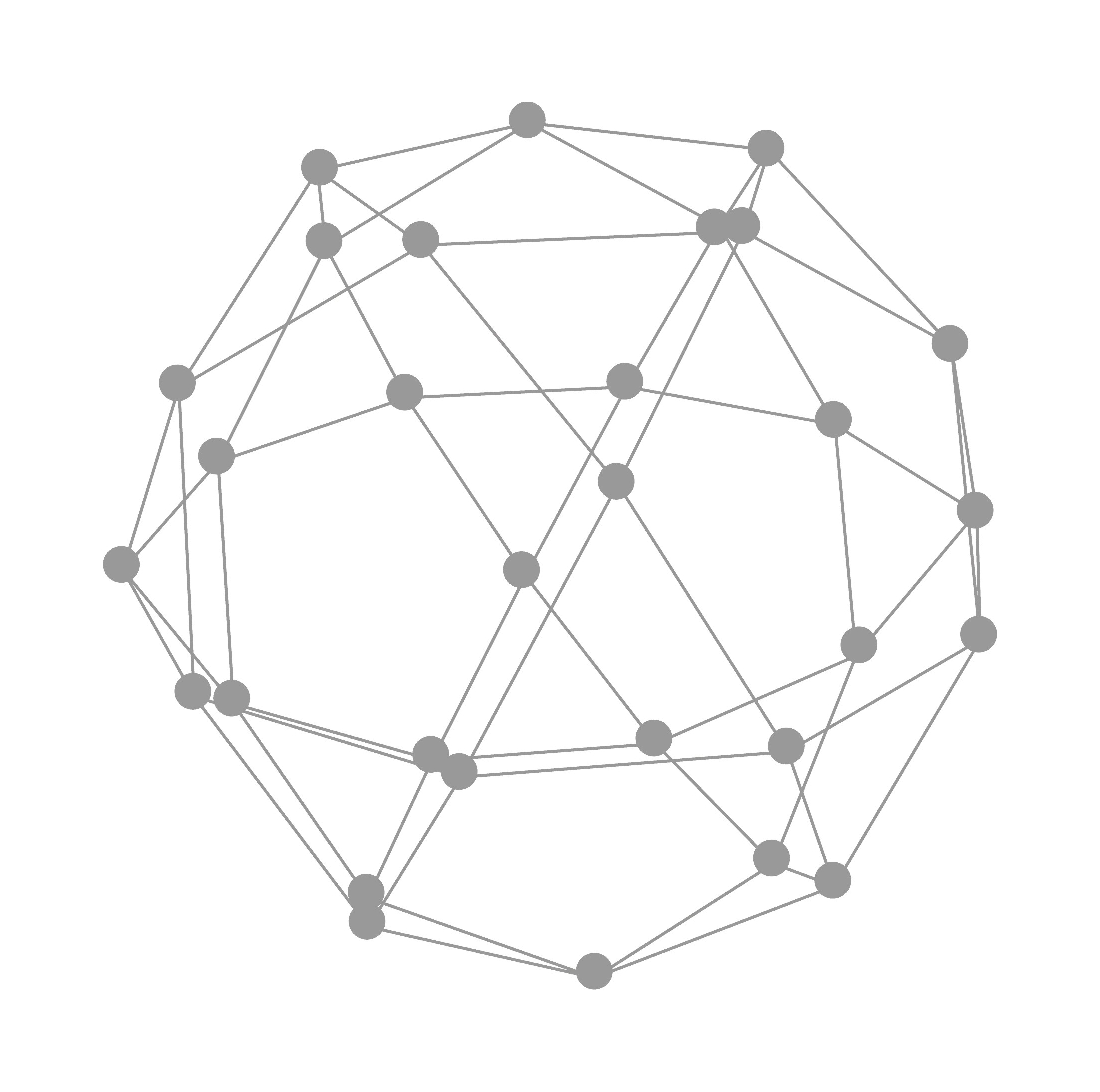}
\caption{Sketch of the \mofe{} molecule (l.h.s., picture taken
  from Ref. \onlinecite{MSS:ACIE99}). The underlying spin system
  exhibits the structure of an icosidodecahedron, i.e. possesses
  icosahedral ($I_h$) point-group symmetry (r.h.s.).} 
\label{fig:beauty}
\end{figure}

Magnetic molecules, for instance those of Archimedian
type,\cite{BGG:JCSDT97,MSS:ACIE99,KTM:DT10} are often -- not
only from a scientific point of view -- perceived to be of a
special beauty (cf. Fig. \ref{fig:beauty}). Certainly, this view
is closely related to the high symmetry which can be found in
the chemical structures and is referred to as point-group
symmetry.\cite{Mue:Sci03} Point-group symmetries do not only
contribute to the beauty of magnetic molecules, but they are
also very instrumental in characterizing the energy levels of
the spectra and thus in extracting physical information from the
underlying spin system. Often a numerical exact diagonalization
remains impossible unless point-group symmetries are used in
order to reduce the dimensionalities of the Hamilton matrices.

It must be emphasized that there is a clear difference between a
group of symmetry operations in real space that map the molecule
on itself and the corresponding group of symmetry operations in
a many-body spin system. Since from a physical point of view a
magnetic molecule is described by a system of interacting spins
with a certain coupling graph (cf. r.h.s of Fig. \ref{F-1}), the
term point-group refers rather to a group of symmetry operations
on this coupling graph than to operations in real space. As a
direct result, the group-theoretical characterization is also
rather based on the symmetry of the coupling graph than on the
molecular symmetry. Of course, since the number of appearing
coupling constants as well as their strengths are estimated from
the chemical structure of the molecule, there is a close
connection between the symmetries of the molecule and the
corresponding coupling graph but not necessarily a one-to-one
correspondence.

In Heisenberg spin systems point-group symmetries can be
included by mapping the symmetry operations on spin
permutations. In order to emphasize this, the term
spin-permutational symmetry instead of point-group symmetry is
often used. However, in the work at hand the term point-group
symmetry is used although the symmetry is always incorporated by
mapping the point-group operations on spin permutations. In
context to this, it has to be mentioned that in systems which
include anisotropies the incorporation of point-group symmetries
is much more complicated since rotations in real space have to
be performed.\cite{Lue:Magnetochemie,GHK:IC09,Sch:CMP09,TBS:DT09}

\begin{figure}[ht!]
\centering \includegraphics[width=80mm]{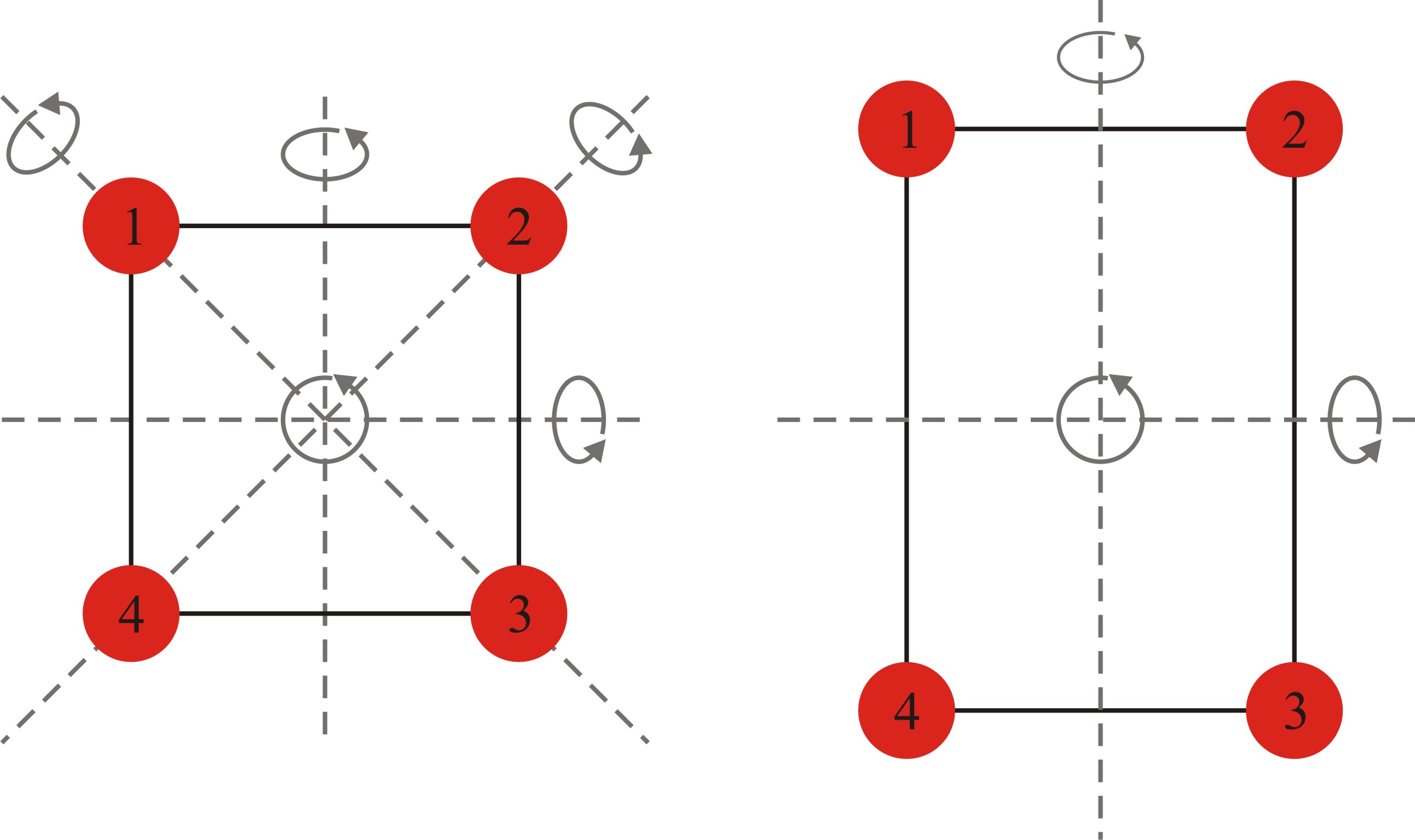}
\caption{Coupling graph of a square of identical spins $s$ with
$D_4$ symmetry axes (l.h.s) and a rectangle with $D_2$ symmetry
axes (r.h.s.).}
\label{F-3}
\end{figure}

Figure \ref{F-3} exemplarily shows the coupling graphs of a
square and a rectangle consisting of identical spins $s$ with
the symmetry axes of $D_4$ and $D_2$ being indicated. The
rectangle can be seen as resulting from the square -- a system
of four interacting spins with only one coupling constant $J$ --
by introducing a second coupling constant of a different
strength $J'$ between spin pairs \mbox{$<1,4>$} and
\mbox{$<2,3>$}. In this case the coupling strength is indicated
by the length of the coupling paths between the spins. The
introduction of the second coupling constant then results in a
reduction of the point-group symmetry from $D_4$ to $D_2$.

The point-group operations on the spin system can be identified
with permutations of the spins that leave the Hamiltonian
invariant.\cite{Gri:SB72} Such a permutation is represented by
an operator $\op{G}(R_i)$ of the point-group $\mathcal{G}$ for
which the commutation relation
\begin{equation}
   \com{\op{H}_\text{Heisenberg}}{\op{G}(R_i)}=0
\end{equation}
holds, where $i=1,\dots,h$ numbers the symmetry operations up to
the order $h$ of $\mathcal{G}$. 

The theory of group representations now provides the theoretical
background for the use of point-group symmetries. The
irreducible representations $\Gamma^{(n)}$ of a point-group
$\mathcal{G}$ can be used to classify the energy eigenstates of
$\op{H}_\text{Heisenberg}$ and to block-factorize the Hamilton
matrices. The dimensionalities of the resulting subspaces,
i.e. blocks, can be calculated with only little information. The
irreducible matrix representations ${\bf{\Gamma}}^{(n)}(R_i)$ of
the group elements, i.e. the permutation operators, are somewhat
arbitrary concerning the choice of the underlying basis. Thus,
these elements are represented by their character, i.e. the
trace of the particular matrix representation. The character
$\chi^{(n)}(R_i)$ is invariant under unitary transformations and
is in general given by
\begin{equation}
   \chi^{(n)}(R) = \text{Tr} \, {\bf{\Gamma}}^{(n)}(R) =
   \sum_{\lambda=1}^{l_n} \Gamma^{(n)}(R)_{\lambda \lambda} 
   \ ,
\end{equation}
where $l_n$ denotes the dimension of the $n$-th irreducible
representation $\Gamma^{(n)}$. 

A given character table of $\mathcal{G}$ now enables one to
calculate the dimensions of the resulting blocks within the
Hamilton matrix, i.e. the dimensions of the subspaces
$\mathcal{H}(S,M,\Gamma^{(n)})$. Character tables for various
point-groups can be found in almost every textbook about the
theory of group representations. The authors would like to refer
to these concerning the construction of character
tables.\cite{Tin:Group_theory,Wig:group,Cot:group}

In order to calculate the dimensions of the subspaces
$\mathcal{H}(S,M,\Gamma^{(n)})$, the reducible matrix
representations ${\bf{G}}(R_i)$ of the operators $\op{G}(R_i)$
have to be considered. With $l,m = 1,\dots,\text{dim} \,
\mathcal{H}(S, M)$ the matrix elements of these matrices are
given by
\begin{equation*}
   G(R_i)_{lm}=\bra{\alpha_l \, S \, M}\op{G}(R_i)\ket{\alpha_m \, S \, M}
   \ ,
\end{equation*}
where the subscripts attached to $\alpha$ indicate that specific
basis states are considered. The decomposition of the character
$\chi (R_i)=\sum_{ll}G(R_i)_{ll}$ with respect to the
irreducible representations $n$ of $\mathcal{G}$ then yields
\begin{equation} \label{eq:factor_repr}
   \chi(R_i)=\sum_{n} a_n \chi^{(n)}(R_i)
   \ .
\end{equation}

Using the great orthogonality theorem\cite{Tin:Group_theory},
the above equation results in the expression 
\begin{equation} \label{eq:dim_irreps} \begin{split}
   a_n &= \text{dim} \, \mathcal{H}(S,M,\Gamma^{(n)}) \\
   & = \frac{1}{h} \sum_{k} N_k [\chi^{(n)}(\mathcal{C}_k)]^*
   \chi(\mathcal{C}_k) 
   \ ,
\end{split}
\end{equation}
where $a_n$ gives the multiplicity of the irreducible
representation $\Gamma^{(n)}$ that is contained in the reducible
representation ${\bf{G}}(\mathcal{C}_k)$. $\mathcal{C}$ refers
to the classes which the group elements can be divided in. Each
class contains equivalent operations, for example n-fold
rotations, which are linked by the same group operation and thus
hold the same character. $N_k$ denotes the number of elements of
$\mathcal{C}_k$.

From Eq.~\fmref{eq:dim_irreps} the dimensions of the subspaces
$\mathcal{H}(S,M,\Gamma^{(n)})$ can be calculated, but no
information about the basis states that span these Hilbert
spaces is given. The required symmetrized basis states can be
determined by the application of the projection
operator\cite{Tin:Group_theory}
\begin{equation} \label{eq:projection_op}
   \mathcal{P}^{(n)}_{\kappa \kappa} = \frac{l_n}{h} \sum_{i}
   [\Gamma^{(n)}(R_i)_{\kappa \kappa}]^* \op{G}(R_i) 
   \ .
\end{equation}
This operator projects out that part of a function $\ket{\phi}$
which belongs to the $\kappa$-th row of the irreducible
representation $\Gamma^{(n)}$. A subsequent application of the
operator in Eq.~\fmref{eq:projection_op} on basis states, for
example of the form $\ket{\alpha \, S \, M}$, is called the
\textit{basis-function generating
machine}.\cite{Tin:Group_theory}

Although Eq.~\fmref{eq:projection_op} provides the information
required to construct symmetrized basis states that serve as a
basis for the irreducible representations of $\mathcal{G}$, it
is important to notice that the matrices of $\Gamma^{(n)}$ have
to be known completely. Of course, this does not cause a problem
as long as $\mathcal{G}$ entirely consists of one-dimensional
irreducible representations. However, if more-dimensional
irreducible representations appear, it is often more convenient
to use the projection operator
\begin{equation} \label{eq:projection_op2}
   \mathcal{P}^{(n)} = \sum_\kappa \mathcal{P}^{(n)}_{\kappa \kappa}  = \frac{l_n}{h} \sum_{i} [\chi^{(n)}(R_i)]^* \op{G}(R_i)
\end{equation}
that only requires information which can be extracted from the
character table of $\mathcal{G}$. The operator of
Eq.~\fmref{eq:projection_op2} projects out that part of a
function $\ket{\alpha \, S \, M}$ which belongs to the
irreducible representation $\Gamma^{(n)}$, irrespective of the
row. As a consequence one has to orthonormalize resulting
functions -- for example by a \textit{Gram-Schmidt
orthonormalization} -- in order to obtain the correct
symmetrized basis functions (cf. App. \ref{sec-A-5}).

\subsection{Point-group symmetry operations acting on vector-coupling states} \label{sec-2-4}
In the previous section \ref{sec-2-3} some general remarks about
the use of point-group symmetries have already been made. In
order to transform the Hamilton matrix to a block-diagonal form
with respect to irreducible representations $\Gamma^{(n)}$ of a
point group $\mathcal{G}$, one has to construct symmetrized
basis functions $\ket{\alpha \, S \, M \,
\Gamma^{(n)}}$. Equation \fmref{eq:projection_op2} provides the
projection operator that projects out that part of a state which
belongs to the $n$-th irreducible representation of
$\mathcal{G}$. However, the main challenge when constructing
symmetrized basis states is to find an expression for the action
of the operators $\op{G}(R_i)$ on states of the form
$\ket{\alpha \, S \, M}$.

The above mentioned operators $\op{G}(R_i)$ which correspond to
operations on the coupling graph (cf. Fig. \ref{F-3}) can be
defined by their action on the product basis composed of the
single-spin eigenstates of $\op{s}^z(i)$. The states of the
product basis of $N$ identical spins can be denoted by
\begin{equation*}
   \ket{s_1 m_1} \otimes \ket{s_2 m_2} \otimes \dots \otimes \ket{s_N m_N} \equiv \ket{m_1 \, m_2 \, \dots \, m_N}
\end{equation*}
and fulfill the eigenvalue equation
\begin{equation*}
   \op{s}^{z}(i) \ \ket{m_1 \, \dots \, m_i \, \dots \, m_N} = m_i \ \ket{m_1 \, \dots \, m_i \, \dots \, m_N}
\end{equation*}
according to their definition.

Now, in a spin system consisting of $N$ spins an operator
$\op{G}(R)$ is considered that is given in the form
$\op{G}(\pi(1) \, \pi(2) \, \dots \, \pi(N))$ and describes a
point-group operation. The notation $\op{G}(\pi)$ indicates for
all $i=1,\dots,N$ that the point-group operation interchanges
the spin at site $i$ with the spin at site $\pi(i)$. The action
of the operator $\op{G}(R)$ on the product basis is given by
\begin{equation} \label{eq:Wirkung_Produktbasis}
   \op{G}(R) \ \ket{m_1 \, m_2 \, \dots \, m_N} = \ket{m_{\pi(1)} \, m_{\pi(2)} \, \dots \, m_{\pi(N)}}.
\end{equation}
In Eq.~\fmref{eq:Wirkung_Produktbasis} the following happens: by
the action of $\op{G}(R)$ the single-spin system at site $i$
takes the $z$-component which the system at site $\pi(i)$ has
taken previously. The result is a different state of the product
basis with permuted $m$-values.

The operators $\op{G}(R)$ are only defined by their action on
product states whereas the details of constructing symmetrized
basis states that are linear combinations of vector-coupling
states of the form $\ket{\alpha \, S \, M}$ are still
unknown. How to find an expression for the action of $\op{G}(R)$
on these vector-coupling states is -- to some extent -- shown in
Refs. \onlinecite{Wal:PRB00} and \onlinecite{SBO:JPA07}.

However, a slightly different perspective concerned with the
application of general symmetry operations to a vector-coupling
basis shall be presented here (for further technical details see
App. \ref{sec-A-5}). In order to clarify the action of an
operator representing a point-group operation on a
vector-coupling state, the states are labeled according to the
coupling scheme they are belonging to. Additionally, the set of
quantum numbers referring to the coupling scheme is abbreviated
by Greek letters. Primes indicate different basis states within
the same coupling scheme. With these conventions one obtains the
following general result for a transition between two coupling
schemes $a$ and $b$ which is induced by an operator $\op{G}$
representing a point-group operation:
\begin{equation} \label{eq:point-group-operation}
    \op{G} \ \ket{\alpha \, S \, M}_{a} = \delta_{\alpha,\beta} \ \ket{\beta \, S \, M}_{b}
    \ .
\end{equation}
The Kronecker symbol $ \delta_{\alpha,\beta}$ on the right hand
side indicates that the values of the spin quantum numbers of
the different sets $\alpha$ and $\beta$ are the
same. Re-expressing the right hand side of
Eq.~\fmref{eq:point-group-operation} within states belonging to
the coupling scheme $a$, i.e. inserting a suitable form of the
identity operator $\EinsOp$, directly leads to the very
important final result
\begin{equation} \label{eq:point-group-operation_final} \begin{split}
    & \op{G} \ \ket{\alpha \, S \, M}_{a} = \\
    & \qquad \sum_{\alpha'} \delta_{\alpha,\beta} \  _{a}\braket{\alpha' \, S \, M}{\beta \, S \, M}_{b} \ \ket{\alpha' \, S \, M}_a
\end{split} \end{equation}
which contains so-called \textit{general recoupling
coefficients} $_{a}\braket{\alpha \, S \, M}{\beta \, S \,
M}_{b}$. A general recoupling coefficient can be seen as a
scalar product between vector-coupling states belonging to
different coupling schemes $a$ and $b$.

It is by no means trivial to find an expression for a recoupling
coefficient relating different coupling schemes if more than
three or four spins are present. By definition the expressions
for the elements of the transition matrix relating two different
coupling schemes result in Wigner-nJ symbols
(cf. App. \ref{sec-A-1}). While for three interacting spins
Wigner-6J symbols occur, the size $n$ of these Wigner
coefficients increases with every additional spin taken into
account. For four interacting spins the recoupling coefficient
is expressed by a Wigner-9J symbol and for five interacting
spins the recoupling is described by Wigner-12J symbols.

From a computational point of view it turns out to be
unfavorable to use algebraic expressions for higher symbols than
Wigner-9J symbols although there exist expressions for 12J
symbols and 15J symbols.\cite{VMK:quantum_theory} In order to
find an effective way to describe general recoupling
coefficients, one should use expressions in which only Wigner-6J
symbols appear. These 6J symbols can be calculated using the
formula\cite{Rah:PR42}
\begin{equation} \begin{split} \label{eq:racah_formula}
   & \sixj{j_1}{j_2}{j_3}{J_1}{J_2}{J_3} =   \Delta(j_1,j_2,j_3) \Delta(j_1,J_2,J_3) \\
   & \times \Delta(J_1,j_2,J_3) \Delta(J_1,J_2,j_3) \sum_t \frac{(-1)^t (t+1)!}{f(t)}
   \ ,
\end{split}
\end{equation}
where the \textit{triangle coefficient} $\Delta(a,b,c)$ reads
\begin{equation*}
   \Delta(a,b,c) = \left( \frac{(a+b-c)! \, (a-b+c)! \, (-a+b+c)!)}{(a+b+c+1)!} \right)^{\frac{1}{2}} .
\end{equation*}
The function $f(t)$ in Eq.~\fmref{eq:racah_formula} is given by
\begin{equation*} \begin{split}
   f(t) &= (t-j_1-j_2-j_3)! \, (t-j_1-J_2-J_3)! \, \\
   & \times (t-J_1-j_2-J_3)! \, (t-J_1-J_2-j_3)! \, \\
   & \times (j_1+j_2+J_1+J_2-t)! \, (j_2+j_3+J_2+J_3-t)! \, \\
   & \times (j_3+j_1+J_3+J_1-t)!
   \ .
   \end{split}
\end{equation*}
The sum in Eq.~\fmref{eq:racah_formula} is running over
nonnegative integer values of $t$ for which no factorial in
$f(t)$ has a negative argument. Since even the evaluation of
Wigner-6J symbols is a rather involved task as can be seen from
Eq.~\fmref{eq:racah_formula}, it is helpful to analyze the
symmetry properties of the appearing symbols
(cf. Fig. \ref{fig:6J-symmetries}) in order to reduce the
computational effort when constructing symmetrized basis
states. In this regard, when expressing the action of a
point-group operation on a basis state according to
Eq.~\fmref{eq:point-group-operation_final}, only those
recoupling coefficients have to be calculated which are
non-zero.

So far, nothing has been said about the generation of a formula
that describes a general recoupling coefficient. Finding a
formula which only contains Wigner-6J symbols and some phase
factors can most efficiently be done by using graph theory (see
App. \ref{sec-B}). However, the choice of the coupling scheme
which underlies the construction process of the basis states is
essential for an effective computational realization of the
numerical exact diagonalization. In this regard the invariance
of the coupling scheme under the applied point-group operations
is a desired property (see Apps. \ref{sec-A-5} and
\ref{sec-A-6}).

\section{Applications} \label{sec-3}

In this section we like to present three applications for
realistic spin systems of unprecedented size that can be treated
using irreducible tensor operator techniques and point-group
symmetries: a cuboctahedron of $N=12$ spins of spin quantum
number $s=3/2$ (Hilbert space dimension 16,777,216), an
icosahedron of $N=12$ spins of spin quantum number $s=3/2$ (same
dimension), and a spin ring of $N=10$ and $s=5/2$ (Hilbert space
dimension 60,466,176) known as the ferric wheel
Fe$_{10}$.\cite{TDP:JACS94} 

\begin{figure}[ht!]
\centering
\includegraphics*[clip,width=35mm]{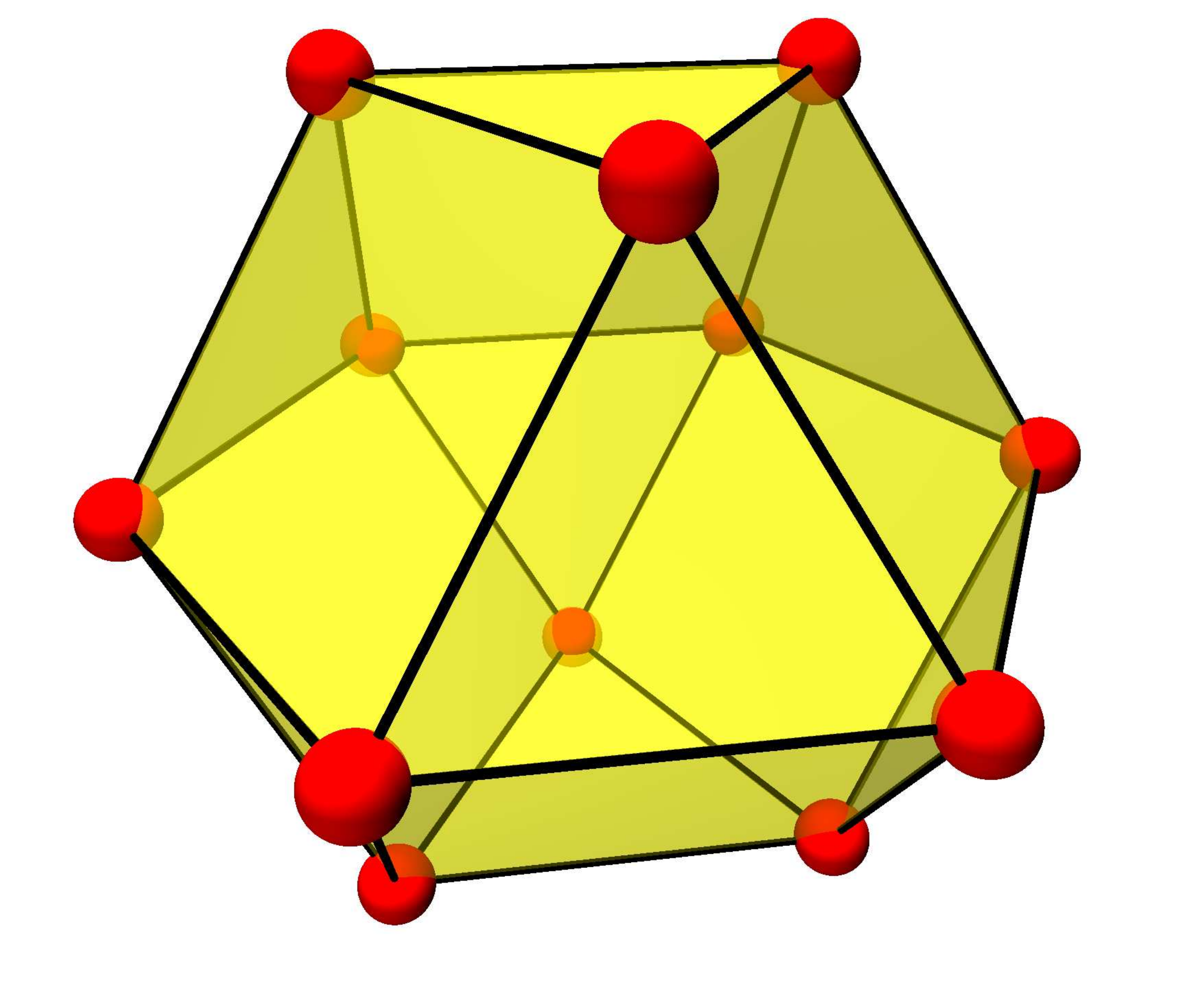}
\includegraphics*[clip,width=45mm]{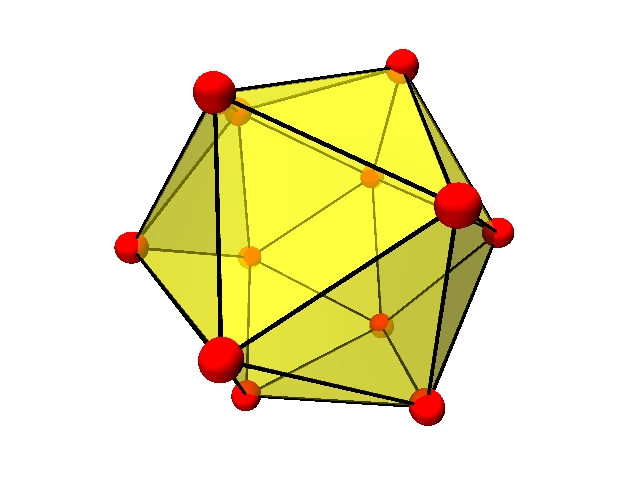}
\caption{Structure of the cuboctahedron (left) and the
  icosahedron (right).}
\label{dt-schnack-F-4}
\end{figure}

It is very important to note that for the case of
antiferromagnetic coupling the cuboctahedron as well as the
icosahedron, see \figref{dt-schnack-F-4}, belong to the class of
geometrically frustrated spin
systems\cite{SSR:JMMM05,SSS:PRL05,SSR:PRB07,ScS:P09} and are
thus hardly accessible by means of Quantum Monte Carlo (QMC)
calculations due to a so-called negative-sign
problem.\cite{SaK:PRB91,San:PRB99,EnL:PRB06} Complete
diagonalization is therefore the only way to study the interesting
behavior of such systems both as function of field and
temperature.\cite{RLM:PRB08,HoZ:JPCS09,KTM:DT10,Sch:DT10}

Density Matrix Renormalization Group (DMRG) techniques provide a
very powerful approximation mainly for one-dimensional spin
systems such as chains.\cite{Whi:PRB93,ExS:PRB03,Sch:RMP05} The
method delivers the relative ground states for orthogonal
subspaces. Extensions to include the approximate evaluation of
excitations have been developed recently,\cite{Jec:PRB02} but are
still primarily useful for one-dimensional systems. For the
example of Fe$_{10}$, this method could provide lowest levels
for each total magnetic quantum number $M$ and thus one could
evaluate the crossing fields of the lowest levels. More or less
the same holds true for the Lanczos diagonalization
technique,\cite{Lan:JRNBS50} with which again extremal energy
eigenvalues and eigenstates can be evaluated. Quantum Monte
Carlo would be able to evaluate observables for Fe$_{10}$, since
this ten-site system is non-frustrated. Nevertheless, QMC cannot
determine higher lying energy levels which would be crucial for
e.g. inelastic neutron scattering studies.

Since the three investigated systems exhibit a highly
symmetric structure they can be modeled with just one nearest
neighbor interaction, therefore the Heisenberg Hamiltonian
simplifies to
\begin{equation} \label{eq:Heisenberg_WW}
   \op{H}_\text{Heisenberg} = -J \sum_{<i,j>} \vecop{s}(i) \cdot
   \vecop{s}(j)
\ .
\end{equation}
The summation is running over pairs $<i,j>$ of nearest neighbors
of single-spin vector operators $\vecop{s}$ at sites $i$ and $j$
counting each pair only once.  For the following examples $J<0$
is assumed.

\subsection{The cuboctahedron}

The magnetism of antiferromagnetically coupled and geometrically
frustrated spin systems is a fascinating subject due to the
richness of phenomena that are
observed.\cite{Ram:ARMS94,Gre:JMC01} One of the most famous spin
systems is the two-dimensional kagom\'{e} lattice.
\cite{Gre:JMC01,Diep94,NKH:EPL04,Zhi:PRL02,SHS:PRL02,Atw:NM02}
It is very interesting and from the point of theoretical
modeling appealing that similar but zero-dimensional spin
systems -- in the form of magnetic
molecules\cite{BGG:JCSDT97,MSS:ACIE99,MTS:AC05,TMB:ACIE07,PLK:CC07}
-- exist that potentially could show many of the special
features of geometrically frustrated antiferromagnets. The
cuboctahedron which consists of squares surrounded by triangles
serves as one zero-dimensional ``little brother" of the kagom\'{e}
antiferromagnet; the icosidodecahedron, which consists of
pentagons surrounded by triangles, is another one. Such finite
size antiferromagnets offer the possibility to discover and
understand properties that are shared by the infinitely extended
lattices. An example is the discovery of localized independent
magnons,\cite{SHS:PRL02,SSR:EPJB01} which explain the unusual
magnetization jump at the saturation field. Also the plateau at
$1/3$ of the saturation magnetization that appears in systems
built of corner sharing triangles could be more deeply
investigated by looking at the cuboctahedron and the
icosidodecahedron.\cite{SNS:PRL05,RLM:PRB08}

\begin{figure}[ht!] \centering
   \includegraphics[width=60mm]{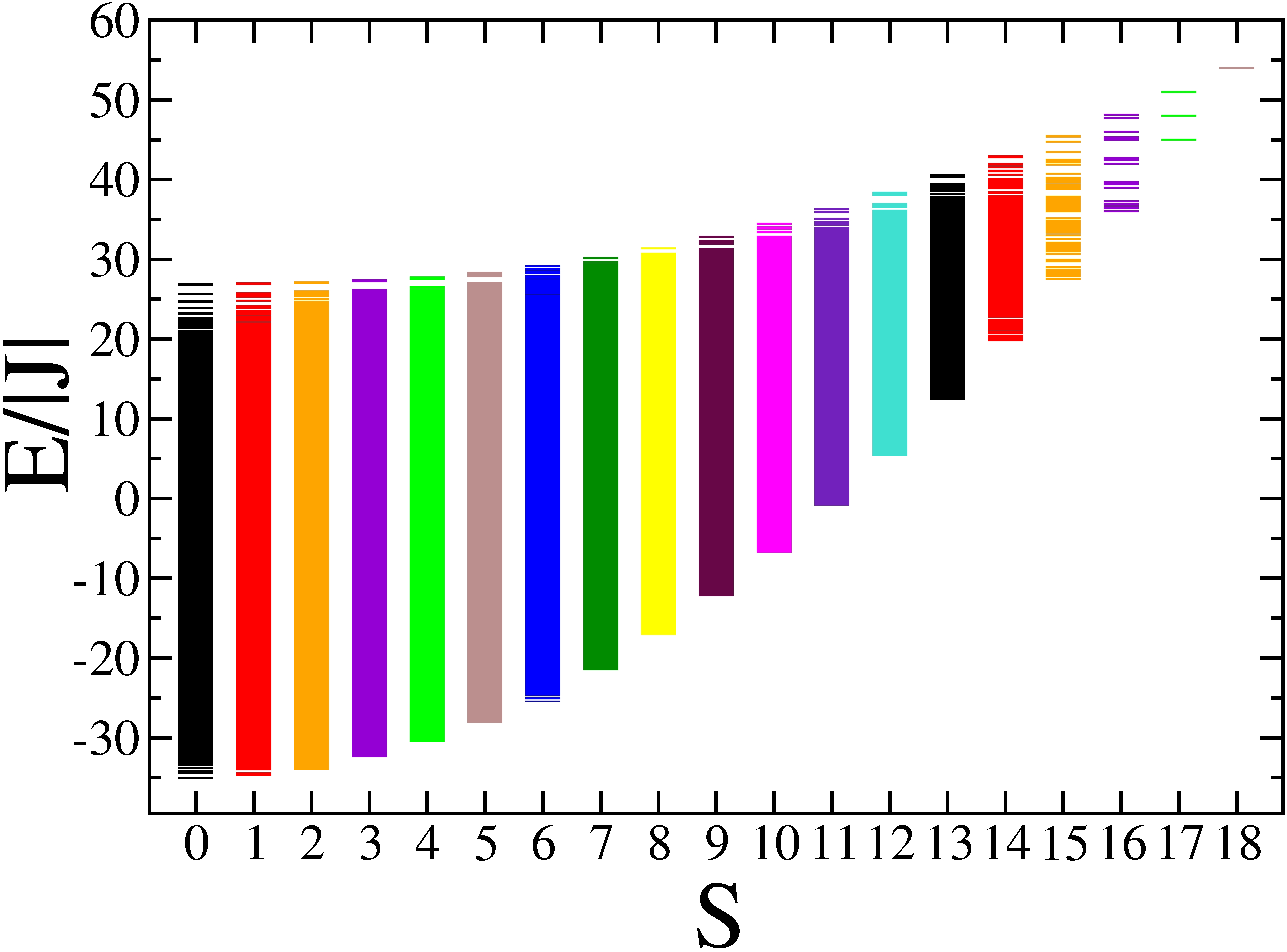}
\caption{Full energy spectrum of a cuboctahedron with
  $s=3/2$. The energy levels have been calculated using $D_2$
  point-group symmetry.}
\label{fig:cubo-3_full} 
\end{figure}

\begin{figure}[ht!] \centering
   \includegraphics[width=60mm]{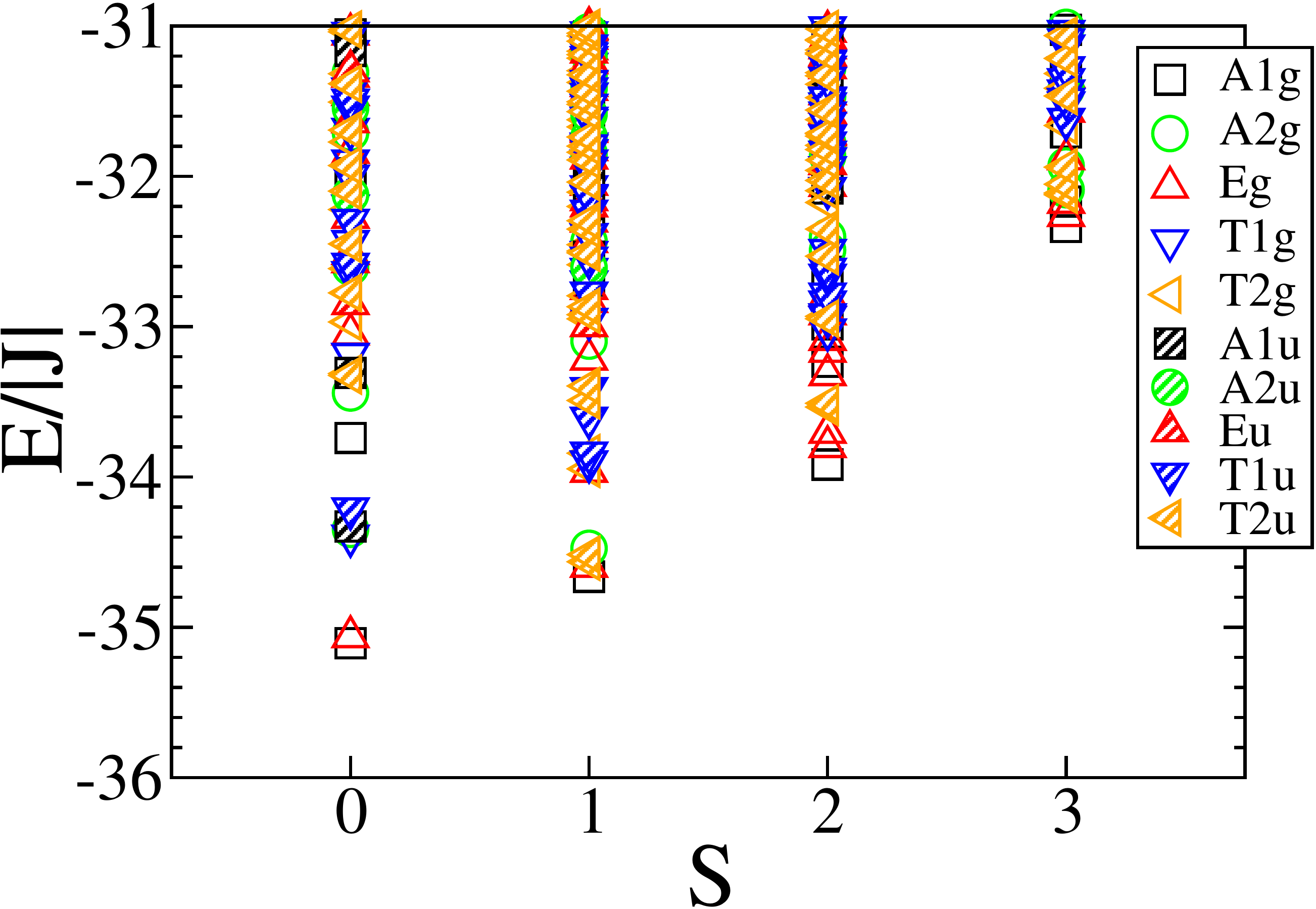}
\caption{Low-lying part of the energy spectrum of a
  cuboctahedron $s=3/2$. The energy levels are labeled
  according to irreducible representations of the full
  octahedral group $O_h$.} 
\label{fig:cubo-3_Oh} 
\end{figure}

The energy levels of the cuboctahedron ($N=12$, $s=3/2$, Hilbert
space dimension 16,777,216), shown in \figref{fig:cubo-3_full},
could be numerically evaluated using the $D_2$ point-group
symmetry. For low-lying sectors of $S=0,1,2,3$ we also
determined the energy levels according to irreducible
representations of the full octahedral group $O_h$, see
\figref{fig:cubo-3_Oh}. One feature that can be clearly seen in
\figref{fig:cubo-3_Oh} is the existence of an additional
low-lying singlet below the first triplet.

\begin{figure}[ht!]
\centering
\includegraphics[clip,width=60mm]{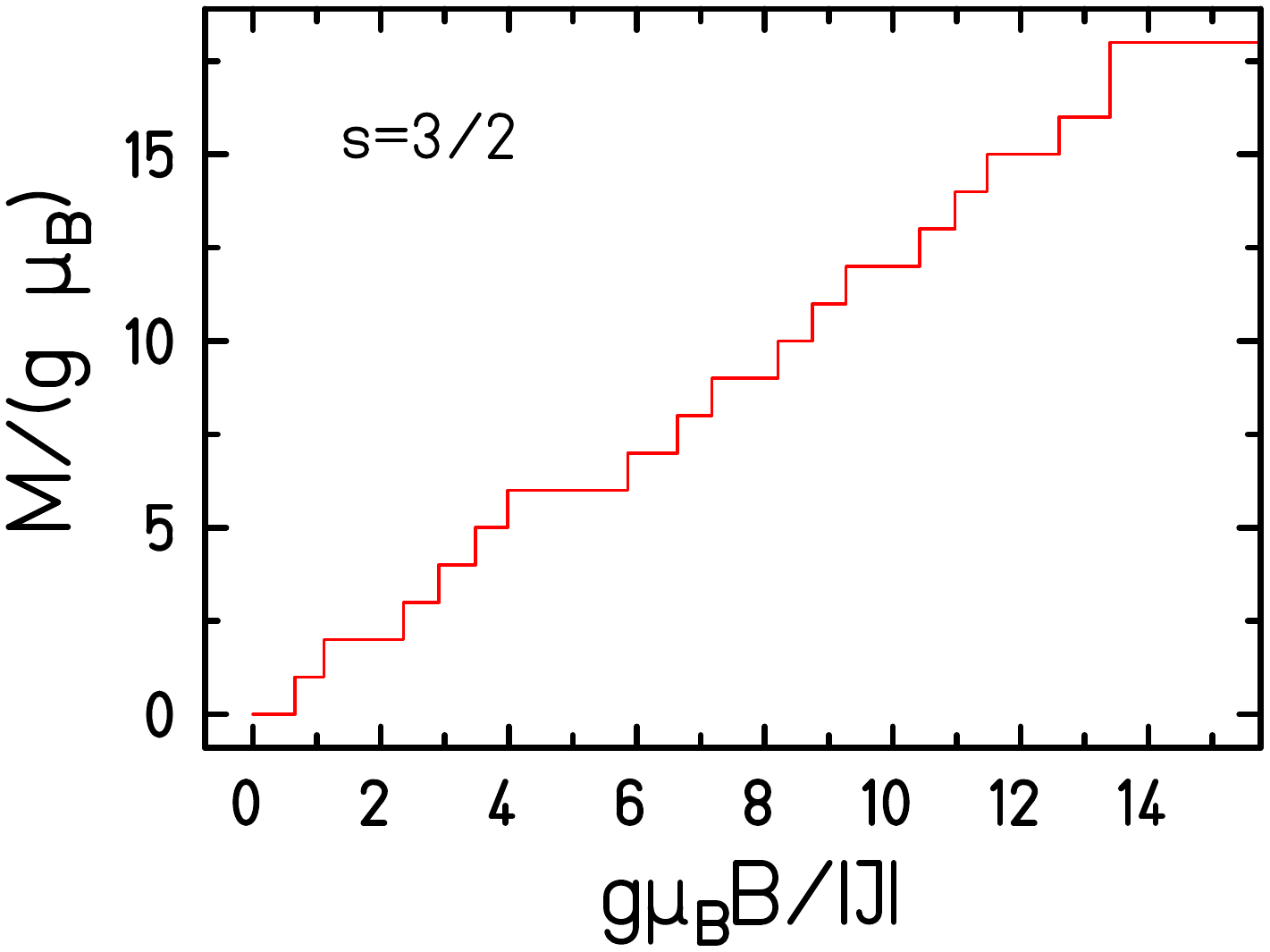}
\caption{Magnetization as a function of applied field at $T=0$
  for the regular cuboctahedron with $s=3/2$. The extended
  plateau at $1/3$ of the saturation magnetization is clearly
  visible.} 
\label{F-cu-2}
\end{figure}

Figure~\xref{F-cu-2} shows the magnetization curve at $T=0$ for
the regular cuboctahedron with $s=3/2$. This curve shows the
aforementioned plateau at $1/3$ of the saturation magnetization
and a jump to saturation of height $\Delta M = 2$.

\begin{figure}[ht!]
\centering
\includegraphics[clip,width=60mm]{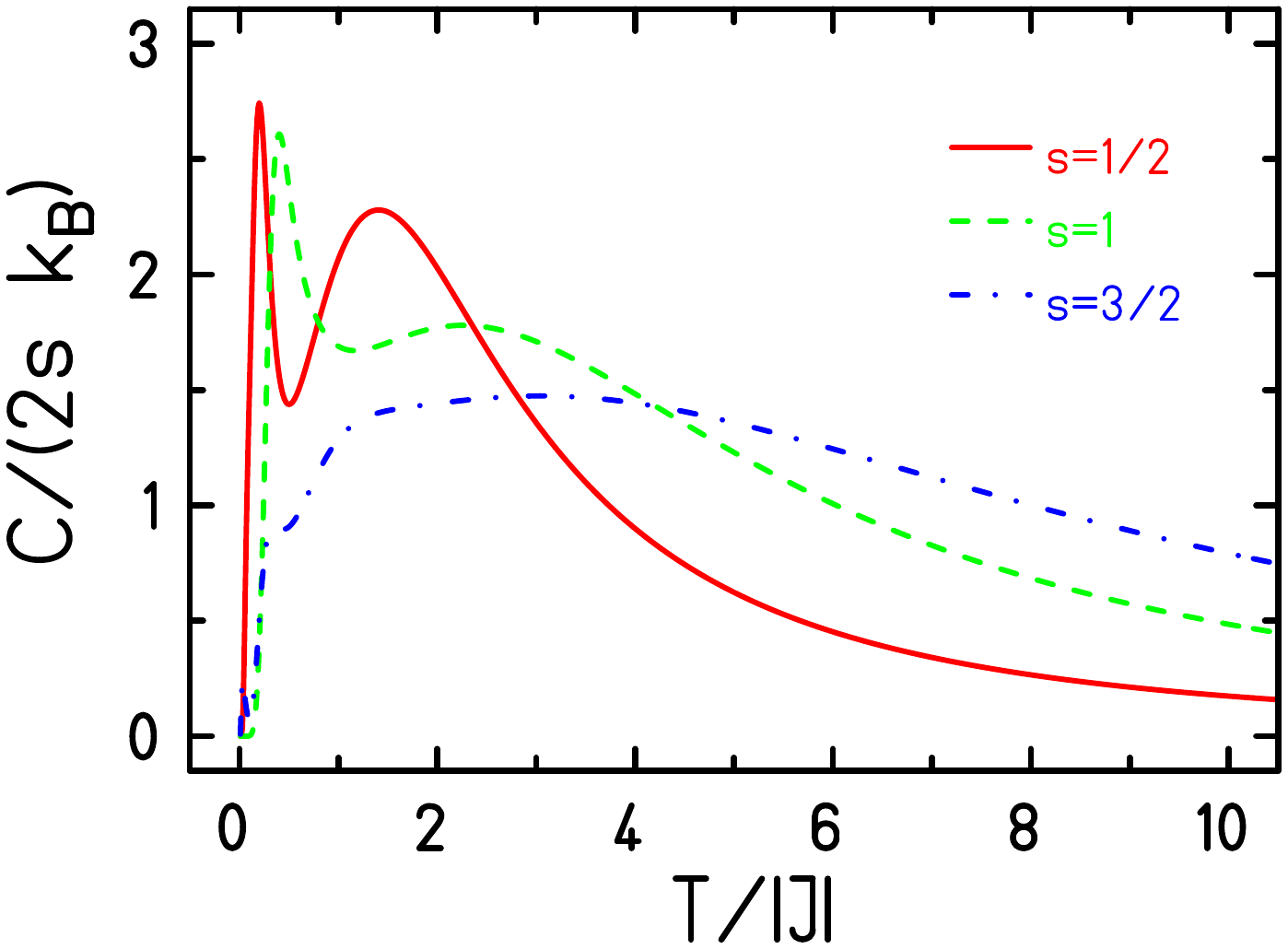}

\includegraphics[clip,width=60mm]{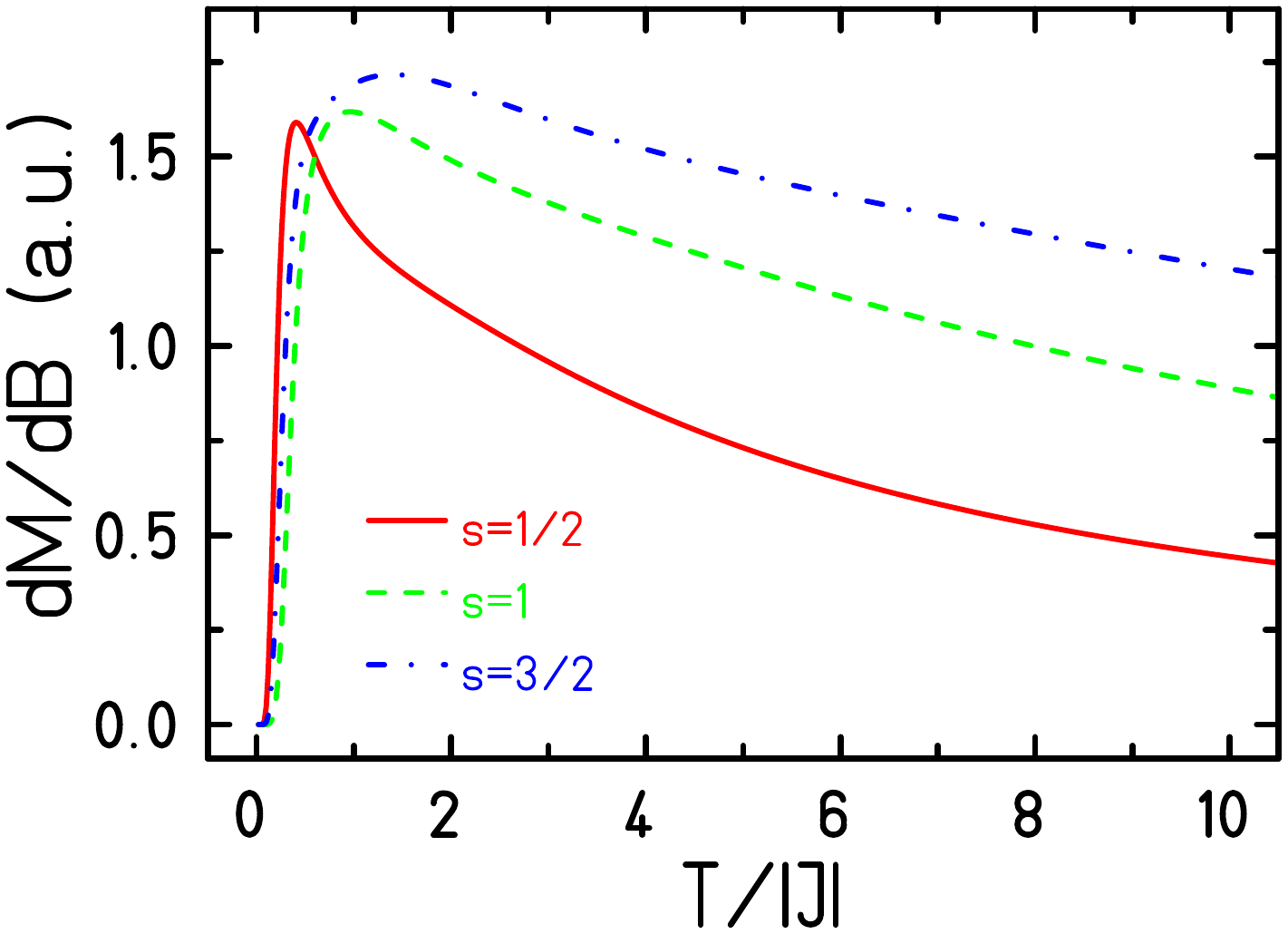}
\caption{Zero-field heat capacity (top) and zero-field susceptibility
  (bottom) for the regular cuboctahedron with $s=1/2$, $s=1$,
  and $s=3/2$.}
\label{F-4}
\end{figure}

Figure  \xref{F-4}  compares the  heat  capacity  (top) and  the
zero-field susceptibility (bottom) for the regular cuboctahedron
with  $s=1/2$, $s=1$,  and $s=3/2$.  The heat  capacity  shows a
pronounced  double peak  structure for  $s=1/2$ and  $s=1$ which
dissolves  into  a  broad  peak  with  increasing  spin  quantum
number. The  broad peak also  moves to higher  temperatures with
increasing  $s$.  The  reason   for  the  first  sharp  peak  is
twofold.  Since there  are  several gaps  between the  low-lying
levels the density of  states has a very discontinuous structure
which  results in the  double peak  structure.  For  $s=1/2$ the
low-lying singlets provide a very low-lying non-magnetic density
of states which is responsible for the fact that the first sharp
peak is at such low temperatures. For $s=1$ the first sharp peak
results  from both  excited singlet  as well  as  lowest triplet
levels. For $s=3/2$  a remnant of the first  sharp peak is still
visible; it is  given by the low-lying singlets,  but since they
are so few,  also influenced by the lowest  triplet levels.  The
behavior   of   the  heat   capacity   is   contrasted  by   the
susceptibility,  bottom of  \figref{F-4}, which  reflects mostly
the  density of  states of  magnetic levels  and is  only weakly
influenced  by low-lying  singlets. Therefore,  the  first sharp
peak, or any other structure at very low temperatures, is
absent.

\subsection{The icosahedron}

A spin system where the spins are mounted at the vertices of an
icosahedron and interact antiferromagnetically along the edges
seems to be rather appealing since it exhibits unusual
frustration properties such as metamagnetic phase
transitions.\cite{CoT:PRL92,SSS:PRL05,Kon:PRB05,Kon:PRB07}
Unfortunately, it appears to be challenging to synthesize such
structures although icosahedra are rather stable cluster
configurations.\cite{TEM:CEJ06} 

\begin{figure}[ht!]
\centering
\includegraphics[clip,width=60mm]{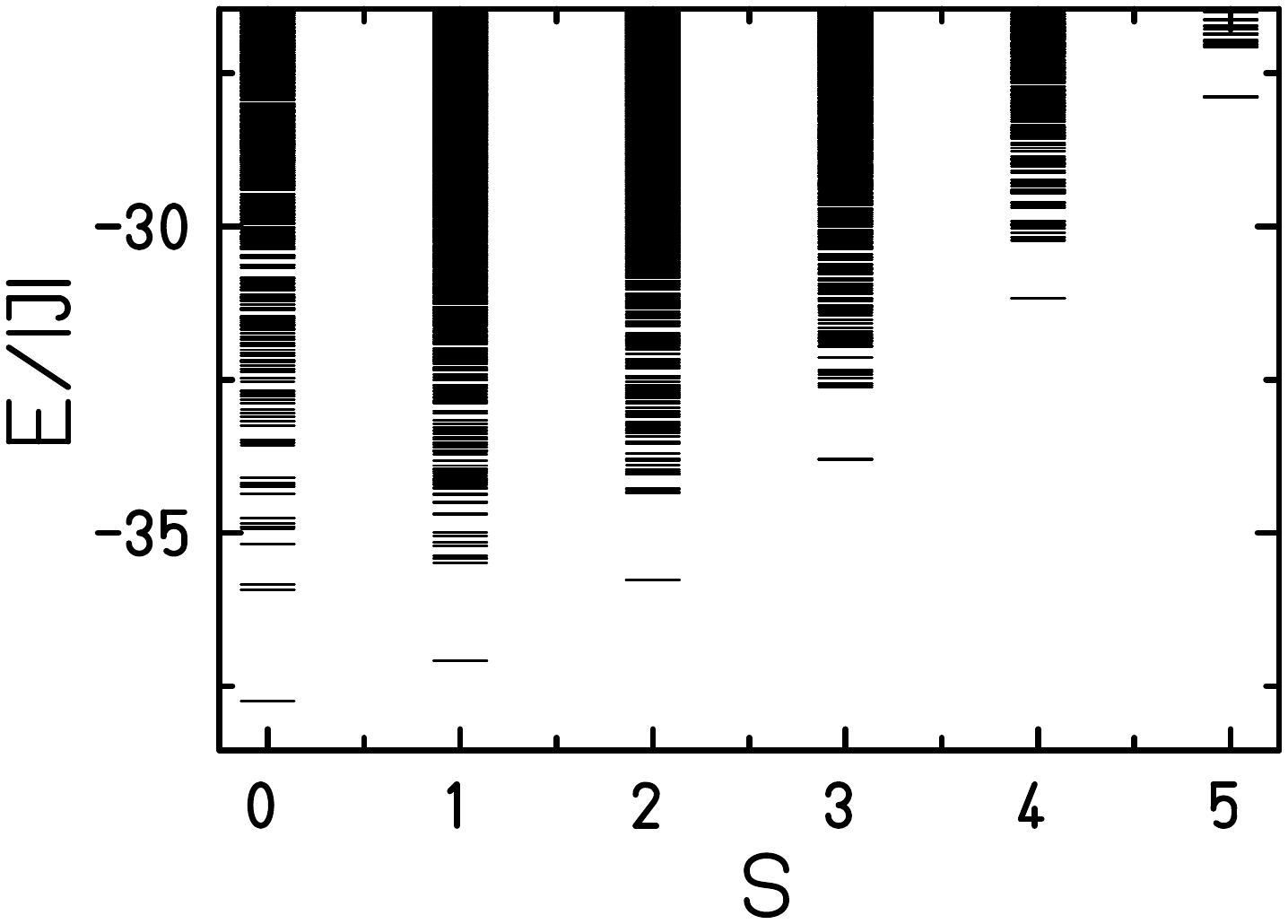}
\caption{Low-lying energy levels of an antiferromagnetically
  coupled icosahedron $N=12$ and $s=3/2$.}
\label{F-I-A}
\end{figure}

Here we investigate the properties of an icosahedron ($N=12$)
with single spin quantum number $s=3/2$ (Hilbert space dimension
16,777,216). The complete set of energy eigenvalues has been
determined using $D_2$ symmetry. Figure~\ref{F-I-A} displays the
low-lying part of all levels. Looking at the data file it turns
out that very many (really unusually many) levels are highly
degenerate, which is in accordance with earlier investigations
of icosahedra with smaller single spin quantum
number.\cite{Kon:PRB05,SSR:PRB07} We find that for every sector
of total spin $S$ and total magnetic quantum number $M$ the
irreducible representations $A_2$, $B_1$, and $B_2$ contain
exactly the same energy eigenvalues, whereas $A_1$ is
different. The non-degenerate ground state belongs to $A_1$. In
addition very often near degeneracies occur.

\begin{figure}[ht!]
\centering
\includegraphics[clip,width=60mm]{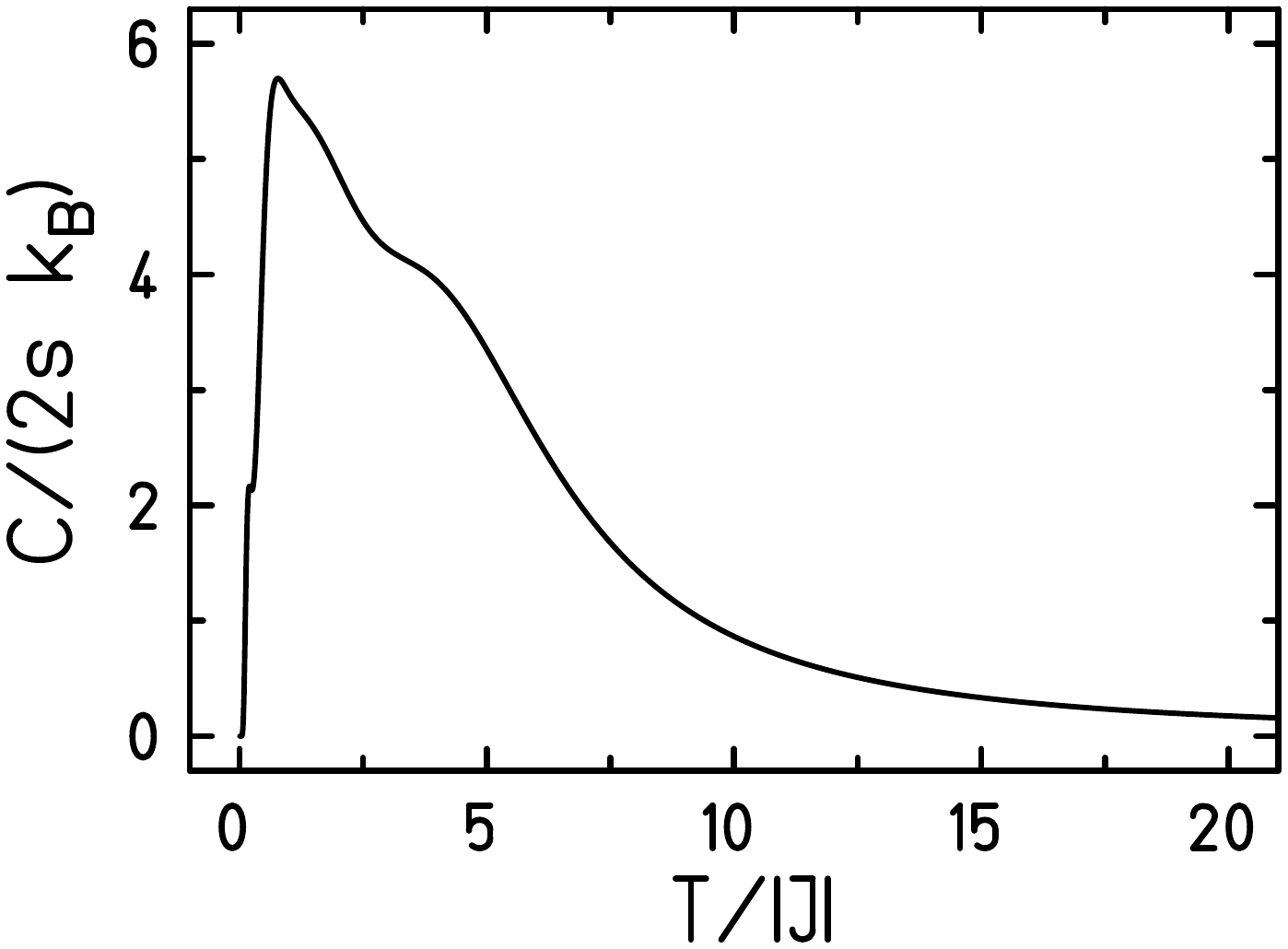}

\includegraphics[clip,width=60mm]{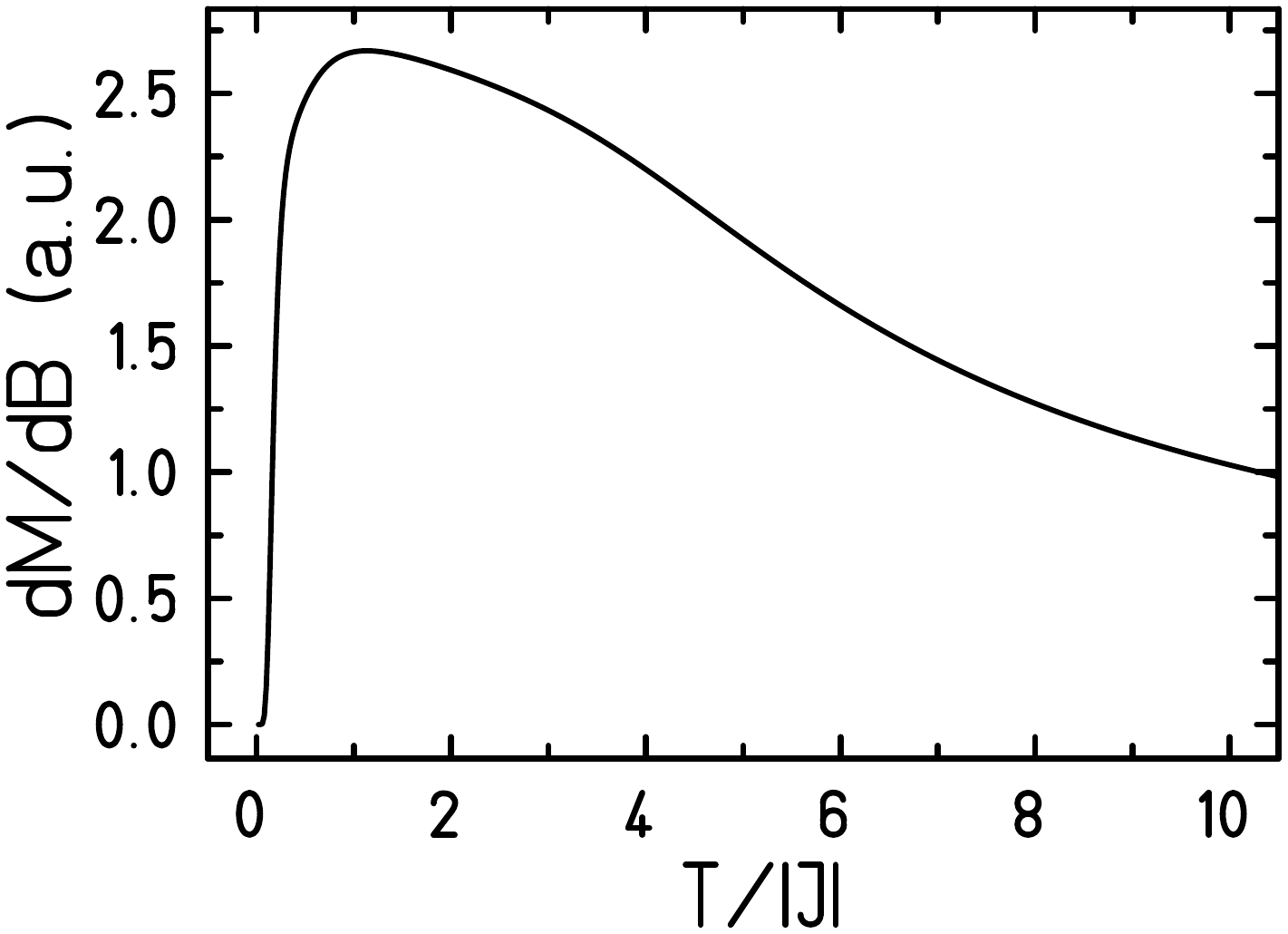}
\caption{Zero-field heat capacity (top) and zero-field
  susceptibility (bottom) for the regular icosahedron with
  $s=3/2$.}
\label{F-I-B}
\end{figure}

Figure~\ref{F-I-B} shows the related zero-field heat capacity
(top) and zero-field susceptibility (bottom). While the
susceptibility does not look so unusual compared to other spin
structures, the heat capacity appears to be really weired. Half
way up the initial rise at very low temperatures there is a
Schottky-like peak that stems from the slightly split ground
state ($S=0$) levels. The further rise is due to the fact that
the lowest $2\times 3$ degenerate $(S=1)$ level is energetically
rather close. Although higher-lying levels are separated by gaps
from the lowest levels they also contribute to the heat capacity
due to their massive degeneracy. Altogether this results in a
low-temperature heat capacity that is much larger than the heat
capacity of the cuboctahedron, compare \figref{F-4}.

\begin{figure}[ht!] \centering
   \includegraphics[width=60mm]{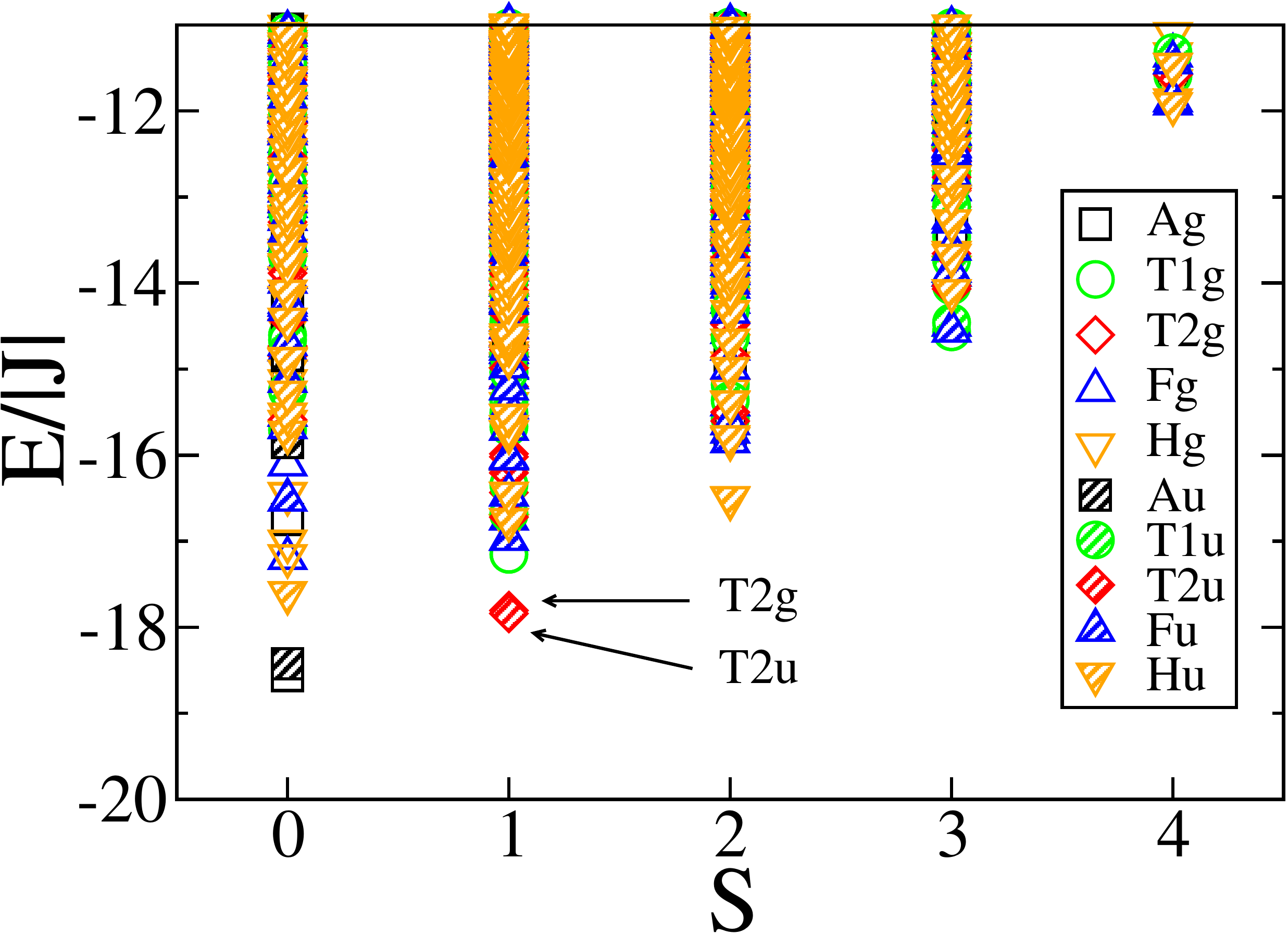}
\caption{Low-lying part of the energy spectrum of an
  icosahedron with $s=1$. The energy levels are labeled
  according to irreducible representations of the full
  icosahedral group $I_h$.} 
\label{F-I-C} 
\end{figure}

The icosahedron may also serve as an example for the technical
complexity due to the evaluation of recoupling
coefficients. When combining $I_h$ symmetry with $SU(2)$ many
different recoupling formulas have to be generated (119 for the
120 group operations minus identity to be precise). This renders
a treatment of the $s=3/2$ icosahedron in $I_h$
impossible. Although the sizes of the Hamilton matrices for the
irreducible representations would be small, it is the
construction of basis states that turns the evaluation of the
needed matrix elements into a very lengthy procedure. Even in a
parallelized job on 256 cores on a supercomputer this task could
not be completed. For the $s=1$ icosahedron we could finish a
decomposition into irreducible representations of $I_h$, but --
to give an example -- the construction and subsequent
diagonalization of the largest subspace, which has only a
dimension of 3315, took about three days on 128 brand new
Nehalem processors. Figure~\xref{F-I-C} shows the low-lying part
of the energy spectrum of an icosahedron with $s=1$. The energy
levels are labeled according to irreducible representations of
the full icosahedral group $I_h$. The exact and near
degeneracies that have been discussed above for the case of
$s=3/2$ and $D_2$ can now be resolved. For instance, the lowest
$S=1$ level belongs to $T_{2u}$ and is thus threefold
degenerate. It is split from the higher-lying $T_{2g}$ by only
$3/1000|J|$, which in any calculation or measurement would look
like a six-fold degeneracy, see also
Ref.~\onlinecite{Kon:PRB05}.

\subsection{Rings}

Molecular ferric wheels are among the very first magnetic
molecules. It appears that they can be synthesized in many even
numbered sizes, e.g. $N=6,8,\dots,18$ Fe(III)
spins.\cite{TDP:JACS94,GCR:S94,CCF:CEJ96,SBU:ACIE:97,ACC:ICA00,SCA:PRB05}
Since the spin of the Fe(III) ions is $s=5/2$ the Hilbert space
dimension grows rapidly from one ring size to the next. For the
ferric wheel Fe$_{10}$ it reaches already 60,466,176 rendering
a complete diagonalization impossible, at least in the
past.\cite{TDP:JACS94} Based on the observation that the lowest
energy eigenvalues $E_{\text{min}}(S)$ obey a quadratic
dependence with respect to total
spin,\cite{JJL:PRL99,ScL:PRB00,Wal:PRB01,WAL:PRB07,ScS:PRB09,SLS:CMP09}
which is Lande's interval rule, approximations could be derived
for the level crossing fields\cite{JJL:PRL99} as well as for
low-lying excitations in for instance the truly giant ferric
wheel Fe$_{18}$. \cite{WSC:PRL09} As mentioned earlier, QMC is
also capable of evaluating observables for even-membered
unfrustrated spin rings.\cite{EnL:PRB06} Nevertheless, none of
the methods is able to determine higher-lying states.

\begin{figure}[ht!]
\centering
\includegraphics[clip,width=60mm]{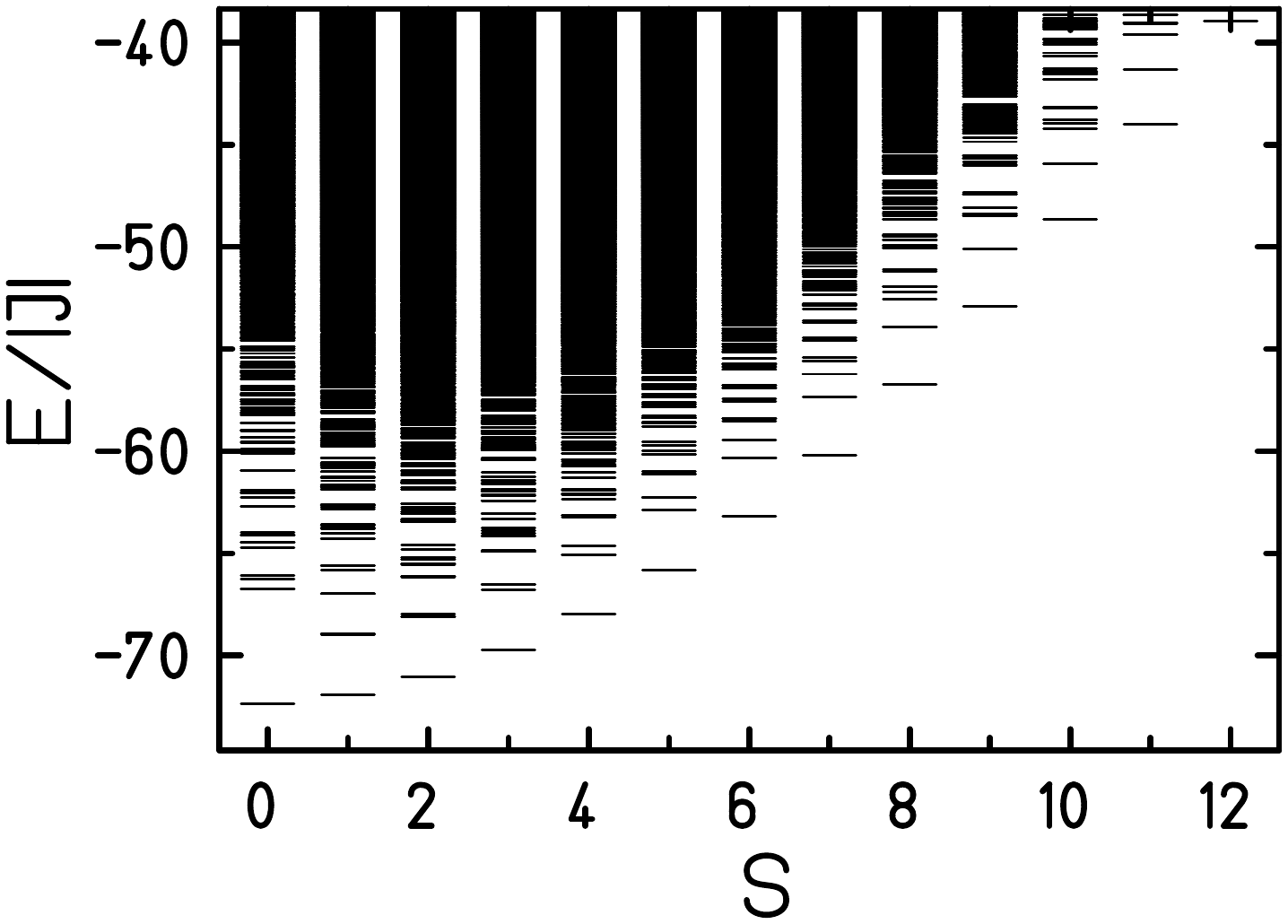}
\caption{Low-lying energy levels of an antiferromagnetically
  coupled spin ring with $N=10$ and $s=5/2$.}
\label{F-Fe10-A}
\end{figure}

In the following we present the first complete diagonalization
study of a spin ring similar to Fe$_{10}$, i.e. with $N=10$
and $s=5/2$. This enables us to subsequently evaluate all
thermodynamic functions, all excited levels, and if needed even
the evolution for time-dependent observables. For the
diagonalization we used only the $D_2$ symmetry because it
reduces the matrices already sufficiently and allows a faster
computation of recoupling coefficients and thus matrix elements
compared to the $C_{10}$ symmetry.

\begin{figure}[ht!]
\centering
\includegraphics[clip,width=60mm]{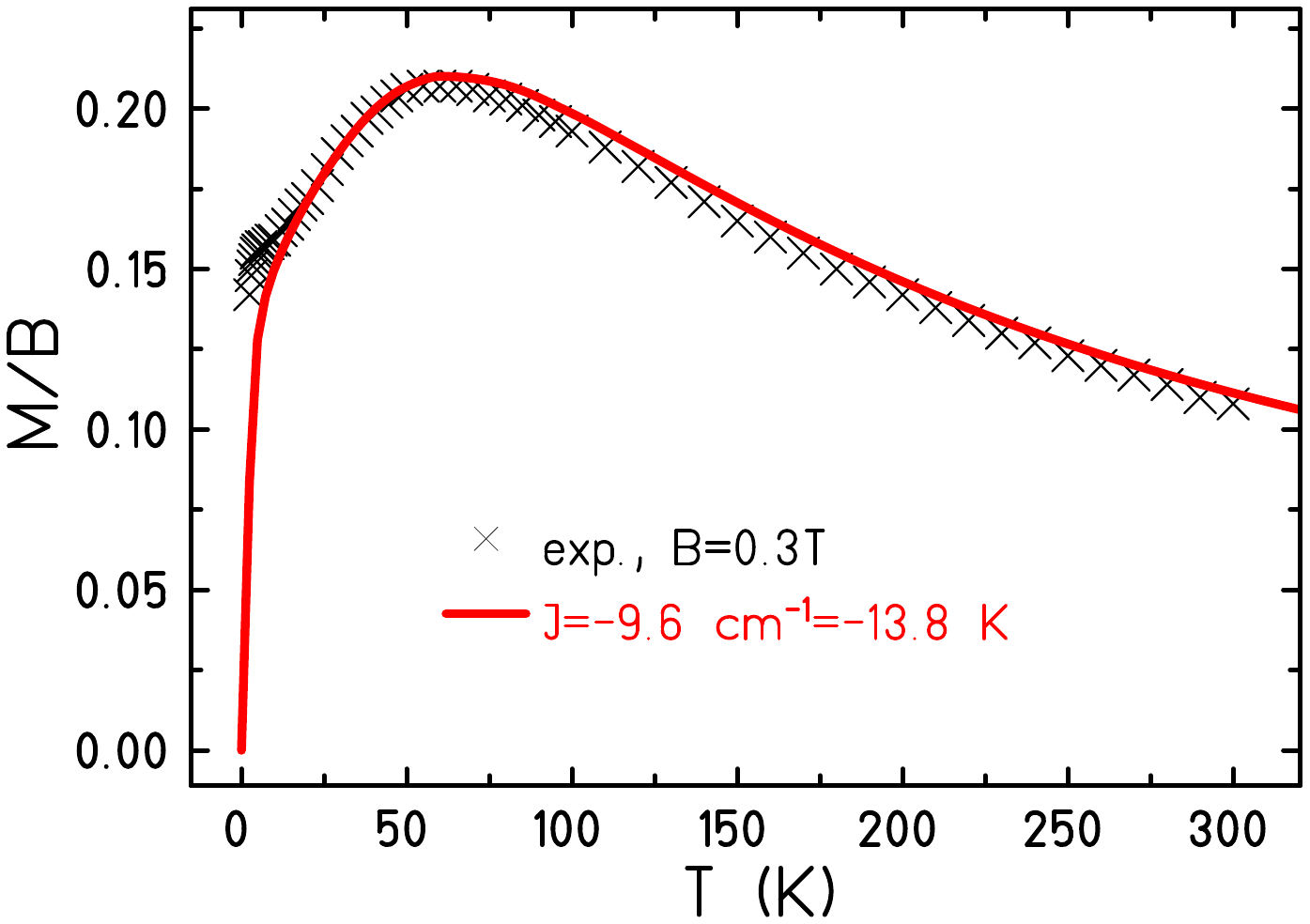}

\includegraphics[clip,width=60mm]{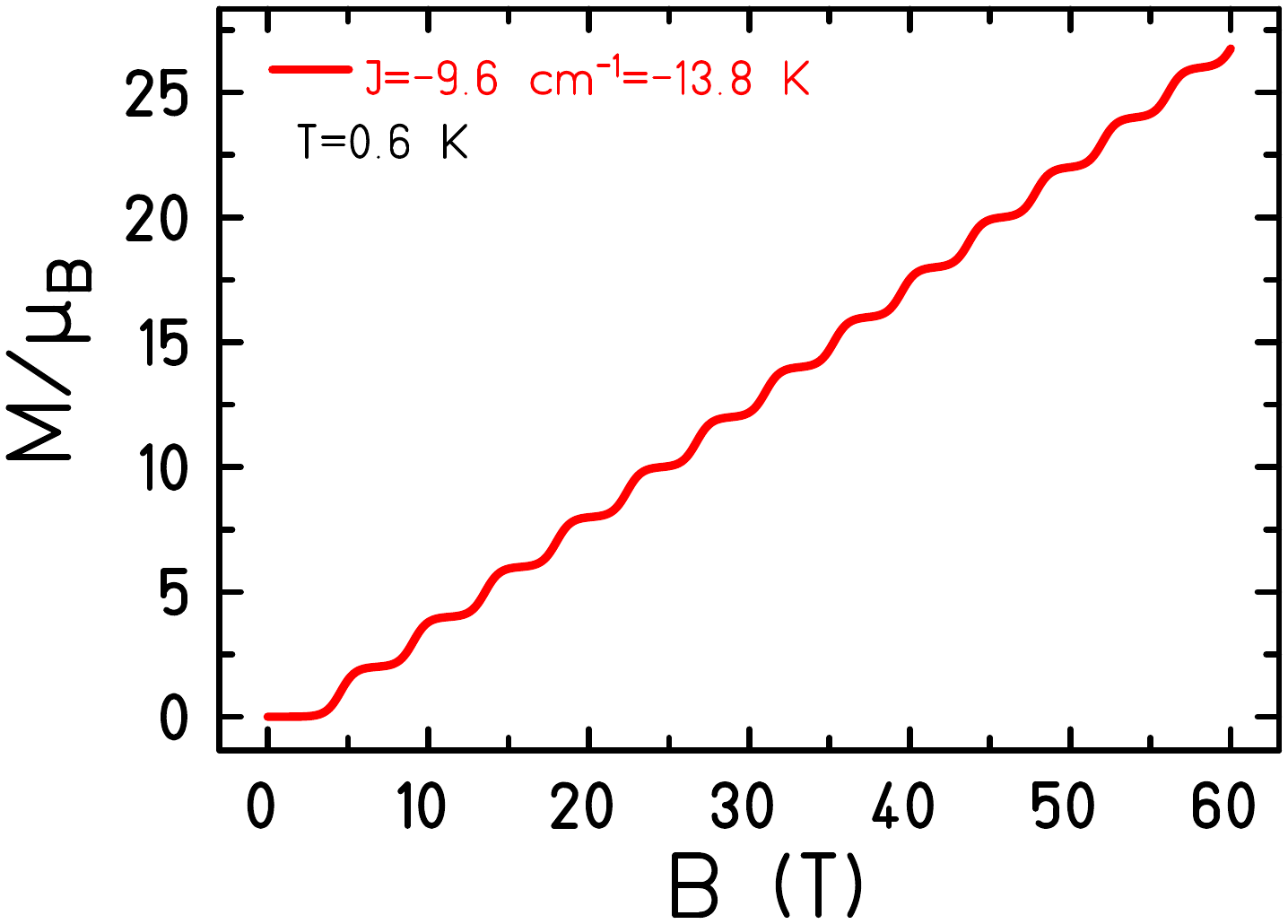}
\caption{Susceptibility (top) and magnetization (bottom) of an
  antiferromagnetically coupled spin ring with $N=10$ and
  $s=5/2$. The exchange parameter $J=-9.6$~cm$^{-1}$ as well as
  the susceptibility data are taken from
  Ref.~\onlinecite{TDP:JACS94}.}  
\label{F-Fe10-B}
\end{figure}

Figure \ref{F-Fe10-A} displays the low-lying energy levels for
various sectors of total spin $S$. The rotational band structure
of energy levels, which is at the heart of the aforementioned
approximation, is clearly visible. Having obtained the
eigenvalues an evaluation of the magnetization as function of
temperature and field is easily possible. Figure \ref{F-Fe10-B}
shows the susceptibility as function of temperature (top) and
the magnetization as function of applied field (bottom) of an
antiferromagnetically coupled spin ring with $N=10$ and
$s=5/2$. The exchange parameter $J=-9.6$~cm$^{-1}$ as well as
the susceptibility data are taken from
Ref.~\onlinecite{TDP:JACS94}. $g=2.0$ was assumed for the
calculation as in the original paper. Since the theoretical
susceptibility curve is now known exactly, a detailed discussion
of the experimental data becomes possible. As conjectured
already in Ref.~\onlinecite{TDP:JACS94} the low-temperature hump
of the susceptibility is not a feature of the ring
Hamiltonian. We found numerically that it can also not be
produced by a (reasonable) next-nearest neighbor
interaction. The authors of the experimental paper suggested
that it might be produced by dimers of Fe(III), that should also
contribute (slightly) to the large temperature behavior, where
the agreement with the theoretical curve is also not
perfect. Our obtained energy eigenvalues would allow now to set
up a model that contains the ring and an unknown amount of
dimer impurities. Open parameters would then be the amount of
impurity and the $g$ value which very likely deviates slightly
from 2.

\section{Outlook}\label{sec-4}

In this review we have undertaken the attempt to explain how
numerical approaches to diagonalize the Heisenberg Hamiltonian
work, that employ the spin-rotational symmetry $SU(2)$ in
combination with point-group symmetries. We hope that we could
provide the reader with detailed knowledge on how to set up such
a diagonalization scheme, especially through the extended
technical appendix.

What remains open for the future is to develop efficient schemes
to evaluate recoupling coefficients, which at the present stage
can be a very time consuming procedure that sometimes renders
the calculation of matrix elements practically impossible
although the complete matrix would be rather small in the end. A
natural next step consists in finding optimal coupling schemes
for a given molecule and point-group symmetry.

%
\section*{Acknowledgments}
We thank Dante Gatteschi, Boris Tsukerblat, Martin H{\"o}ck, and
J{\"o}rg Ummethum for carefully reading the manuscript. This
work was supported by the German Science Foundation (DFG)
through the research group 945 and a Ph.D. program of the State
of Lower Saxony at Osnabr\"uck University. Computing time at the
Leibniz Computing Center in Garching is also gratefully
acknowledged. Last but not least we like to thank the State of
North Rhine-Westphalia and the DFG for financing our local SMP
supercomputer as well as the companies BULL and ScaleMP for
constant support.

%

\appendix

\section{Realization of the irreducible tensor operator technique} \label{sec-A}
In this section the theoretical foundations presented in
Sec. \ref{sec-2} shall be specified. To this end, after having
developed a basic idea of the coupling of angular momenta a spin
square deals as a small example system. It is demonstrated how
an appropriate basis can be constructed and how the Heisenberg
Hamiltonian looks like when expressed in terms of irreducible
tensor operators. Further on, it is shown how the matrix elements
can be evaluated by decoupling the irreducible tensor operator
that describes the system. A central aspect is the use of
point-group symmetries in combination with irreducible tensor
operators that leads to the occurrence of general recoupling
coefficients.

\subsection{Coupling of angular momenta and Wigner-nJ symbols} \label{sec-A-1}
As a first step the coupling of two general angular momenta
$\vecop{j}_1$ and $\vecop{j}_2$ shall be discussed. Regarding
this, the work at hand mainly refers to the use of definitions
and name conventions that have been introduced by
Wigner.\cite{Wig:group}

The vector coupling rule for the addition of angular momenta,
known from elementary atomic physics, states that the resulting
angular momenta $\vecop{J}$ can be characterized by a quantum
number $J$. $J$ assumes all values
\begin{equation} \label{eq:vector_addition_rule}
   |j_1 - j_2|,|j_1-j_2|+1,\dots, j_1+j_2 -1, j_1 + j_2 \ ,
\end{equation}
where $j_1$ and $j_2$ denote the quantum numbers of the angular
momenta $\vecop{j}_1$ and $\vecop{j}_2$. Equation
\fmref{eq:vector_addition_rule} is also referred to as
triangular condition for the coupling of two angular momenta.

From a group-theoretical point of view the eigenstates $\ket{j
\, m}$ of $\op{j}^z$ and $\vecop{j}^2$ span a
$(2j+1)$-dimensional irreducible representation $D^{(j)}$ of the
group $R_3$, i.e. the group of all rotations within
three-dimensional space. Keeping this in mind, the above vector
addition rule Eq.~\fmref{eq:vector_addition_rule} is equivalent
to
\begin{equation} \label{eq:coup}
   D^{(j_1)} \otimes D^{(j_2)} = \sum_{J=|j_1 - j_2|}^{j_1+j_2} D^{(J)} 
\ .
\end{equation}
According to this equation, the direct product between two
irreducible representations of the dimensions $(2j_1+1)$ and
$(2j_2+1)$, which is in general reducible, decays into
$(2J+1)$-dimensional irreducible representations with respect to
$J$.

Now, the operators $\op{D}^{(j_1)}(R)$ and $\op{D}^{(j_2)}(R)$
are associated with an arbitrary coordinate rotation $R$ of
$R_3$ and operate in the Hilbert spaces spanned by $\ket{j_1 \,
m_1}$ and $\ket{j_2 \, m_2}$. What does Eq.~\fmref{eq:coup} mean
for the direct product $\mat{D}^{(j_1)}(R) \otimes
\mat{D}^{(j_2)}(R)$ of the corresponding matrix representations?
The product can be transformed into a block-diagonal form
$\mat{U}(R)$ by a unitary matrix $\mat{A}$, i.e.
\begin{equation} \label{eq:transformop}
   \mat{D}^{(j_1)}(R) \otimes \mat{D}^{(j_2)}(R) = \mat{A}^\dagger \mat{U}(R) \mat{A}
\ ,
\end{equation}
where $\mat{U}(R)$ has the form
\begin{equation*} \begin{split}
   &\mat{U}(R)= \\
   & \quad \begin{bmatrix}
   \mat{D}^{(|j_1-j_2|)}(R) & \mat{0}  & \hdots & \mat{0}\\
   \mat{0} & \mat{D}^{(|j_1-j_2|+1)}(R) & \hdots & \mat{0} \\
   \vdots & \vdots & \hdots & \vdots \\ 
   \mat{0} & \mat{0} & \hdots & \mat{D}^{(j_1 + j_2)}(R)
   \end{bmatrix}
\ . \end{split}
\end{equation*}
The matrices $\mat{D}^{(J)}(R)$ appearing therein comprise
matrix elements with respect to those functions that span the
irreducible representations $D^{(J)}$ in Eq.~\fmref{eq:coup}.

The determination of the elements of the transformation matrix
$\mat{A}$ has been a central task in the theory of group
representations. The elements of $\mat{A}$ are the so-called
\textit{Clebsch-Gordan coefficients} $C_{m_1 \, m_2 \, M}^{j_1
\, j_2 \, J}$. They appear in a more familiar form as scalar
products between a state of the form $\ket{j_1 \, j_2 \, J \,
M}$ and the product states $\ket{j_1 \, m_1 \, j_2 \, m_2}$
leading to the decomposition
\begin{eqnarray} \label{eq:clebsch}
  \ket{j_1 \, j_2 \, J \, M} &= & \sum_{m_1,m_2} \braket{j_1 \, m_1 \, j_2 \, m_2}{j_1 \, j_2 \, J \, M} \, \ket{j_1 \, m_1 \, j_2 \, m_2} \nonumber \\
  &=& \sum_{m_1,m_2} C_{m_1 \, m_2 \, M}^{j_1 \, j_2 \, J} \ket{j_1 \, m_1 \, j_2 \, m_2}
\ .
\end{eqnarray}
The Clebsch-Gordan coefficients therefore relate two different
orthonormal sets of basis vectors. It should be emphasized that
these sets are obviously not orthogonal to each other because
they span the same space. The Clebsch-Gordan coefficients are
non-zero only if the vector addition rule from
Eq.~\fmref{eq:vector_addition_rule} and simultaneously the
equation $m_1+m_2=M$ hold. A very important symmetry of the
Clebsch-Gordan coefficients is
\begin{equation} \label{eq:symmetry_clebsch}
   C_{m_1 \, m_2 \, M}^{j_1 \, j_2 \, J} = (-1)^{j_1+j_2-J} C_{m_2 \, m_1 \, M}^{j_2 \, j_1 \, J}
   \ ,
\end{equation}
with $(-1)^{j_1+j_2-J} = \pm 1$ according to Eq.~\fmref{eq:vector_addition_rule}.

In order to reveal the symmetry properties of the Clebsch-Gordan
coefficients, they are reformulated in a straightforward
manner. A proper reformulation leads to the \textit{Wigner
coefficients} or \textit{Wigner-3J symbols}
\begin{equation*}
   \threej{j_1}{j_2}{J}{m_1}{m_2}{M}
   \ ,
\end{equation*}
which are related to the Clebsch-Gordan coefficients by
\begin{equation} \begin{split} \label{eq:clebsch-wigner3j}
   &\threej{j_1}{j_2}{J}{m_1}{m_2}{M} = \\
   & \qquad (-1)^{j_1-j_2-M} (2J+1)^{-\frac{1}{2}} \, C_{m_1 \, m_2 \, -M}^{j_1 \, j_2 \, J}
\ .
\end{split} \end{equation}
The relation between Clebsch-Gordan coefficients and the
Wigner-3J symbols from Eq.~\fmref{eq:clebsch-wigner3j} directly
leads to non-zero values of the Wigner-3J symbols only if
$m_1+m_2+M=0$ holds and the vector addition rule from
Eq.~\fmref{eq:vector_addition_rule} is fulfilled.

Expressed in terms of Wigner-3J symbols the symmetry property of
the Clebsch-Gordan coefficients given in
Eq.~\fmref{eq:symmetry_clebsch} takes the form
\begin{equation} \label{eq:wigner3J_symmetry}
   \threej{j_1}{j_2}{J}{m_1}{m_2}{M} = (-1)^{j_1+j_2+J} \threej{j_2}{j_1}{J}{m_2}{m_1}{M}
   \ .
\end{equation}
Thus, the Wigner-3J symbols are invariant under an even number
of permutations of the columns whereas under a single
permutation they obey Eq.~\fmref{eq:wigner3J_symmetry}. A
further evaluation of the symmetry properties of the
Clebsch-Gordan coefficients yields an additional symmetry of the
Wigner-3J symbols, i.e.
\begin{equation} \begin{split} \label{eq:wigner3J_symmetry2}
   & \threej{j_1}{j_2}{J}{m_1}{m_2}{M} = \\
   & \qquad (-1)^{j_1+j_2+J} \threej{j_1}{j_2}{J}{-m_1}{-m_2}{-M}
   \ .
\end{split} \end{equation}

So far the coupling of only two angular momenta has been
considered. A procedure similar to the one which has led to the
Wigner-3J symbols now leads to the occurence of
\textit{Wigner-6J symbols}. In the case of a coupling of three
angular momenta, $\vecop{J}=\vecop{j}_1 + \vecop{j}_2 +
\vecop{j}_3$, a basis can be constructed in which the
representations of the operators $\vecop{J}^2$ and $\op{J}^z$ as
well as $\vecop{j}_1^2$, $\vecop{j}_2^2$, and $\vecop{j}_3^2$
are diagonal. Obviously, there exists a certain freedom of
choice in the construction of this basis. The resulting
$\vecop{J}$ can be constructed in three different ways, namely
\begin{eqnarray}
\vecop{J}= (\vecop{j}_1 + \vecop{j}_2) + \vecop{j}_3 \ , \label{eq:coup_no1} \\
\vecop{J}= \vecop{j}_1 + (\vecop{j}_2 + \vecop{j}_3) \ , \label{eq:coup_no2} \\
\vecop{J}= (\vecop{j}_1 + \vecop{j}_3) + \vecop{j}_2 \ . \label{eq:coup_no3}
\end{eqnarray}
This leads to three different basis sets, each with the square
of one of the operators $\vecop{j}' = \vecop{j}_1 +
\vecop{j}_2$, $ \vecop{j}'' = \vecop{j}_2 + \vecop{j}_3 $, and
$\vecop{j}''' = \vecop{j}_1 + \vecop{j}_3$ given in a diagonal
form. The matrix elements of the unitary transformation matrix
which connects two of these sets of basis states can be found as
scalar products between states belonging to two different
\textit{coupling schemes}. Expressing a state belonging to a
coupling scheme resulting from a coupling according to
Eq.~\fmref{eq:coup_no1} in terms of states belonging to the
scheme resulting from Eq.~\fmref{eq:coup_no2} yields
\begin{equation} \begin{split} \label{eq:recoupling_intro}
  &\ket{j_1 \,  j_2 \, j_{12} \, j_3 \, J \, M} = \\
  &\sum_{j_{23}} \braket{j_1 \, j_2 \, j_3 \, j_{23} \, J}{j_1 \, j_2 \, j_{12} \, j_3 \, J} \, \ket{j_1 \, j_2 \, j_3 \, j_{23} \, J \, M}
  \ ,
\end{split} \end{equation}
with the quantum numbers $j_{12}$ and $j_{23}$ referring to
${\vecop{j}'}^2$ and ${\vecop{j}''}^2$, respectively. 

Scalar products of the kind found in
Eq.~\fmref{eq:recoupling_intro} -- often called
\textit{recoupling coefficients} -- are independent of any
magnetic quantum number. They can be evaluated by decomposing
the vector-coupling states into a sum of product states with the
help of an extended version of Eq.  \fmref{eq:clebsch}. For
example, the decomposition of the ket on the left hand side of
the aforementioned recoupling coefficients yields
\begin{equation} \begin{split} \label{eq:clebsch_extended}
  & \ket{j_1 \, j_2 \, j_3 \, j_{23} \, J \, M} = \\
  & \sum_{m_1, m_2, m_3} \braket{j_1 \, m_1 \, j_2 \, m_2 \, j_3
  \, m_3}{j_1 \, j_2 \, j_3 \, j_{23} \, J \, M} \\ 
  & \qquad \times \ket{j_1 \, m_1 \, j_2 \, m_2 \, j_3 \, m_3}
\ .
\end{split} \end{equation}
The scalar products in Eq.~\fmref{eq:clebsch_extended}, i.e. the
matrix elements of the transformation matrix that connects the
vector-coupling state with the product states, are called in
analogy to the former name convention \textit{generalized
Clebsch-Gordan coefficients} $C_{m_1 \, m_2 \, m_3 \, m_{23} \,
M}^{j_1 \, j_2 \, j_3 \, j_{23} \, J}$ for the coupling of three
angular momenta. They can be simplified to a product of
Clebsch-Gordan coefficients according to
\begin{equation} \begin{split} \label{eq:gen_clebsch}
   C_{m_1 \, m_2 \, m_3 \, m_{23} \, M}^{j_1 \, j_2 \, j_3 \,
     j_{23} \, J} = \sum_{m_{23}} C_{m_2 \, m_3 \, m_{23}}^{j_2
     \, j_3 \, j_{23}} \cdot C_{m_{1} \, m_{23} \, M}^{j_1 \,
     j_{23} \, J} 
\ .
\end{split} \end{equation}
As one can see, generating generalized Clebsch-Gordan
coefficients for the coupling of more than three angular momenta
and -- in addition -- an expression of it in terms of
Clebsch-Gordan coefficients is then a straightforward task. Here
it should be mentioned that according to
Eq.~\fmref{eq:gen_clebsch} generalized Clebsch-Gordan
coefficients can also be expressed as a sum over products of
Wigner-3J symbols using the relation from
Eq.~\fmref{eq:clebsch-wigner3j}. Coefficients of this kind are
then called \textit{generalized Wigner
coefficients}.\cite{YLV:angular62}

The recoupling coefficients that appear in
Eq.~\fmref{eq:recoupling_intro} can now be reformulated in terms
of Wigner-6J symbols in order to reveal their symmetry
properties. For the transition between the basis sets belonging
to $j_{12}$ and $j_{23}$ the corresponding Wigner-6J symbol is
related to the recoupling coefficient by\cite{Wig:group}
\begin{equation} \begin{split}
   &\sixj{j_1}{j_2}{j_{12}}{j_3}{J}{j_{23}} =
   (-1)^{j_1+j_2+j_3+J} (2j_{12}+1)^{-\frac{1}{2}} \\ 
   & \quad \times (2j_{23}+1)^{-\frac{1}{2}} \, \braket{j_1 \,
   j_2 \,  j_3 \, j_{23} \, J}{j_1 \, j_2 \, j_{12} \, j_3 \, J} 
\ .
\end{split} \end{equation}
Equations \fmref{eq:clebsch_extended} and \fmref{eq:gen_clebsch}
directly show that a Wigner-6J symbol can be expressed as a sum
over products of Clebsch-Gordan coefficients or, by using
Eq.~\fmref{eq:clebsch-wigner3j}, over products of Wigner-3J
symbols.

\begin{figure}[ht!]
\centering
\includegraphics[width=60mm]{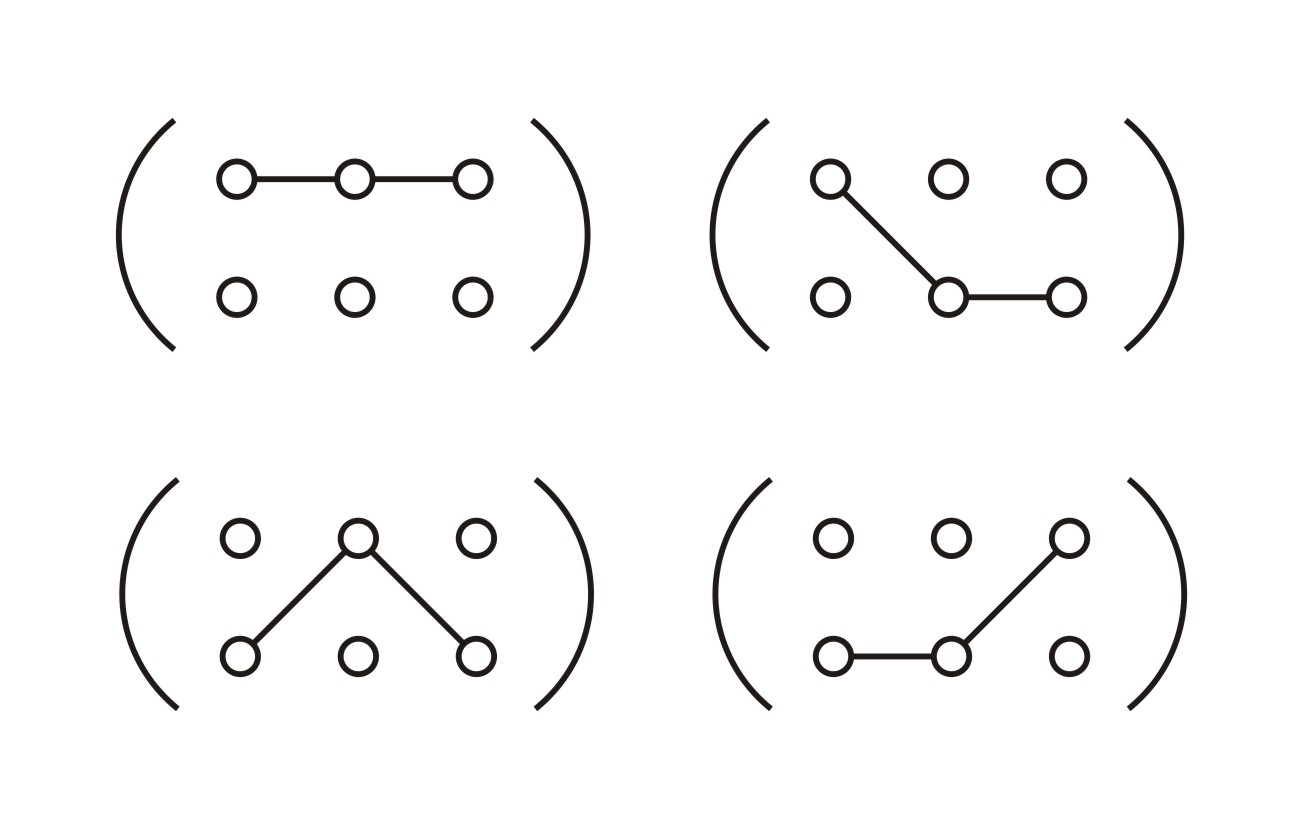}
\caption{Graphical visualization of the four triangular
  conditions occurring in a Wigner-6J symbol which have to be
  valid in order to yield a nonzero result.} 
\label{fig:6J-symmetries}
\end{figure}

For the sake of completeness, also the Wigner-9J
symbols\cite{Wig:group} shall be given which result as elements
of the transition matrices when recoupling four angular
momenta. For example the transition between two different sets
of basis states yields a Wigner-9J symbol like
\begin{equation} \begin{split}
   &\ninej{j_1}{j_2}{j_{12}}{j_3}{j_4}{j_{34}}{j_{13}}{j_{24}}{J} = \\
   & \; \; [(2j_{12}+1)(2j_{34}+1)(2j_{13}+1)(2j_{24}+1)]^{-\frac{1}{2}} \\
   & \; \; \times \braket{j_1 \, j_2 \, j_{12} \, j_3 \, j_4 \,
   j_{34} \, J}{j_1 \, j_3 \, j_{13} \, j_2 \, j_4 \, j_{24} \,
   J} 
\ .
\end{split} \end{equation}
A very comprehensive overview of algebraic expressions for
Wigner-nJ symbols as well of their symmetry properties is given
in Ref. \onlinecite{VMK:quantum_theory}. Regarding the use of
Wigner symbols in connection with irreducible tensor operators,
here only a graphical visualization of the triangular conditions
for a Wigner-6J symbol is shown. The Wigner-6J symbol is nonzero
only if for certain triads of quantum numbers the triangular
condition (Eq.~\fmref{eq:vector_addition_rule}) holds. For which
combination of quantum numbers of angular momenta the triangular
condition has to be valid in order to yield a nonzero Wigner-6J
symbol, is visualized graphically in \figref{fig:6J-symmetries}.

Since a Wigner-9J symbol can be written as a sum over products
of Wigner-6J symbols\cite{VMK:quantum_theory}, these triangular
conditions play an important role in reducing the computational
effort when matrix elements of the Heisenberg Hamiltonian are
calculated using irreducible tensor operators.

\subsection{The construction of basis states} \label{sec-A-2}
The reduction of the dimensionalities of the Hilbert spaces in
which the Hamilton matrices are set up in order to solve the
eigenvalue problem is always a desirable, but also numerically
involved task. Especially if the basis a priori reflects
symmetry properties of the system, an appropriate choice can be
of great help. As mentioned in Sec. \ref{sec-2}, a basis that
consists of vector-coupling states and incorporates full
spin-rotational ($SU(2)$) symmetry would be the first choice. In
isotropic spin systems as described by a Heisenberg Hamiltonian
the Hamilton matrix is then block-diagonalized with respect to
$S$ and $M$ without further calculations since there are no
transition elements between states of a different total magnetic
quantum number $M$ and different total-spin quantum number $S$.

The vector-coupling states used in the present work appear in
the form $\ket{\alpha \, S \, M}$. $\alpha$ denotes a set of
intermediate quantum numbers resulting from the chosen coupling
scheme according to which the spins are coupled. As mentioned in
the previous section the choice of the coupling scheme is
somewhat arbitrary since it only reflects the bracketing in the
expression for the total spin operator of the system $\vecop{S}
= \sum_i \vecop{s}_i$.

The simplest choice of a coupling scheme would probably be a
successive coupling of the single-spin vector operators
$\vecop{s}_i$. In the case of a spin square the set of
intermediate quantum numbers, if coupled according to a
successive coupling scheme, looks like
\begin{equation*}
   \alpha = \{ s_{12}=\overline{S}_1, s_{123}=\overline{S}_2 \}
\end{equation*}
leading to a vector-coupling state of the form $\ket{s_1 s_2
\overline{S}_1 s_3 \overline{S}_2 s_4 S M}$. Here, the notation
of the intermediate spin quantum numbers, i.e. $s_{12}$ and
$s_{123}$, is changed in comparison to
App. \ref{sec-A-1}. Intermediate spins are now numbered with
respect to their order of appearance in the coupling scheme and
additionally overlined. This notation has a clear advantage if
larger spin systems are investigated and is used in the
following. In order to clarify which spins are coupled, the
single-spin quantum numbers $s_i$ can also be found in the
ket. It would not be necessary to include them since they appear
as fixed numbers, but it turns out to be more convenient.

\begin{figure}[ht!]
\centering
\includegraphics[width=80mm]{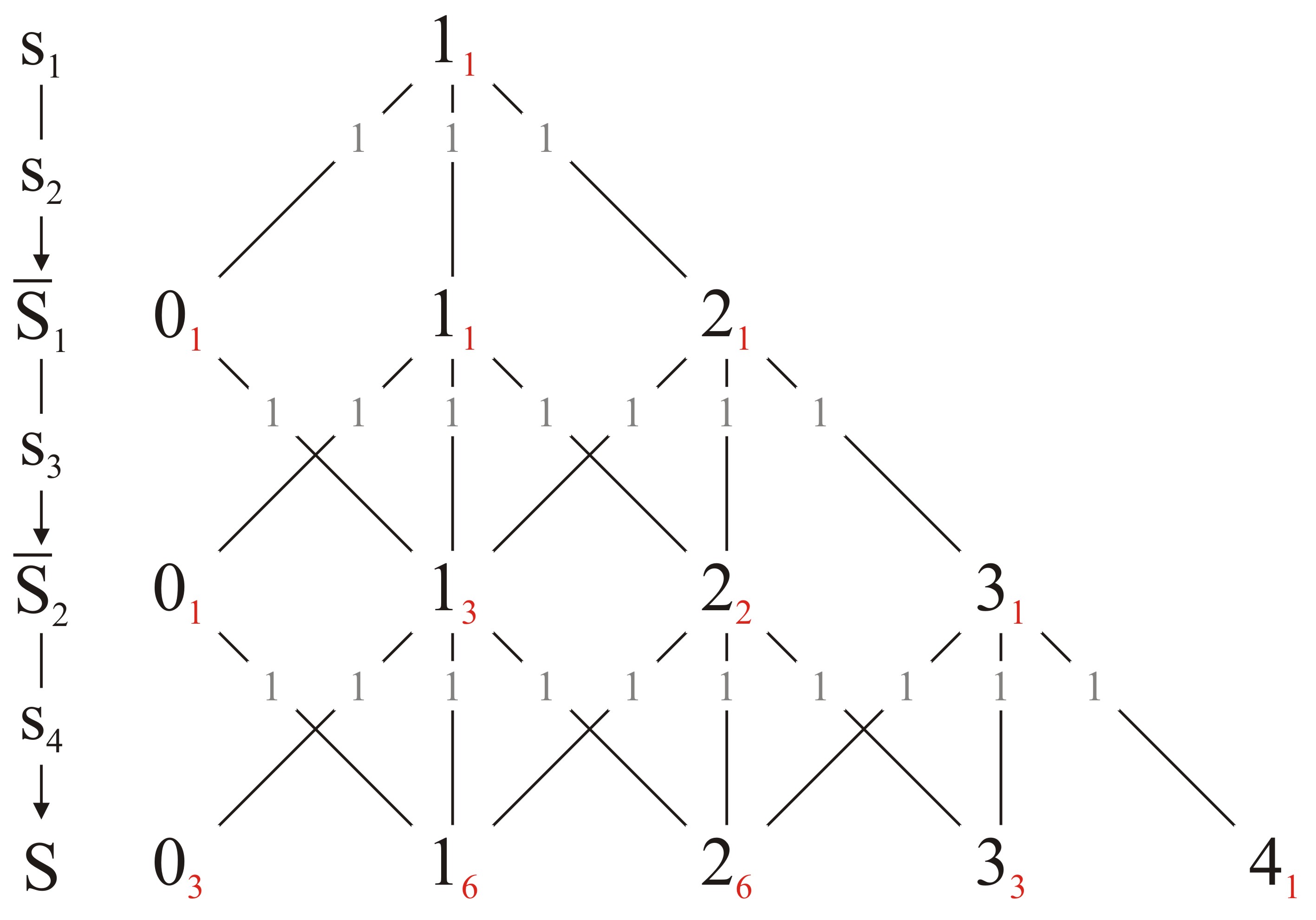}
\caption{Successive coupling of four spins $s=1$. The (red)
  subscripts denote the multiplicity, i.e. the number of paths
  leading to the spin quantum number. The gray numbers refer to
  the spin quantum numbers of the coupled single spins. On the
  left the coupling scheme is indicated.} 
\label{F-6}
\end{figure}

Now, if the coupling scheme is chosen and the framework of the
resulting basis states is fixed, one has to construct the basis
states by finding those values of the intermediate spins that
are valid according to Eq.~\fmref{eq:vector_addition_rule}. This
procedure can be visualized by constructing a coupling pyramid
as it is shown in Fig. \ref{F-6}. In Fig. \ref{F-6} four spins
with $s=1$ are successively coupled in order to yield the values
of the total-spin quantum number $S$. The (red) subscripts next
to the quantum numbers of the intermediate spins denote the
number of different paths leading to the same value for an
intermediate spin quantum number, i.e. the multiplicity. The
small (gray) numbers interrupting the lines connecting different
intermediate spin quantum numbers indicate the quantum numbers
of the single spins, i.e. $s_i=1$. For the sake of clarity, on
the left of Fig. \ref{F-6} the underlying coupling scheme is
given once more.

\begin{figure}[ht!]
\centering
\includegraphics[width=80mm]{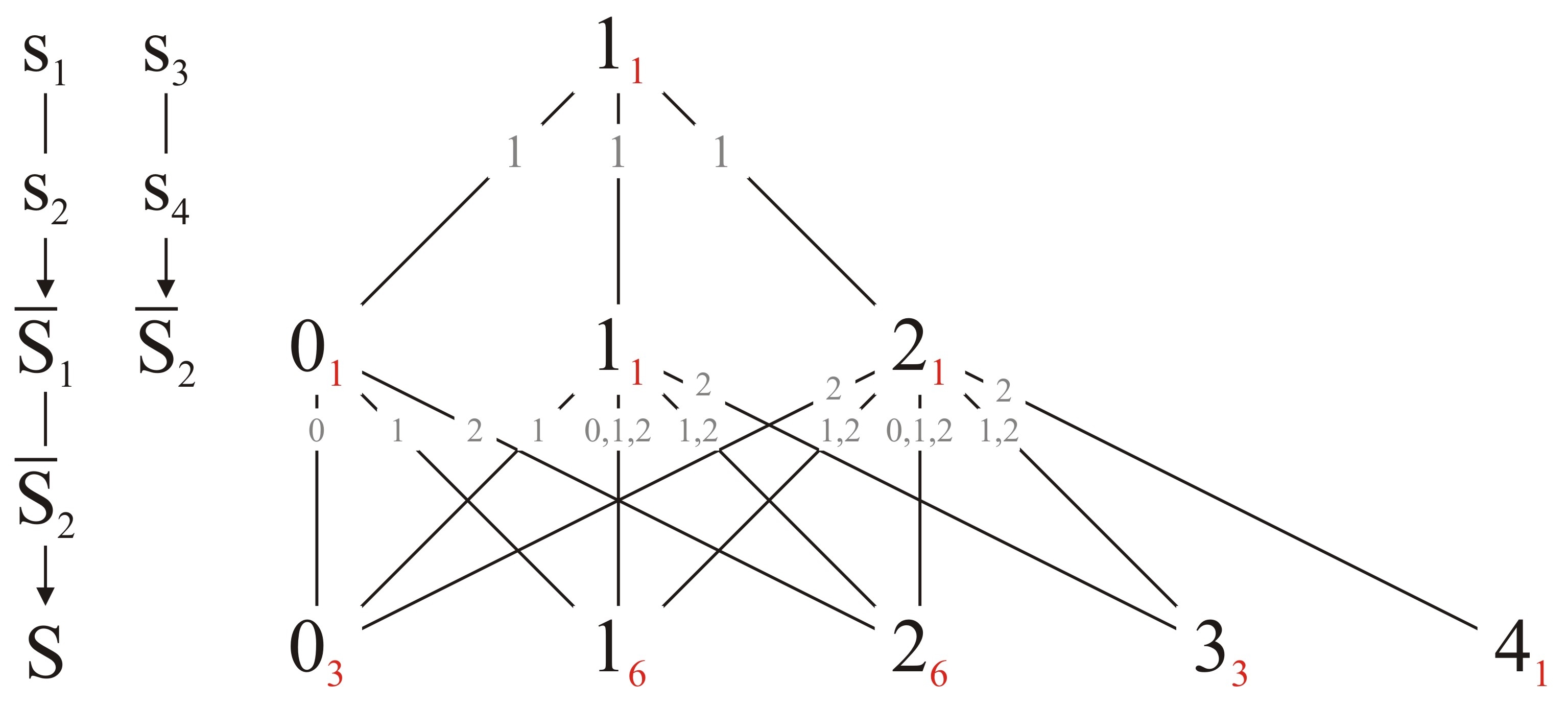}
\caption{Pairwise coupling of four spins $s=1$. The (red)
  subscripts denote the multiplicity, i.e. the number of paths
  leading to the spin quantum number. The gray numbers refer to
  the spin quantum numbers of the coupled single or intermediate
  spins. On the left the coupling scheme is indicated.} 
\label{F-7}
\end{figure}

Of course, a successive coupling scheme is not the only possible
way to couple the single spins $\vecop{s}_i$ to a total spin
$\vecop{S}$. For example, the construction of quasi-classical
states as described in Refs. \onlinecite{ScS:PRB09} and
\onlinecite{ScS:P09} requires a different coupling
scheme. There, a coupling scheme is chosen in which spins
belonging to a certain sublattice are coupled in order to get
the total sublattice spin that is afterwards coupled to the
total spin of the system. In Fig. \ref{F-7} a coupling pyramid
for a different coupling scheme of a spin square is shown. The
spins $\vecop{s}_1$, $\vecop{s}_2$ and $\vecop{s}_3$,
$\vecop{s}_4$ are coupled to yield intermediate spins
$\overline{\vecop{S}}_1$ and $\overline{\vecop{S}}_2$,
respectively. The intermediate spins are then coupled to the
total spin $\vecop{S}$.

The resulting multiplicity of states with the same quantum
number $S$ is obviously independent of the chosen coupling
scheme. At this point it is important to realize once again that
these states, although resulting from different coupling
schemes, form basis sets of the same Hilbert space
$\mathcal{H}$. In the case of a square the coupling of four
single spins $s=1$ discussed above results in basis states that
span the subspaces $\mathcal{H}(S)$ with
$S=0,1,\dots,4$. Writing this result in a different way using
Eq.~\fmref{eq:factor_repr}, one obtains for the direct product
of the irreducible representations $D^{(1)}$ of the single spins
the following expression:
\begin{equation*} \begin{split}
     & D^{(1)} \otimes D^{(1)} \otimes D^{(1)} \otimes D^{(1)} = \\
     & \qquad 3 \cdot D^{(0)} + 6 \cdot D^{(1)} + 6 \cdot D^{(2)} + 3 \cdot D^{(3)} + 1 \cdot D^{(4)}
     \ .
\end{split} \end{equation*}

The knowledge of the dimensions of the resulting irreducible
representations, i.e. subspaces $\mathcal{H}(S)$, is a central
task whenever performing numerical exact
diagonalization. Deduced from a successive coupling of the
spins, the dimensions $\text{dim}\,\mathcal{H}(S)$ can be
calculated by a simple recurrence formula.

The number of paths leading to a certain combination $(S,n)$ is
denoted by $d_S$. $n$ refers to the number of participating
spins in each step and runs from $1$ to $N$. $n$ can also be
seen as labeling the rows of the coupling pyramid for a
successive coupling (see Fig. \ref{F-6}). In the case of a
homonuclear system with $N$ spins $s$, the multiplicity
$d_S(S,n+1)$ is given by
\begin{equation} \label{eq:dim_recurrence}
   d_S(S,n+1) = \sum^{\text{min}(S+s,n \cdot s)}_{S'=|S-s|} d_S(S',n)
   \ ,
\end{equation}
where $S$ lies in the interval that is bounded by
\begin{equation} \label{eq:S_restr}
   (n+1)s \geq S \geq
   \left\{
   \begin{array}{cl}
   0 & \text{if} \; 2s(n+1) \; \text{even}\\
   \frac{1}{2} & \text{if} \; 2s(n+1) \; \text{odd}
   \end{array}
   \right.
\end{equation}
according to a vector coupling rule
(cf. Eq.~\fmref{eq:vector_addition_rule}). The number of paths
leading to a certain quantum number $S$ for one spin is just
\begin{equation}
   d_S(S,1) =
   \left\{
   \begin{array}{cl}
   1 & \text{if} \; S=s \\
   0 & \text{else}
   \end{array}
   \right.
   \ .
\end{equation}
The dimension of the Hilbert space $\mathcal{H}(S)$ is then
given by $\text{dim}\,\mathcal{H}(S)=d_S(S,N) \cdot (2S+1)$. 

The dimensions within a heteronuclear spin system, i.e. a system
with different values of single-spin quantum numbers, are
obtained along the same route. However, the range of valid
values for $S$ has to be calculated for each step of recurrence
separately, in contrast to the use of
Eq.~\fmref{eq:S_restr}. Furthermore, the sum in
Eq.~\fmref{eq:dim_recurrence} would run from $S'=|S-s_i|$ to
$\text{min}(S+s_i,\sum^{n}_{i=1} s_i)$ where the spin quantum
number of every single spin $s_i$ is individually labeled by
the index $i$.

\subsection{The Heisenberg Hamiltonian expressed as irreducible tensor operator} \label{sec-A-3}
In order to determine the spectra of magnetic molecules as it is
done throughout this work, the Heisenberg Hamiltonian of
Eq.~\fmref{eq:Heisenberg_WW} has to be expressed as irreducible
tensor operator. In this section a general expression for the
Heisenberg Hamiltonian is presented which can be used as a
starting point for the calculation of the energy spectrum using
the irreducible tensor operator approach.

\textbf{spin dimer} - The first step in finding an expression
for the Heisenberg Hamiltonian in the form of an irreducible
tensor operator is done by considering a spin dimer. The
Hamiltonian of the dimer takes the simple form
\begin{equation} \label{eq:Heisenberg_dimer}
   \op{H}_\text{dimer} = -J \, \vecop{s}(1) \cdot \vecop{s}(2)
   \ .
\end{equation}
Now, using Eq.~\fmref{eq:prod_ito} the above
Eq.~\fmref{eq:Heisenberg_dimer} can easily be reproduced. Since
the Heisenberg term is given by a scalar product, the
corresponding compound irreducible tensor operator is of rank
$k=0$. Using Eqs. \fmref{eq:spintensors} --
\fmref{eq:scalar_prod_tensor} one finds the expression
\begin{equation*} \begin{split}
   & \left\{ \vecito{s}{1}(1) \otimes \vecito{s}{1}(2) \right\}^{(0)} \\
   & \quad = \sum_{q_1,q_2} C^{1 \, 1 \, 0}_{q_1 \, q_2 \, 0} \cdot \ito{s}{1}{q_1}(1) \ito{s}{1}{q_2}(2) \\
   & \quad = \frac{1}{\sqrt{3}} \left( \ito{s}{1}{-1}(1) \cdot \ito{s}{1}{1}(2) + \ito{s}{1}{1}(1) \cdot \ito{s}{1}{-1}(2) \right. \\
   & \left. \quad \quad - \ito{s}{1}{0}(1) \cdot \ito{s}{1}{0}(2) \right) \\
   & \quad = - \frac{1}{\sqrt{3}} \, \vecop{s}(1) \cdot \vecop{s}(2) 
   \ .
\end{split} \end{equation*}
Thus, the tensorial form of the Heisenberg Hamiltonian of a spin dimer is
\begin{eqnarray} \label{eq:tensor_dimer}
   \op{H}_\text{dimer} = \sqrt{3} J \, \left\{ \vecito{s}{1}(1) \otimes \vecito{s}{1}(2) \right\}^{(0)}
   \ .
\end{eqnarray}

\textbf{spin triangle} - Since the tensorial form of
$\op{H}_\text{dimer}$ in Eq.~\fmref{eq:tensor_dimer} describes a
simple bilinear spin-spin interaction, an expression for a
general Heisenberg Hamiltonian can now be developed. As a first
extension a spin triangle is considered. The Hamiltonian that
has to be converted to an irreducible tensor operator is
\begin{equation} \begin{split}
   \op{H}_\triangle =& -J \, \left( \vecop{s}(1) \cdot \vecop{s}(2) + \vecop{s}(2) \cdot \vecop{s}(3) \right. \\
   & \left. +\vecop{s}(3) \cdot \vecop{s}(1) \right)
   \ .
\end{split} \end{equation}
In a very general form the successive coupling of three
single-spin irreducible tensor operators of ranks $k_1$, $k_2$
and $k_3$ leads to
\begin{equation} \begin{split} \label{eq:tensor_trimer}
   &\vecito{T}{k}_\therefore (k_1,k_2,k_3,\overline{k}_1) = \\
   & \qquad  \left\{ \left\{ \vecop{s}^{(k_1)}(1) \otimes \vecop{s}^{(k_2)}(2) \right\}^{(\overline{k}_{1})} \otimes \vecop{s}^{(k_3)}(3) \right\}^{(k)}
   \ .
\end{split} \end{equation}
Here $k$ denotes the rank of the resulting irreducible tensor
operator and $\overline{k}_1$ the rank of the intermediate
(coupled) one. The ranks of the many-particle tensor operators
are given by the coupling rules for spin quantum numbers known
from the spin vector coupling. For example, the rank
$\overline{k}_1$ is given by $\overline{k}_1 =
|k_1-k_2|,|k_1-k_2|+1,\dots,k_1+k_2$ with $k$ being determined
accordingly. It must be emphasized that
$\vecito{T}{k}_\therefore$ includes all spin-spin interactions
of a trimeric spin system, and thus it has to be specified in
order to give the desired tensorial formulation of
$\op{H}_\triangle$.

With $\vecito{s}{0}=\EinsOp$ and the tensorial expression found
for a bilinear coupling in Eq.~\fmref{eq:tensor_dimer} one
arrives at
\begin{eqnarray} \label{eq:tensor_dreieck}
   \op{H}_\triangle &=& \sqrt{3} J \left( \ito{T}{0}{\therefore} (1,1,0,0) + \ito{T}{0}{\therefore} (1,0,1,1) \right. \nonumber \\
   && \left. + \ito{T}{0}{\therefore} (0,1,1,1) \right)  \nonumber \\
   &=& \sqrt{3} J \sum_{<i,j>} \ito{T}{0}{\therefore} (\{ k_i \},\{ \overline{k}_i \} | k_i=k_j=1).
\end{eqnarray}
The notation of $\ito{T}{0}{\therefore} (\{ k_i \},\{
\overline{k}_i \} | k_i=k_j=1)$ indicates that only the ranks of
single-spin tensor operators $i$ and $j$ are chosen to equal $1$
whereas the other tensor operators are of zero rank. The rank of
the intermediate tensor operator $\overline{k}_1$ is fixed by
the contributions of $k_1$ and $k_2$ to the zero-rank tensor
operator $\vecito{T}{k=0}_\therefore$.

Following Eqs. \fmref{eq:tensor_dimer} and
\fmref{eq:tensor_dreieck}, the Heisenberg Hamiltonian of a
general spin system is given by
\begin{equation} \label{eq:Tensor_Heisenberg} \begin{split}
   &\op{H}_\text{Heisenberg} =\\
   & \qquad \sqrt{3} \sum_{<i,j>} J_{ij} \, \ito{T}{0}{}(\{ k_i \},\{ \overline{k}_i \}|k_i=k_j=1)
   \ ,
\end{split} \end{equation}
where the irreducible tensor operator $\vecito{T}{k}(\{ k_i
\},\{ \overline{k}_i \})$ directly depends on the investigated
system and the chosen coupling scheme for the coupling of the
single-spin tensor operators $\vecito{s}{k_i}(i)$.

As an example, the resulting expression for the general,
tetrameric tensor operator $\vecito{T}{k}_{: :}$ in the case of
a spin square shall be presented. The underlying coupling scheme
of the single-spin tensor operators is chosen to be pairwise
because it is often advantageous to couple this way when
additionally point-group symmetries are used (see
App. \ref{sec-A-5}). Then, the general irreducible tensor
operator of a tetrameric system takes the form
\begin{equation} \begin{split} \label{eq:Tensor_Quadrat}
      & \vecito{T}{k}_{: :}(k_1,k_2,k_3,k_4,\overline{k}_1,\overline{k}_2) = \\
      & \qquad \left\{ \left\{ \vecop{s}^{(k_1)}(1) \otimes \vecop{s}^{(k_2)}(2) \right\}^{(\overline{k}_{1})} \right. \\
      & \qquad \left. \otimes \left\{ \vecop{s}^{(k_3)}(3) \otimes \vecop{s}^{(k_4)}(4) \right\}^{(\overline{k}_{2})} \right\}^{(k)}
      \ .
\end{split} \end{equation}
The values of the ranks appearing in
Eq.~\fmref{eq:Tensor_Quadrat} for the spin square modeled by a
Heisenberg Hamiltonian are shown in
Tab. \ref{tab:square_ks}. The ranks of the single-spin tensor
operators are fixed by the particular spin-spin
interaction. Then, the ranks of the intermediate tensor
operators $\overline{k}_1$ and $\overline{k}_2$ can be
constructed from the known coupling rules with the given ranks
$k_i$, $i=1,\dots,4$, so as to yield an irreducible tensor
operator with $k=0$.

\begin{table}[t!] \centering
   \extrarowheight3pt
   \begin{tabular}{| c | c | c | c | c | c | c | c |}
      \hline
      & $k_1$ & $k_2$ & $k_3$ & $k_4$ & $\overline{k}_1$ & $\overline{k}_2$ & $k$ \\
      \hline
      \hline
      $<1,2>$ & 1 & 1 & 0 & 0 & 0 & 0 & 0 \\
      $<2,3>$ & 0 & 1 & 1 & 0 & 1 & 1 & 0 \\
      $<3,4>$ & 0 & 0 & 1 & 1 & 0 & 0 & 0 \\
      $<4,1>$ & 1 & 0 & 0 & 1 & 1 & 1 & 0 \\
      \hline
   \end{tabular}
   \caption{Values of the ranks for a spin square described by a Heisenberg Hamiltonian. The rows refer to the nearest-neighbor interaction $<i,j>$ between single spins $i$ and $j$.}
   \label{tab:square_ks} 
\end{table}

\subsection{Matrix elements - decoupling} \label{sec-A-4}
The calculations of the matrix elements of the Hamiltonian in
Eq.~\fmref{eq:Tensor_Heisenberg} are performed with the help of
the Wigner-Eckart theorem
.\cite{GaP:GCI93,BCC:IC99,BeG:EPR,Tsu:group_theory,Tsu:ICA08,BBO:PRB07}
For this reason, the coupling scheme of the basis states
$\ket{\alpha \, S \, M}$ cannot be chosen independently from
$\vecito{T}{k}$. Those quantum numbers which appear within the
set $\alpha$ should also appear in the sets $\{ k_i \}$ and $\{
\overline{k}_i \}$, i.e. the couplings for the generation of the
basis states and the general irreducible tensor operator of the
system should be chosen to be equal. Otherwise, transformations
between states of different coupling schemes would be necessary.

In order to show how matrix elements can be calculated, the
application of the \textit{decoupling} procedure is discussed
for a square system. The Heisenberg Hamiltonian in the tensorial
form can be derived from Eq.~\fmref{eq:Tensor_Heisenberg} and is
given by
\begin{equation} \begin{split} \label{eq:Hamiltonian_square}
   &\op{H}_\square =\\
   & \qquad \sqrt{3} J \sum_{<i,j>} \ito{T}{0}{: :}(\{ k_i \},\{ \overline{k}_i \}|k_i=k_j=1)
   \ .
\end{split} \end{equation}
The general irreducible tensor operator of the tetrameric system
$\vecito{T}{k}_{: :}$ based on a pairwise coupling scheme was
already presented in Eq.~\fmref{eq:Tensor_Quadrat}. The values
of the ranks appearing in $\ito{T}{k=0}{: :}(\{ k_i \},\{
\overline{k}_i \})$ are tabulated in
Tab. \ref{tab:square_ks}. The pairwise coupling scheme in the
used construction of $\vecito{T}{k}_{: :}$ corresponds to basis
states of the form
\begin{equation*}
   \ket{\alpha \, S \, M} = \ket{s_1 s_2 \overline{S}_1 s_3 s_4 \overline{S}_2 S M}
   \ .
\end{equation*}

By the application of the Wigner-Eckart theorem the calculation
of the matrix elements of $\op{H}_\square$ is now -- apart from
the prefactor and the summation over the nearest-neighbor
interactions according to Eq.~\fmref{eq:Hamiltonian_square} --
reduced to the evaluation of terms like
\begin{equation} \begin{split} \label{eq:square_element}
     & \bra{\alpha \, S \, M} \ito{T}{0}{: :} (\{ k_i \},\{
     \overline{k}_i \}) \ket{\alpha' \, S' \, M'} = \\ 
     & \qquad (-1)^{S-M} \threej{S}{0}{S'}{-M}{0}{M'} \\
     & \qquad \times \reduced{s_1 s_2 \overline{S}_{1} s_3 s_4
     \overline{S}_{2} S}{\vecito{T}{0}_{: :}}{s_1 s_2
     \overline{S}'_{1} s_3 s_4 \overline{S}'_{2} S'} 
     \ .
\end{split} \end{equation}
In the case of a square four such terms appear in
$\op{H}_\square$ differing from each other by the values of $\{
k_i \}$ and $\{ \overline{k}_i \}$.

Since the Wigner-3J symbol in Eq.~\fmref{eq:square_element} is
reduced to 
\begin{equation*}
   \threej{S}{0}{S'}{-M}{0}{M'} = \frac{(-1)^{S-M}}{\sqrt{2S+1}}
   \cdot \delta_{S,S'} \delta_{M,M'} 
\end{equation*}
Eq.~\fmref{eq:square_element} itself is reduced to
\begin{equation} \begin{split} \label{eq:matrix_square}
   & \bra{\alpha \, S \, M} \ito{T}{0}{: :} (\{ k_i \},\{
   \overline{k}_i \}) \ket{\alpha' \, S \, M} = \\ 
   & \frac{(-1)^{2(S-M)}}{\sqrt{2S+1}} \cdot \reduced{s_1 s_2
   \overline{S}_{1} s_3 s_4 \overline{S}_{2} S}{\vecito{T}{0}_{:
   :}}{s_1 s_2 \overline{S}'_{1} s_3 s_4 \overline{S}'_{2} S} 
    \ .
\end{split} \end{equation}
Since all matrix elements between states of different $S$ and
$M$ vanish, it becomes obvious that all calculations can be
performed in subspaces $\mathcal{H}(S,M)$. Furthermore, because
$\op{H}_\square$ is independent of an external magnetic field
and thus no $M$-dependence of the energies is given, the
spectrum can be evaluated in subspaces $\mathcal{H}(S,M=S)$.

Following Eq.~\fmref{eq:matrix_square} the matrix elements can
directly be obtained by determining the reduced matrix elements
of $\vecito{T}{0}_{: :}$. By a successive application of
Eq.~\fmref{eq:red_compound}, i.e. a successive decoupling of
$\vecito{T}{0}_{: :}$, the reduced matrix elements are traced
back to the reduced matrix elements of single-spin tensor
operators that are given in Eqs. \fmref{eq:red_spin1} and
\fmref{eq:red_spin2} for $k=0,1$.

For the reduced matrix element of $\vecito{T}{k}_{: :}$ the decoupling yields
\begin{widetext}
\begin{equation} \begin{split} \label{eq:decoupling}
   & \reduced{s_1 s_2 \overline{S}_{1} s_3 s_4 \overline{S}_{2} S}{\vecito{T}{k}_{: :}}{s_1 s_2 \overline{S}'_{1} s_3 s_4 \overline{S}'_{2} S'} = \\
   & \qquad \big[ (2S + 1) (2S' + 1) (2k +1) \big]^\frac{1}{2}
       \big[ (2\overline{S}_2 + 1) (2\overline{S}'_2 + 1) (2\overline{k}_2 + 1) \big]^\frac{1}{2}
       \big[ (2\overline{S}_1 + 1) (2\overline{S}'_1 + 1) (2\overline{k}_1 + 1 ) \big]^\frac{1}{2} \\
   & \qquad \times \ninej{\overline{S}_1}{\overline{S}'_1}{\overline{k}_1}{\overline{S}_2}{\overline{S}'_2}{\overline{k}_2}{S}{S'}{k}
       \ninej{s_3}{s_3}{k_3}{s_4}{s_4}{k_4}{\overline{S}_2}{\overline{S}'_2}{\overline{k}_2}
       \ninej{s_1}{s_1}{k_1}{s_2}{s_2}{k_2}{\overline{S}_1}{\overline{S}'_1}{\overline{k}_1}
       \reduced{s_1}{\vecito{s}{k_1}}{s_1} \reduced{s_2}{\vecito{s}{k_2}}{s_2} \reduced{s_3}{\vecito{s}{k_3}}{s_3} \reduced{s_4}{\vecito{s}{k_4}}{s_4}
\end{split} 
\ .
\end{equation}
\end{widetext}
In the case of $\op{H}_\square$, the appearing ranks of the
tensor operators can be found in Tab. \ref{tab:square_ks} and
$S'$ is forced to be $S'=S$ by the Wigner-3J symbol in
Eq.~\fmref{eq:square_element}.

The clear structure of the resulting expression for a reduced
matrix element in Eq.~\fmref{eq:decoupling} now allows one to
write a highly flexible and structured computer program that
takes over the calculation and diagonalization of the Hamilton
matrix. Regarding the calculation of the matrix elements, a
recurrence formula can be implemented which decouples the
irreducible tensor operator of the system
step-by-step.\cite{BCC:JCC99}

\subsection{Using point-group symmetries} \label{sec-A-5}
\begin{figure}[ht!]  \centering
   \includegraphics[width=50mm]{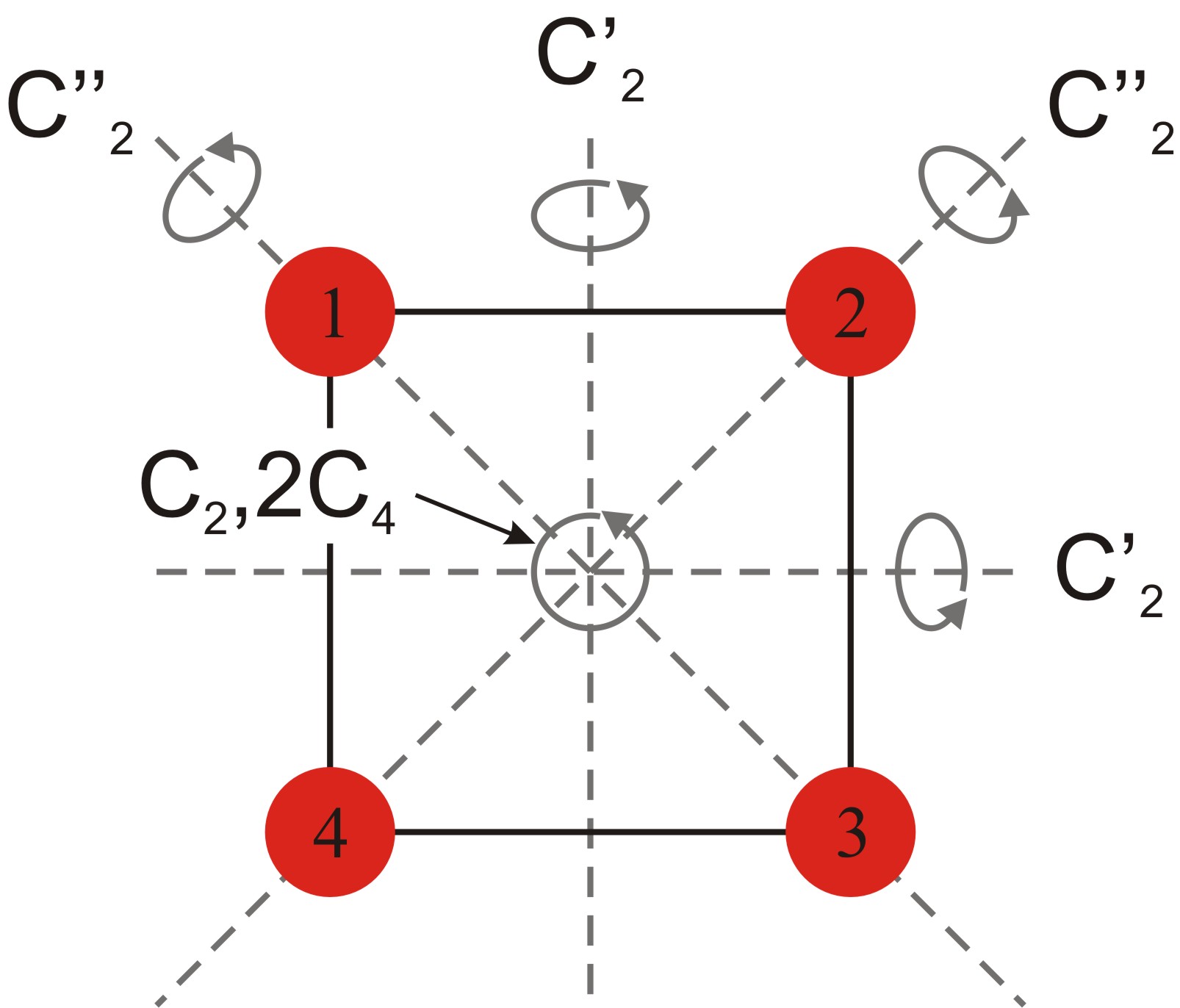}
   \caption{Sketch of a spin square with $D_4$-symmetry operations.}
   \label{F-8}
\end{figure}

General considerations concerning the use of point-group
symmetries in Heisenberg spin systems have already been
presented in Secs. \ref{sec-2-3} and \ref{sec-2-4}. Now, as a
clarification of these considerations, a $\pi$-rotation around
the central $C_2$-axis of a spin square is considered
(cf. Fig. \ref{F-8}). According to a successive coupling scheme,
the vector-coupling basis states are given in the form $\ket{s_1
s_2 \overline{S}_1 s_3 \overline{S}_2 s_4 S M}$. It must be
emphasized here that the underlying coupling scheme, that
determines the way how basis states are constructed, can be
chosen independently from any symmetry considerations, although
a suitable choice will reduce the calculations as is
discussed below.

Figure \ref{F-8} shows the coupling graph of the square with
$D_4$-symmetry operations. The operations are labeled with
respect to n-fold rotations around the given axes. Operations
belonging to the same class are marked with identical labels
while identical operations belonging to different classes can be
distinguished by the number of primes.

\setlength{\extrarowheight}{0.2cm}
\begin{table}[t!] \centering
   \begin{tabular}{| c | c | c | c | c | c | }
      \hline
      \multirow{2}{0.8cm}{\centering $D_4$} & \multicolumn{5}{c|}{\textbf{Classes}} \\
      \cline{2-6}
      & E & $C_2$ & $2C_4$ & $2C'_2$ & $2C''_2$ \\ 
      \hline
      \hline
      \multirow{2}{*}{op's} & $\op{G}(1\,2\,3\,4)$ & $\op{G}(3\,4\,1\,2)$ & $\op{G}(4\,1\,2\,3)$ & $\op{G}(2\,1\,4\,3)$ & $\op{G}(1\,4\,3\,2)$ \\
      &  & & $\op{G}(2\,3\,4\,1)$ & $\op{G}(4\,3\,2\,1)$ & $\op{G}(3\,2\,1\,4)$ \\
      \hline
   \end{tabular}
   \caption{Classes of the $D_4$ point group with corresponding operators for the spin square.}
   \label{tab:D4_quadrat} 
\end{table}

In Tab. \ref{tab:D4_quadrat} the operators corresponding to the
$D_4$-symmetry operations on the square are given. The operators
are classified with respect to the classes they belong to. The
mentioned operator that acts in spin space and corresponds to a
$\pi$-rotation around the central axis takes the form
$\op{G}(3\,4\,1\,2)$.

According to the generalized Clebsch-Gordan coefficients for the
coupling of four spins and Eq.~\fmref{eq:gen_clebsch}, unfolding
the state $\ket{s_1 s_2 \overline{S}_1 s_3 \overline{S}_2 s_4 S
M}$ into a linear combination of product states results in the
expression
\begin{equation} \begin{split} \label{eq:Uebergang_S_zu_m}
   & \ket{s_1 s_2 \overline{S}_1 s_3 \overline{S}_2 s_4 \, S \, M} = \\
   & \sum_{\sum_i m_i=M} C_{m_1 \, m_2 \, \overline{M}_1}^{s_1 \, s_2 \, \overline{S}_1} \cdot
   C_{\overline{M}_1 \, m_3 \, \overline{M}_2}^{\overline{S}_1 \, s_3 \, \overline{S}_2} \cdot
   C_{\overline{M}_2 \, m_4 \, M}^{\overline{S}_2 \, s_4 \, S} \\
   & \quad \times \ket{m_1 \, m_2 \, m_3 \, m_4}
   \ .
\end{split} \end{equation}
The summation indices are entirely determined by the constraint
$\sum_i m_i=M$. The values of the intermediate magnetic quantum
numbers $\overline{M}_i$ can be deduced from the magnetic
quantum numbers of the involved single spins,
i.e. $\overline{M}_1 = m_1 + m_2$ and $\overline{M}_2 =
\overline{M}_1+m_3$.

Following Eqs. (\ref{eq:Wirkung_Produktbasis}) and
(\ref{eq:Uebergang_S_zu_m}), performing the $\pi$-rotation
described by the operator $\op{G}(3 \, 4 \, 1 \, 2)$ results in
the expression
\begin{equation} \label{eq:Wirkung_Drehung} \begin{split}
   & \op{G}(3 \, 4 \, 1 \, 2) \ \ket{s_1 s_2 \overline{S}_1 s_3 \overline{S}_2 s_4 \, S \, M} = \\
   & \sum_{\sum_i m_i=M} C_{m_1 \, m_2 \, \overline{M}_1}^{s_1 \, s_2 \, \overline{S}_1} \cdot
   C_{\overline{M}_1 \, m_3 \, \overline{M}_2}^{\overline{S}_1 \, s_3 \, \overline{S}_2} \cdot
   C_{\overline{M}_2 \, m_4 \, M}^{\overline{S}_2 \, s_4 \, S} \\
   & \quad \times \ket{m_3 \, m_4 \, m_1 \, m_2}
   \ .
\end{split} \end{equation}
Due to the performed permutation on the product states the
resulting state cannot easily be represented as a
vector-coupling state belonging to the former coupling scheme
given by the successive addition of spin operators:
$\vecop{s}(1) + \vecop{s}(2) = \overline{\vecop{S}}(1)$,
$\overline{\vecop{S}}(1) + \vecop{s}(3) =
\overline{\vecop{S}}(2)$, and $\overline{\vecop{S}}(2) +
\vecop{s}(4) = \vecop{S}$.

At this point, it becomes obvious that the operator $\op{G}$ is
inducing a transition from the coupling scheme, according to
which the basis states have initially been constructed, to
another one. A proper re-labeling of the summation indices in
the sum of Eq.~\fmref{eq:Wirkung_Drehung} with respect to a
point-group operation $R^{-1}$, i.e. a $(-\pi)$-rotation around
the central $C_2$-axis, reveals the resulting coupling scheme in
which $\op{G}(3 \, 4 \, 1 \, 2) \ket{s_1 s_2 \overline{S}_1 s_3
\overline{S}_2 s_4 \, S \, M}$ can be represented as a
vector-coupling state.\cite{Wal:PRB00} In this special case one
finds that $\op{G}(3 \, 4 \, 1 \, 2)$ is inducing a transition
to a coupling scheme given by $\vecop{s}(3) + \vecop{s}(4) =
\overline{\vecop{S}}(1')$, $\overline{\vecop{S}}(1') +
\vecop{s}(1) = \overline{\vecop{S}}(2')$, and
$\overline{\vecop{S}}(2') + \vecop{s}(2) = \vecop{S}$. As a
shorthand notation one can write
\begin{equation*}
   \ket{s_1 s_2 \overline{S}_1 s_3 \overline{S}_2 s_4 S  M} \xrightarrow{\op{G}(3 \,4 \,1 \,2)} \ket{s_3 s_4 \overline{S}_{1'} s_1 \overline{S}_{2'} s_2  S  M}
   \ ,
\end{equation*}
with the limitation that this expression does not give the
concrete values of the appearing quantum numbers of the
states. Nevertheless, it illustrates the transition between
vector-coupling states of different and thus independent
coupling schemes.

Following Eq.~\fmref{eq:point-group-operation_final} the action
of the $\pi$-rotation on a vector-coupling state of the chosen
form results in
\begin{equation} \label{eq:Wirkung_Drehung_final} \begin{split}
   & \op{G}(3 \, 4 \, 1 \, 2) \ \ket{s_1 s_2 \overline{S}_1 s_3 \overline{S}_2 s_4 S M} = \\
   & \quad \sum_{\overline{S}'_1, \overline{S}'_2} \delta_{\overline{S}_1,\overline{S}_{1'}} \delta_{\overline{S}_2,\overline{S}_{2'}} \ket{s_1 s_2 \overline{S}'_1 s_3 \overline{S}'_2 s_4 S M} \\
   & \quad \times \ \braket{s_1 s_2 \overline{S}'_1 s_3 \overline{S}'_2 s_4 S M}{s_3 s_4 \overline{S}_{1'} s_1 \overline{S}_{2'} s_2 S M}
   \ .
\end{split} \end{equation}

As mentioned in Sec. \ref{sec-2-4} the main task when
calculating the action of a point-group operation on a
vector-coupling state is the determination of the general
recoupling coefficients connecting states of the initial and the
resulting coupling scheme. Generating a formula for general
recoupling coefficients can only be performed in a rather
advanced procedure (see Sec. \ref{sec-B}).

\begin{figure}[ht!] \centering
   \includegraphics[width=62mm]{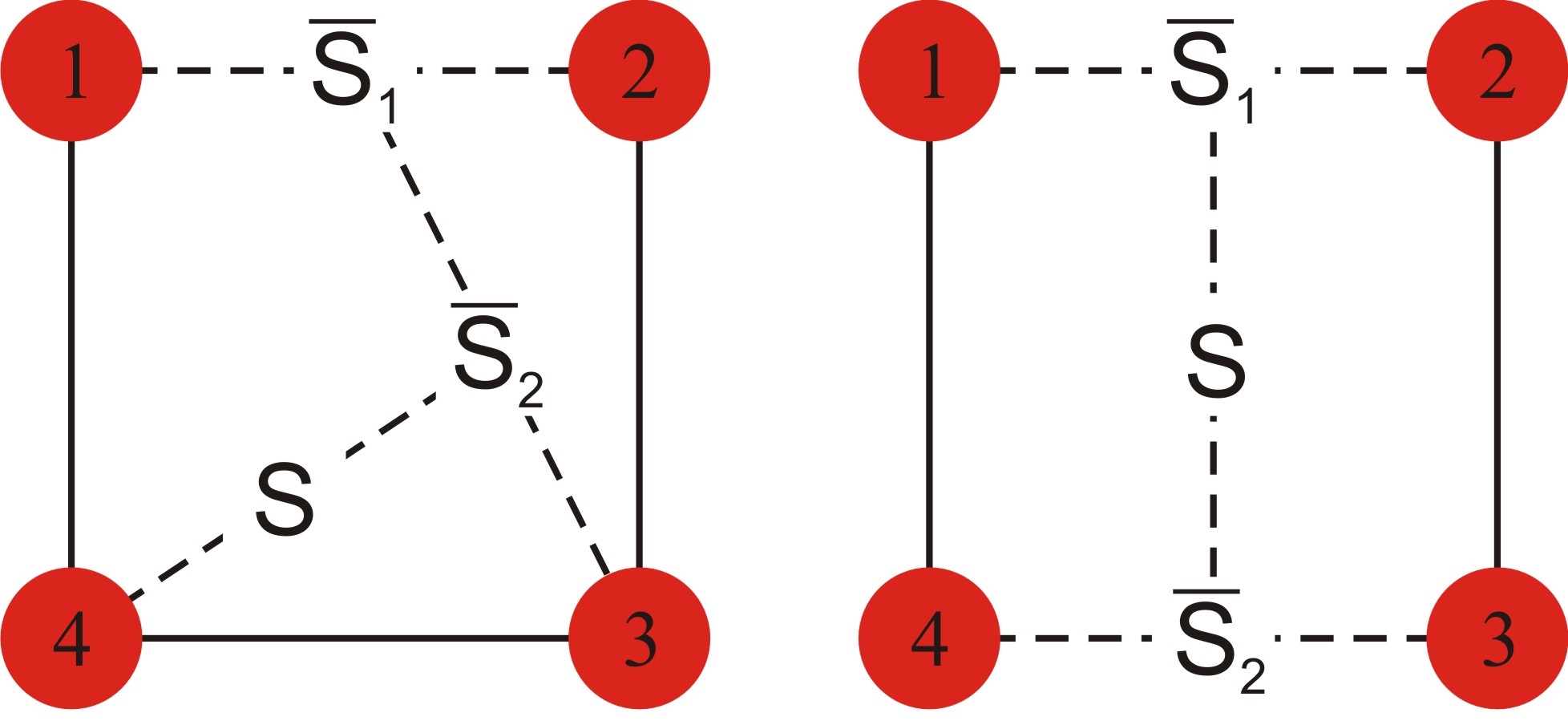}
\caption{Visualization of two possible couplings for the square:
  successive coupling scheme (l.h.s.) and pairwise coupling
  scheme (r.h.s.).} 
\label{fig:Schema_quadrat} 
\end{figure}

Regarding the recoupling coefficients that appear when
performing point-group operations, a very helpful simplification
shall be mentioned here. From Eq.~\fmref{eq:Wirkung_Drehung} it
can be seen that the action of the operator $\op{G}(3 \, 4 \, 1
\, 2)$ on the product states prevents the re-expression of the
resulting linear combination of product states as a simple
vector-coupling state belonging to the initial coupling
scheme. However, the choice of the coupling scheme according to
which the initial basis was constructed is somewhat
arbitrary. In order to minimize the computational effort, which
is directly related to the number of states with non-zero
recoupling coefficient in Eq.~\fmref{eq:Wirkung_Drehung_final},
one has to choose a non-successive coupling scheme. A favorable
coupling scheme of this kind is shown in
Fig. \ref{fig:Schema_quadrat}. This scheme is referred to as
pairwise coupling scheme and the basis states look like
$\ket{s_1 s_2 \overline{S}_1 s_3 s_4 \overline{S}_2 S M}$. The
$\pi$-rotation around the central $C_2$-axis of the square now
induces a transition that can be symbolized by
\begin{equation*}
   \ket{s_1 s_2 \overline{S}_1 s_3 s_4 \overline{S}_2 S  M} \xrightarrow{\op{G}(3 \,4 \,1 \,2)} \ket{s_3 s_4 \overline{S}_{1'} s_1 s_2 \overline{S}_{2'} S  M}
\end{equation*}
and leads according to Eq.~\fmref{eq:point-group-operation} to a recoupling coefficient of the form
\begin{equation*}
   \braket{s_1 s_2 \overline{S}_1 s_3 s_4 \overline{S}_2 S  M}{s_3 s_4 \overline{S}_{1'} s_1 s_2 \overline{S}_{2'} S  M}
   \ .
\end{equation*}
Now, the calculation of a formula for this recoupling
coefficient is trivial since the intermediate spin operators of
the initial and the resulting coupling scheme are mutually the
same, i.e. $\overline{\vecop{S}}(1') = \overline{\vecop{S}}(2)$
and $\overline{\vecop{S}}(2') =
\overline{\vecop{S}}(1)$. Unfolding $\ket{s_3 s_4
(\overline{S}_{1'}=\overline{S}_{2}) s_1 s_2
(\overline{S}_{2'}=\overline{S}_{1}) S M}$ into a linear
combination of product states in analogy to
Eq.~\fmref{eq:Uebergang_S_zu_m} leads to
\begin{equation*} \begin{split}
   & \ket{s_3 s_4 \overline{S}_{2} s_1 s_2 \overline{S}_{1} S  M} = \\
   & \sum_{\sum_i m_i=M} C_{m_3 \, m_4 \, \overline{M}_2}^{s_3 \, s_4 \, \overline{S}_2} \cdot
   C_{m_1 \, m_2 \, \overline{M}_1}^{s_1 \, s_2 \, \overline{S}_1} \cdot
   C_{\overline{M}_2 \, \overline{M}_1 \, M}^{\overline{S}_2 \, \overline{S}_1 \, S} \\
   & \quad \times \ket{m_1 \, m_2 \, m_3 \, m_4}
   \ .
\end{split} \end{equation*}
In order to convert the Clebsch-Gordan coefficients to a form
that appears when unfolding states of the form $\ket{s_1 s_2
\overline{S}_1 s_3 s_4 \overline{S}_2 S M}$ one simply has to
use the symmetry property of the Clebsch-Gordan coefficients
from Eq.~\fmref{eq:symmetry_clebsch}. This leads to
\begin{equation*}
    C_{\overline{M}_2 \, \overline{M}_1 \, M}^{\overline{S}_2 \, \overline{S}_1 \, S} = (-1)^{\overline{S}_1+\overline{S}_2-S} 
    C_{\overline{M}_1 \, \overline{M}_2 \, M}^{\overline{S}_1 \, \overline{S}_2 \, S}
    \ ,
\end{equation*}
and thus an expression for the recoupling coefficient is obtained which only contains one simple phase factor:
\begin{equation*}
   \braket{s_1 s_2 \overline{S}_1 s_3 s_4 \overline{S}_2 S  M}{s_3 s_4 \overline{S}_{1'} s_1 s_2 \overline{S}_{2'} S  M}=(-1)^{\overline{S}_1+\overline{S}_2-S}
   \ .
\end{equation*}
The action of the operator performing a $\pi$-rotation on a
state belonging to the pairwise coupling scheme mentioned above
therefore directly results in a state belonging to the same
coupling scheme with an attached phase factor. Thus, it has been
shown that with a cleverly chosen coupling scheme the
computational effort, that is required for the calculation of
symmetrized basis states according to
Eq.~\fmref{eq:projection_op2}, can be minimized. The graphical
visualization of possible coupling schemes in a square shown in
Fig. \ref{fig:Schema_quadrat} makes it obvious that one will
find a simple recoupling formula depending only on phase factors
whenever one can find a coupling scheme that is invariant under
the performed symmetry operation.\cite{Wal:PRB00} Since one has
to sum over all symmetry operations of the group in order to
construct symmetrized basis states of a given irreducible
representation (see Eq.~\fmref{eq:projection_op2}), the
underlying coupling scheme should be chosen in such a way as to
simplify all or at least most of the resulting recoupling
coefficients. This means that the coupling scheme should be
invariant under all or at least most of the symmetry operations.

However, one will not always be able to find a coupling scheme
that simplifies the calculation of the recoupling coefficients
as shown above. Especially if the system under consideration is
exhibiting three-fold symmetry axes, such a procedure turns out
to be impossible by means of a pairwise coupling scheme. In this
case a generalization of finding a recoupling formula
independent of the choice of the coupling scheme becomes
necessary.

\subsection{Computational effects of the choice of the coupling scheme} \label{sec-A-6}
The computational realization of the theoretical background
presented in this work has been a central task. The performed
calculations would not have been possible without developing a
highly parallelized computer program that is well adapted to the
use in a high performance computing environment. In this section
some remarks on the computational effects of the choice of the
underlying coupling scheme shall be given. In general, as long
as a proper scaling is achieved the most intuitive way to speed
up calculations using high performance computers is to
distribute calculations among many processing
units. Nevertheless, as will be seen below the right choice of
initial parameters like the coupling scheme can help to ease the
problem of calculating energy spectra and thermodynamic
properties of magnetic molecules.

Since several terms appear in this section which might be
unknown to the reader, their particular meaning as well as
related aspects shall be discussed first. The term
\textit{computation time} refers to the cumulative time that is
needed in order to perform a certain number of floating point
operations (FLOPs). The \textit{execution time} refers to the
runtime of the considered part of the program. Assuming a
parallel execution of the program with optimal performance, the
computation time remains unchanged although the operations are
performed in parallel. A reduction of the computation time can
be achieved by reducing the number of FLOPs that have to be
performed. In contrast to this, the execution time usually
decreases with increasing number of processing units. The change
of the execution time as a function of the used processing units
is called \textit{scaling behavior}. It is referred to as
optimal if the execution time is divided by two whenever the
number of used processing units is doubled. The \textit{speed
up} $S_p$ using $p$ processing units is defined as
\begin{equation}
   S_p = \frac{T_1}{T_p}
   \ ,
\end{equation}
where $T_1$ and $T_p$ refer to the execution times of the
sequential and the parallelized algorithm, respectively. An
optimal speed up corresponds to $S_p=p$. The optimal speed up
can be achieved if the whole source code can be parallelized
without dependencies between the operations which are executed
in parallel. Practically, sequential regions and communication
between the processing units often limit the speed up to a value
that is lower than the number of used processing units.

In general, the computational realization of the presented
framework can be divided into two completely independent
parts. On the one hand a matrix representation of the
Hamiltonian is set up with the help of the irreducible tensor
operator approach. On the other hand this matrix or independent
blocks of it are diagonalized, i.e. the eigenvalues and
eigenvectors are determined numerically.

If point-group symmetries are used, a very decisive role
concerning the computation time is played by the construction of
symmetrized basis states. These functions appear as linear
combinations of the initial basis states. The weight of the
states that are included in these linear combinations is
determined by general recoupling coefficients. If a coupling
scheme can be found that is invariant under all point-group
operations, the number of states that contribute to a linear
combination representing a symmetrized basis state is
minimized. As already mentioned, a reduction of computation time
is achieved by choosing a coupling scheme that minimizes the
number of appearing summation indices and Wigner-6J symbols.

The construction of symmetrized basis states plays an important
role for extending the limits of numerical exact diagonalization
with the help of the concepts presented in this work. Whenever
the dimensions of the appearing matrices are to be reduced by
the incorporation of point-group symmetries, a certain amount of
additional computation time has to be spent on the construction
of symmetrized basis states. Since the construction procedure
cannot easily be parallelized, the use of more processing units
within this particular region does not always lead to the
desired reduction of execution time. In any case, one has to
ensure that the recoupling formulas, which determine the number
of performed FLOPs, are the simplest in order to reduce
computation time. As already mentioned, this can be achieved by
choosing a coupling scheme that is invariant under the
operations of the assumed point-group. The resulting recoupling
formulas do then not contain Wigner symbols and are optimal.

\begin{figure}[ht!]
  \centering
  \includegraphics[width=60mm]{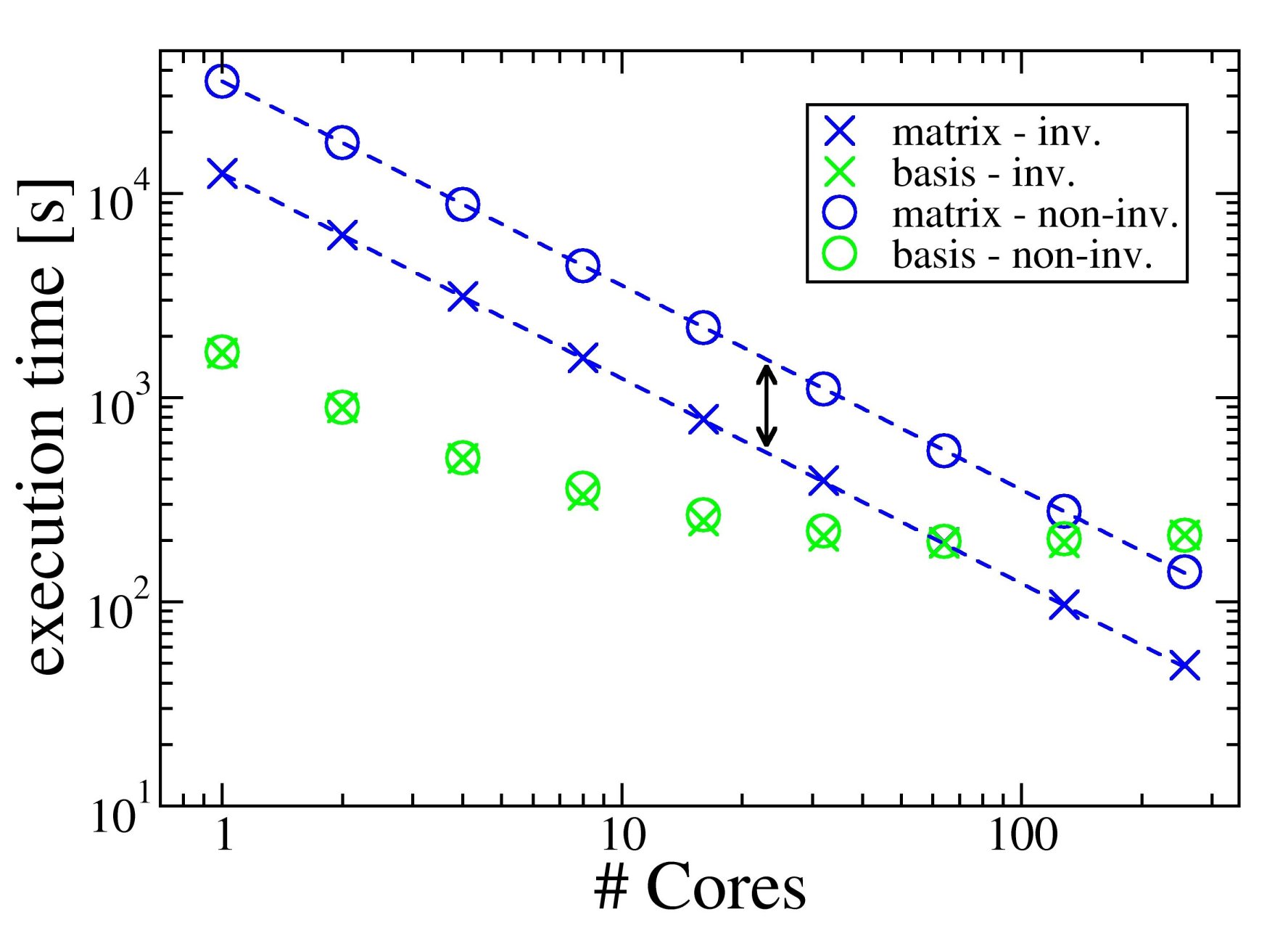} 
  \caption{Scaling of OpenMP parallelized regions. Exemplary
  calculations have been performed using different coupling
  schemes (cuboctahedron $s=3/2$, $D_2$,
  $\mathcal{H}(S=0,M=0,A_1)$). The crosses refer to an invariant
  coupling scheme while the circles refer to a non-invariant
  coupling scheme. The black arrow indicates the different
  execution times for the set up of the matrices and the dashed
  lines show the optimal scaling behavior.} 
  \label{fig:scaling2}
\end{figure}

In Fig. \ref{fig:scaling2} the execution times for the
determination of the energy eigenvalues of a cuboctahedron with
$s=3/2$ in the subspace $\mathcal{H}(S=0,M=0,A_1)$ are shown in
dependence on the chosen coupling scheme. The used point-group
symmetry has been $D_2$ and execution times are given for the
the choice of a completely invariant and a non-invariant
coupling scheme, respectively. Comparing the scaling behavior,
one can see that the performance of both calculations is limited
by the construction of the symmetrized basis
states. Furthermore, it becomes obvious that the set up of the
matrices is heavily influenced by the particular form of the
symmetrized basis states. In the case of the non-invariant
coupling scheme the set up of the matrix has been much slower
than in the case of the invariant coupling scheme because the
symmetrized basis states involve more states of the initial
(vector-coupling) basis.

\section{The calculation of general recoupling coefficients} \label{sec-B}
In this section graph-theoretical considerations are presented
that allow to determine the action of point-group operations on
vector-coupling states. It is shown how a general recoupling
formula can be developed from mapping general Wigner
coefficients on binary trees or Yutsis graphs.

In general, the problem of calculating recoupling coefficients
has to be divided into two parts. The first part is the
generation of a formula that describes the transition between
two different coupling schemes. The second -- and much easier --
part is the evaluation of a given formula using a specific set
of quantum numbers.\cite{FPV:CPC95} This section is exclusively
focused on the first part, i.e. the generation of a recoupling
formula that links different coupling schemes in systems with an
arbitrary number of participating spins which turns out to be
more difficult.

\subsection{Binary trees} \label{sec-B-1}
\begin{figure}[ht!] \centering
   \includegraphics[width=20mm]{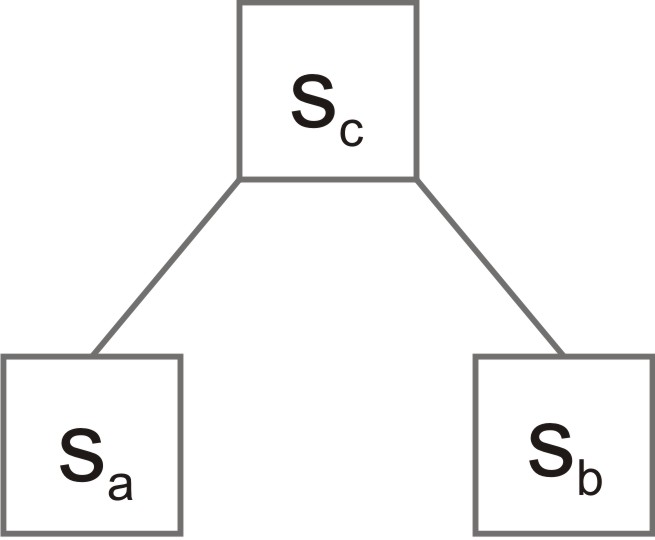}
\caption{Binary tree that corresponds to a single coupling of
  two spins $s_a$ and $s_b$ to a compound spin $s_c$.} 
\label{fig:tree_triad} 
\end{figure}

In the literature one can find successful implementations that
deal with the generation of formulas for general recoupling
coefficients which only involve a series of phase factors and
Wigner-6J symbols.\cite{BCC:JCP96,CBC:IC09,BCC:JCC10} The most
intuitive way of generating a recoupling formula is to operate
on so-called binary trees.\cite{Bur:CPC70,FPV:CPC94,ErS:JPA10}
The correspondence between a binary tree and a given coupling
scheme is rather obvious. Each coupling of two spins $s_a$ and
$s_b$ to a compound spin $s_c$ forms a triad that corresponds to
an elementary binary tree shown in
Fig. \ref{fig:tree_triad}. This tree is composed of only three
angular momenta and can simultaneously be seen as representing a
Clebsch-Gordan coefficient. From such elementary trees a binary
tree can be built up step-by-step that represents the chosen
coupling scheme. The tree constructed this way then contains all
Clebsch-Gordan coefficients that result from the decomposition
of a vector-coupling state belonging to the particular coupling
scheme into product states (cf. Eqs. \fmref{eq:clebsch_extended}
and \fmref{eq:gen_clebsch}).

\begin{figure}[ht!] \centering
   \includegraphics[width=80mm]{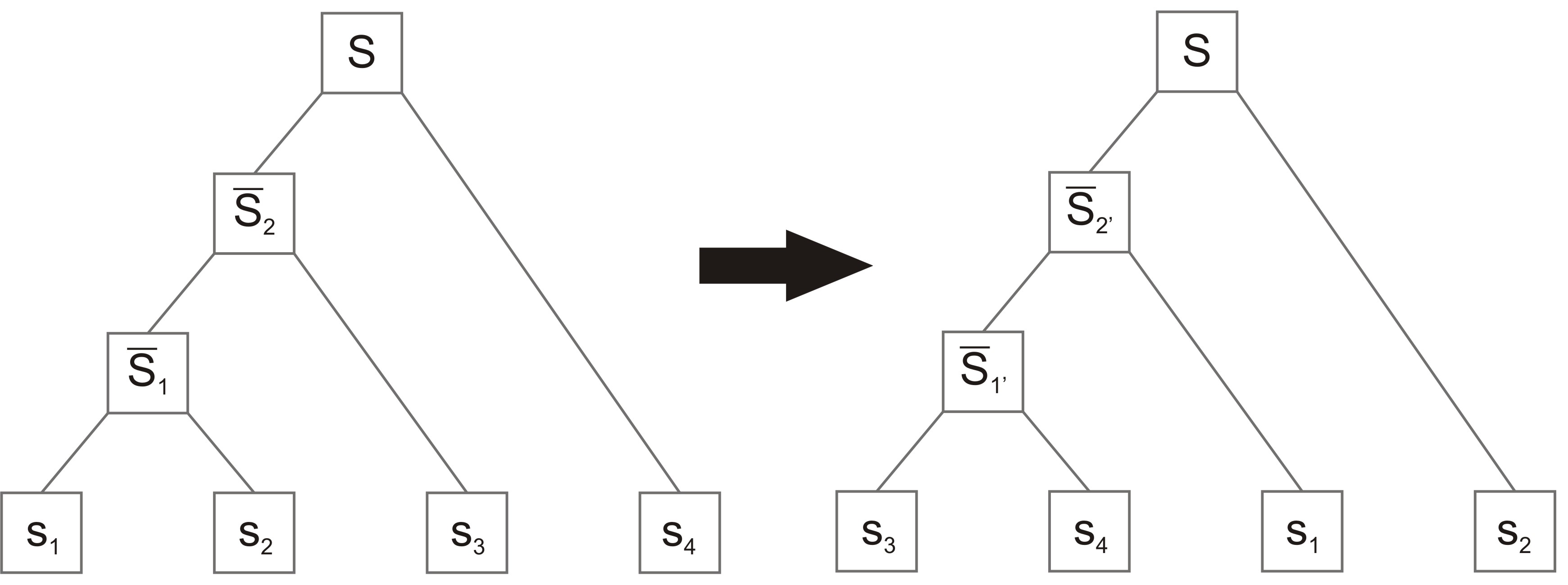}
\caption{Transition between two binary trees that has to be
  performed in order to calculate the recoupling coefficient
  $\braket{s_1 s_2 \overline{S}_{1} s_3 \overline{S}_2 s_4 S
    M}{s_3 s_4 \overline{S}_{1'} s_1 \overline{S}_{2'} s_2 S
    M}$.} 
\label{fig:tree1-tree2} 
\end{figure}

Operating on binary trees in order to generate a formula for a
general recoupling coefficient directly leads to the procedure
that limits the resulting expressions to 6J symbols. Here, the
generation of a formula for the recoupling coefficient
$\braket{s_1 s_2 \overline{S}_{1} s_3 \overline{S}_2 s_4 S
M}{s_3 s_4 \overline{S}_{1'} s_1 \overline{S}_{2'} s_2 S M}$,
that appears in Eq.~\fmref{eq:Wirkung_Drehung_final}, shall be
presented. In this case, generating a recoupling formula
corresponds to finding a transition between the binary trees
that are shown in Fig. \ref{fig:tree1-tree2}. In other words,
one transforms the set of Clebsch-Gordan coefficients that is
represented by the initial tree (l.h.s. of
Fig. \ref{fig:tree1-tree2}) to the set that is represented by
the targeted tree (r.h.s. of Fig. \ref{fig:tree1-tree2}).

Following the usual graph-theoretical name convention the
single-spin quantum numbers are referred to as \textit{leave
nodes} while the intermediate spin quantum numbers are called
\textit{coupled nodes}. The total-spin quantum number appears as
a coupled node of a special kind and is called \textit{root}.

\begin{figure}[t!]
  \centering
  \subfigure[\ Exchange]{
    \label{fig:binary_ops_ex}
    \includegraphics*[width=55mm]{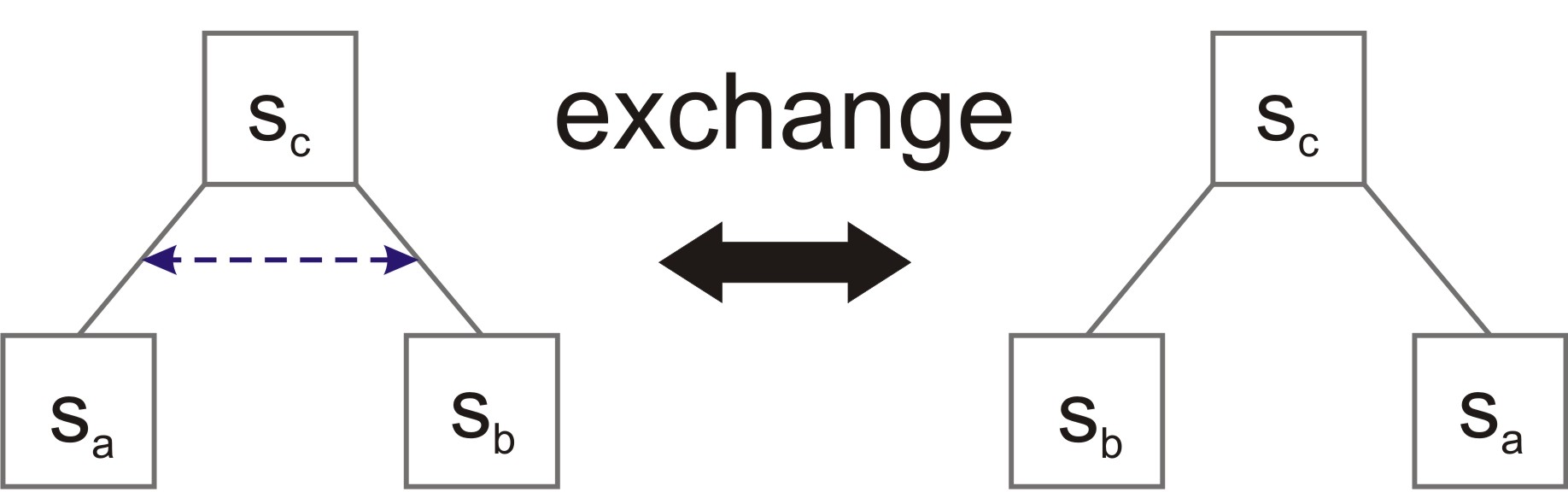} 
  }
  \subfigure[\ Flop]{
    \label{fig:binary_ops_fl}
    \includegraphics*[width=80mm]{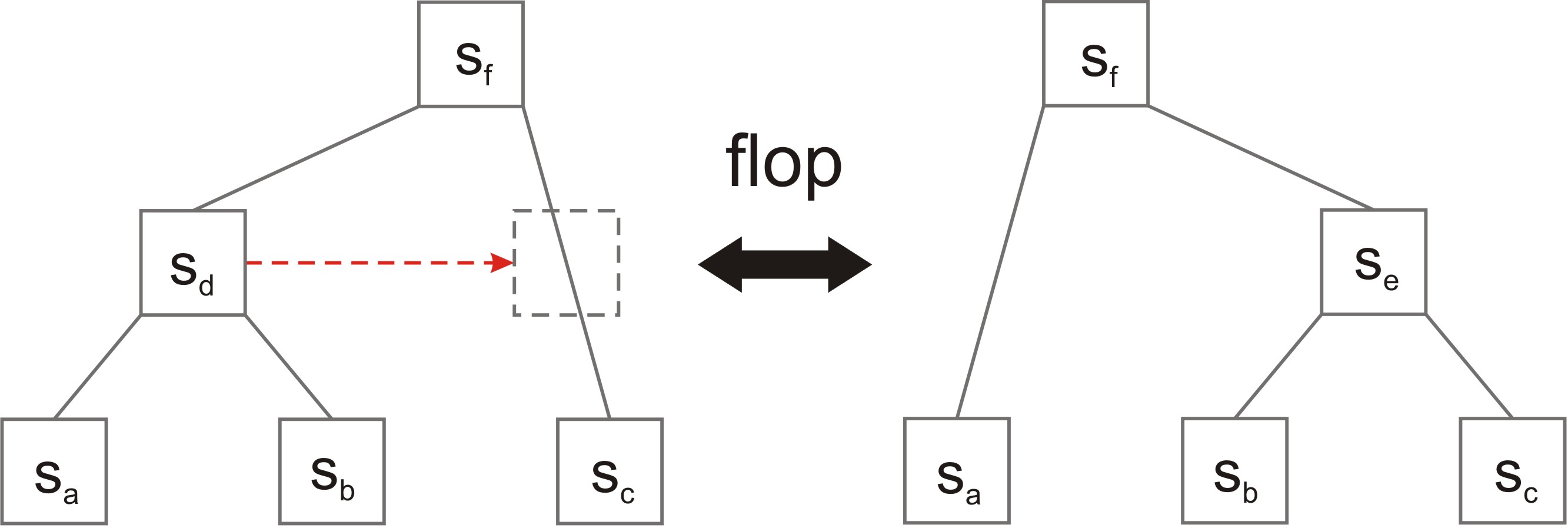} 
  }
  \caption{Operations on a binary tree that lead to a formula for a general recoupling coefficient.}
  \label{fig:binary_ops}
\end{figure}

There are in general two types of operations that have to be
performed in a certain manner in order to yield the desired form
of the recoupling coefficient. These operations are shown in
Fig. \ref{fig:binary_ops} and are called an \textit{exchange}
operation and a \textit{flop} operation. Both operations are
only performed on subtrees of the initial tree, thus only
leading to changes in the particular subtree while leaving the
rest of the tree unchanged. With every operation a certain
contribution to the recoupling formula is obtained.

An exchange operation refers to a recoupling coefficient that
appears when considering the recoupling of two spins $s_a$ and
$s_b$ within a single triad. Obviously, the only way of
recoupling these spins is to perform an exchange between
them. The effect of this operation can easily be derived from
unfolding the vector-coupling states $\ket{s_a s_b s_{c}}$ and
$\ket{s_b s_a s_{c}}$ in terms of product states according to
Eq.~\fmref{eq:clebsch}. A state of the form $\ket{s_a s_b
s_{c}}$ can be written as a state of the form $\ket{s_b s_a
s_{c}}$ by using the symmetry property of the Clebsch-Gordan
coefficients from Eq.~\fmref{eq:symmetry_clebsch}, leading to a
recoupling coefficient
\begin{equation}
   \braket{s_a s_b s_{c}}{s_b s_a s_{c}} = (-1)^{\pm (s_a+s_b-s_{c})}
   \ .
\end{equation}

In analogy, a flop operation refers to the recoupling of three
spins $s_a$, $s_b$, and $s_c$. Denoting the intermediate spin by
$s_d$ and the total spin by $s_f$, a successive coupling scheme
would lead to states that can be written as $\ket{s_a s_b s_d
s_c s_f}$. However, a second coupling scheme can be designed
that results in states of the form $\ket{s_a s_b s_c s_e
s_f}$. By definition a transition between these coupling schemes
is described by a Wigner-6J symbol resulting in
\begin{equation} \begin{split}
   & \braket{s_a s_b s_{d} s_c s_{f}}{s_a s_b s_c s_{e} s_{f}} = \\
   & \quad (-1)^{\pm (s_a+s_b+s_c+s_{f})} \sqrt{(2s_{d}+1)(2s_{e}+1)} \\
   & \quad \times \sixj{s_a}{s_b}{s_{d}}{s_c}{s_{f}}{s_{e}}
   \ .
\end{split} \end{equation}
It has to be mentioned that the flop operation shown in
Fig. \ref{fig:binary_ops_fl} was assumed to create a node,
i.e. a spin quantum number that already exists in the targeted
coupling scheme, namely $s_e$. Whenever a flop operation is
performed that creates a node which is unknown in the targeted
coupling scheme, a summation variable has to be introduced
within the resulting contribution to the recoupling
formula. This summation variable is completely determined by the
symmetry of the appearing Wigner-6J coefficient. The
contribution resulting from a flop operation that creates an
unknown node $s_{e'}$ within the binary tree would look like
\begin{equation*} \begin{split}
   & (-1)^{\pm (s_a+s_b+s_c+s_{f})} \sum_{s_{e'}} \sqrt{(2s_{d}+1)(2s_{e'}+1)} \\
   & \quad \times \sixj{s_a}{s_b}{s_{d}}{s_c}{s_{f}}{s_{e'}}
   \ .
\end{split} \end{equation*}

\begin{figure}[ht!]
  \centering
  \subfigure[\ Flop]{
    \label{fig:binary_1}
    \includegraphics[width=35mm]{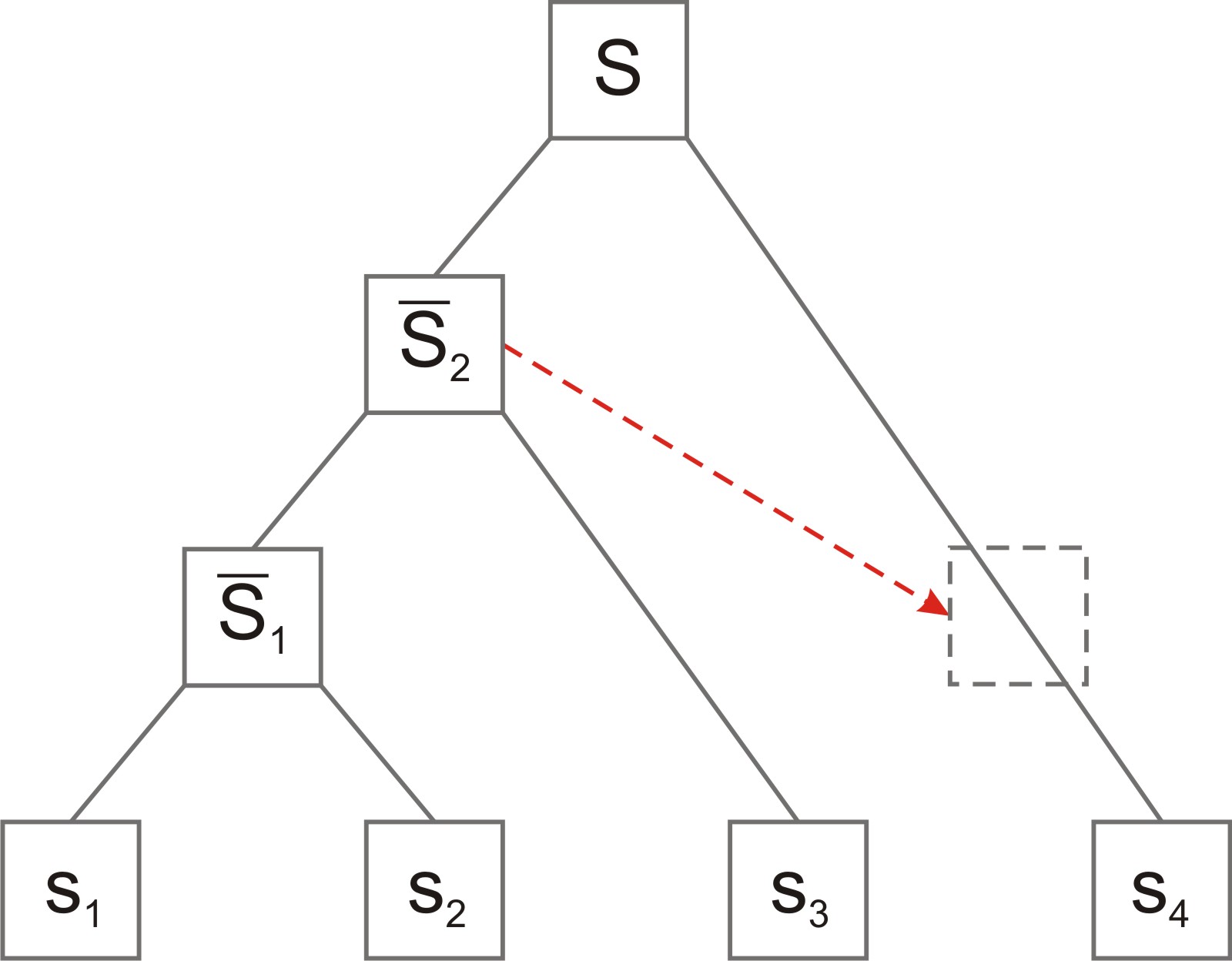} 
  }
  \subfigure[\ Exchange]{
    \label{fig:binary_2}
    \includegraphics[width=35mm]{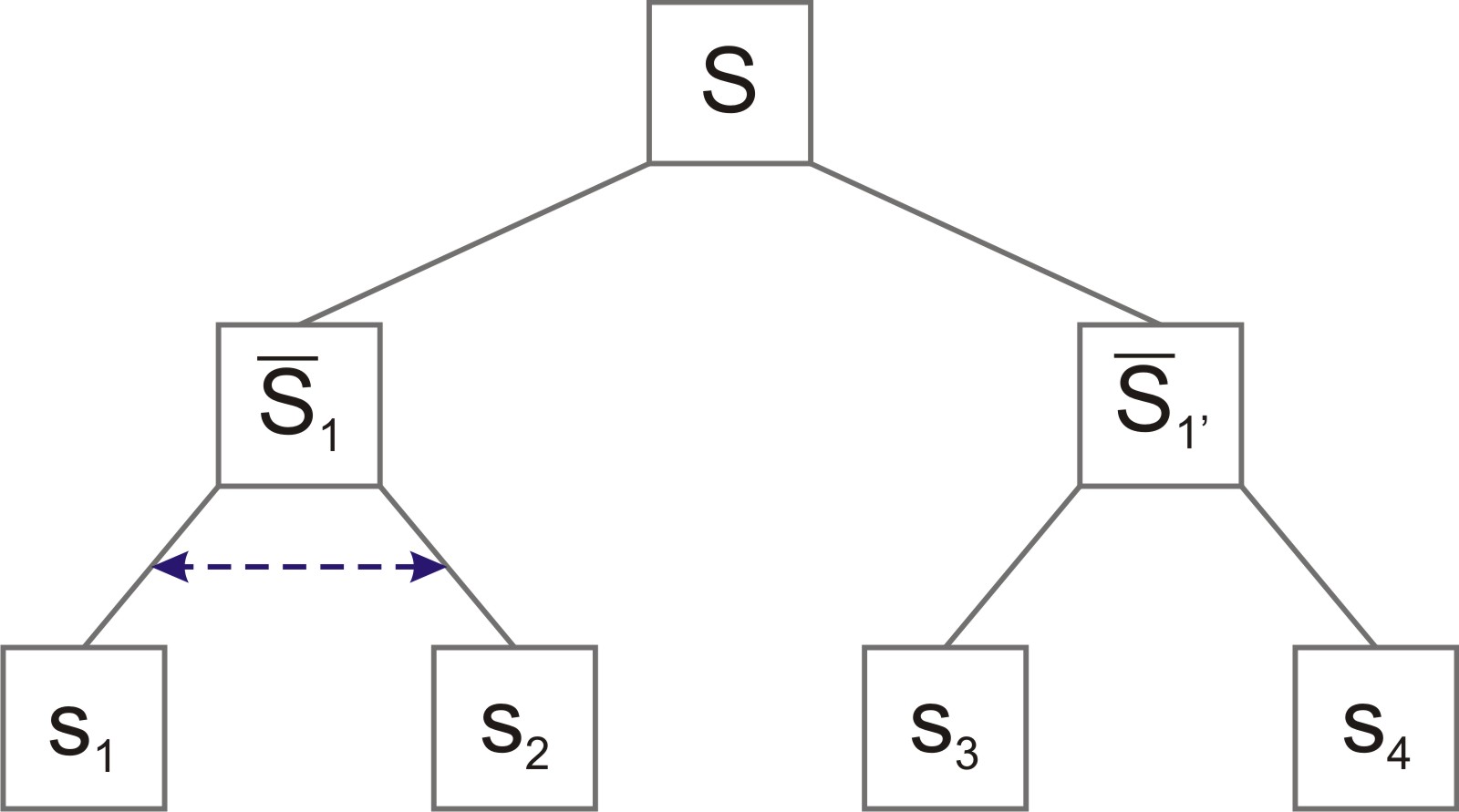} 
  }
    \subfigure[\ Flop]{
    \label{fig:binary_3}
    \includegraphics[width=35mm]{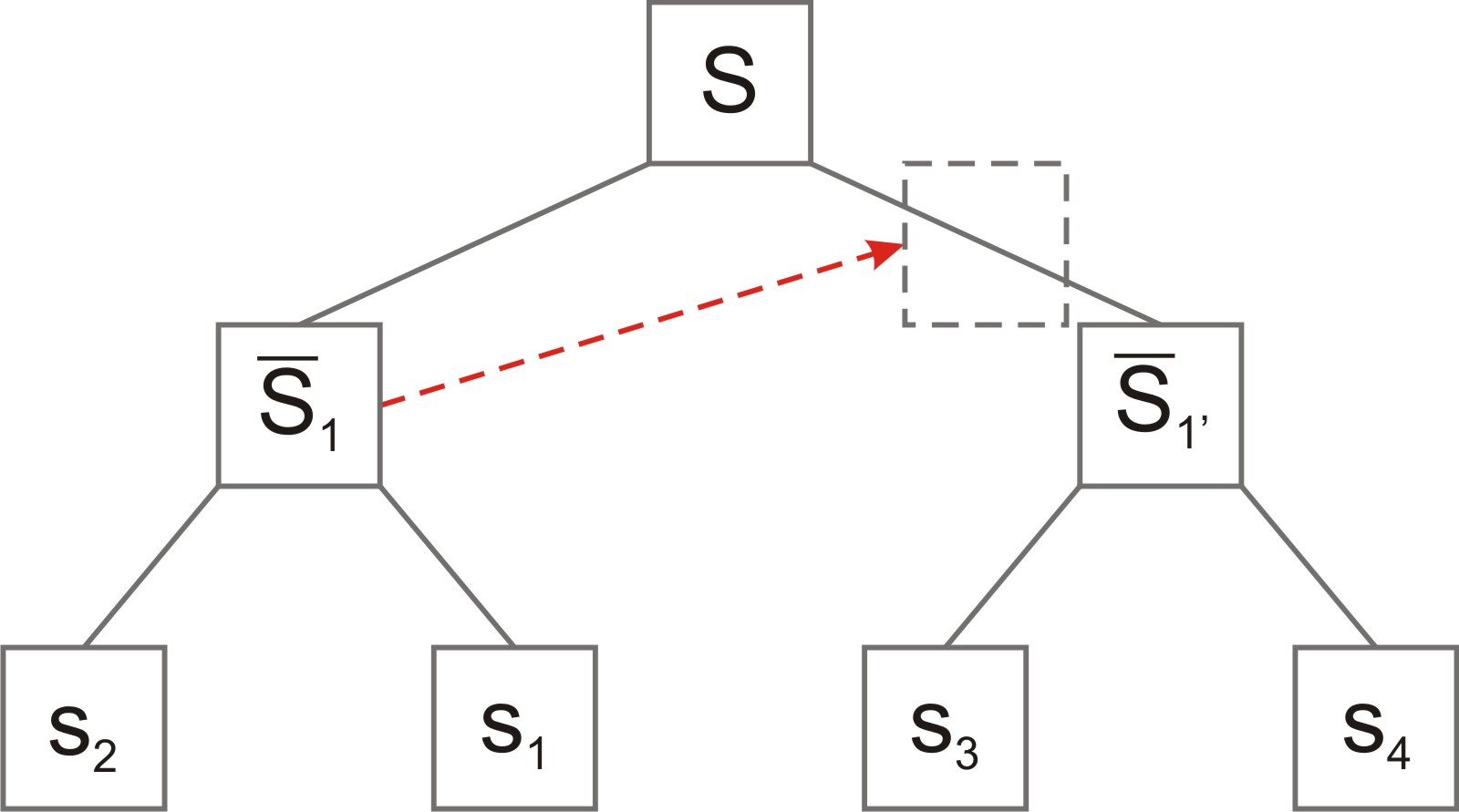} 
  }
      \subfigure[\ Exchange]{
    \label{fig:binary_4}
    \includegraphics[width=35mm]{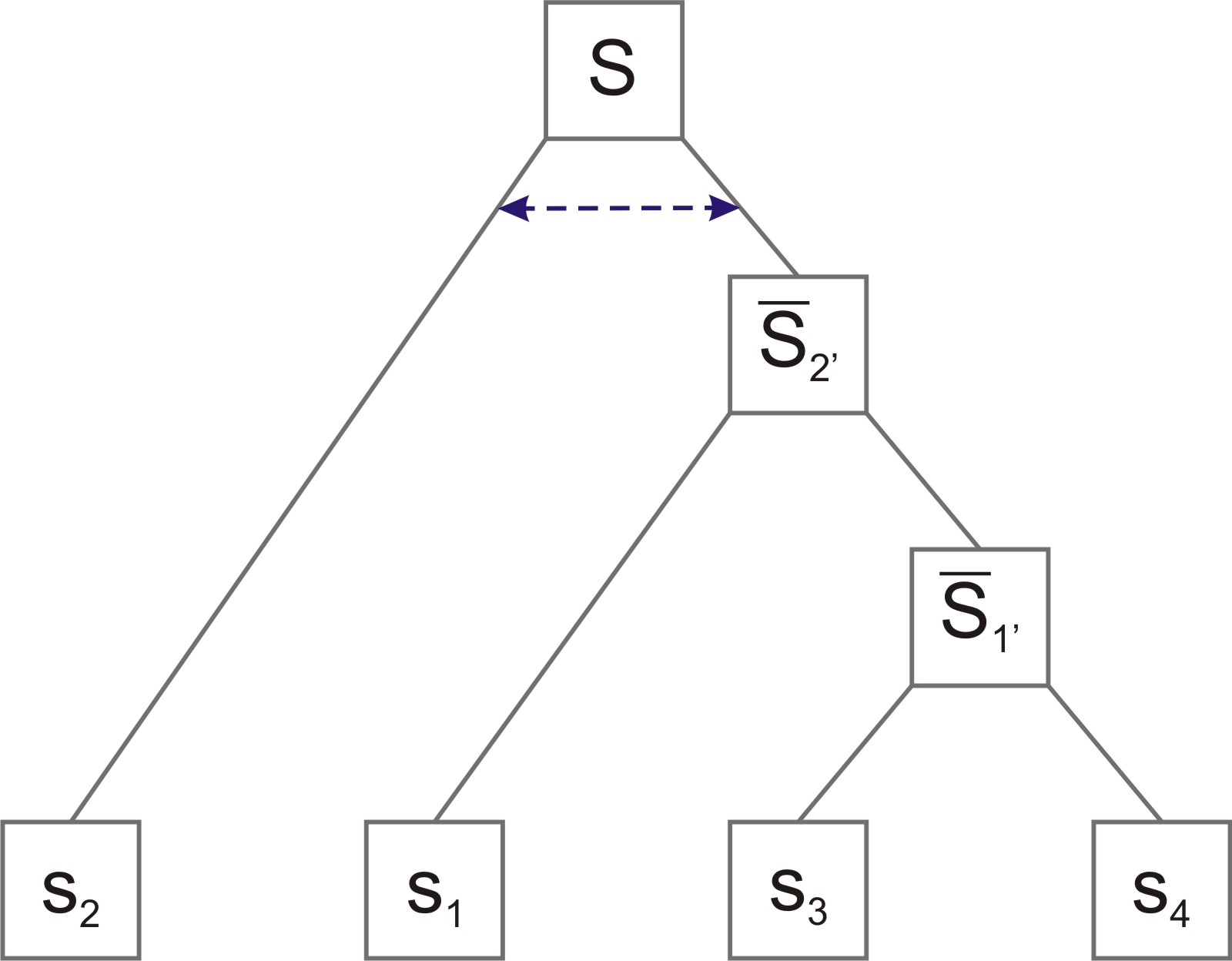} 
  }
      \subfigure[\ Exchange]{
    \label{fig:binary_5}
    \includegraphics[width=35mm]{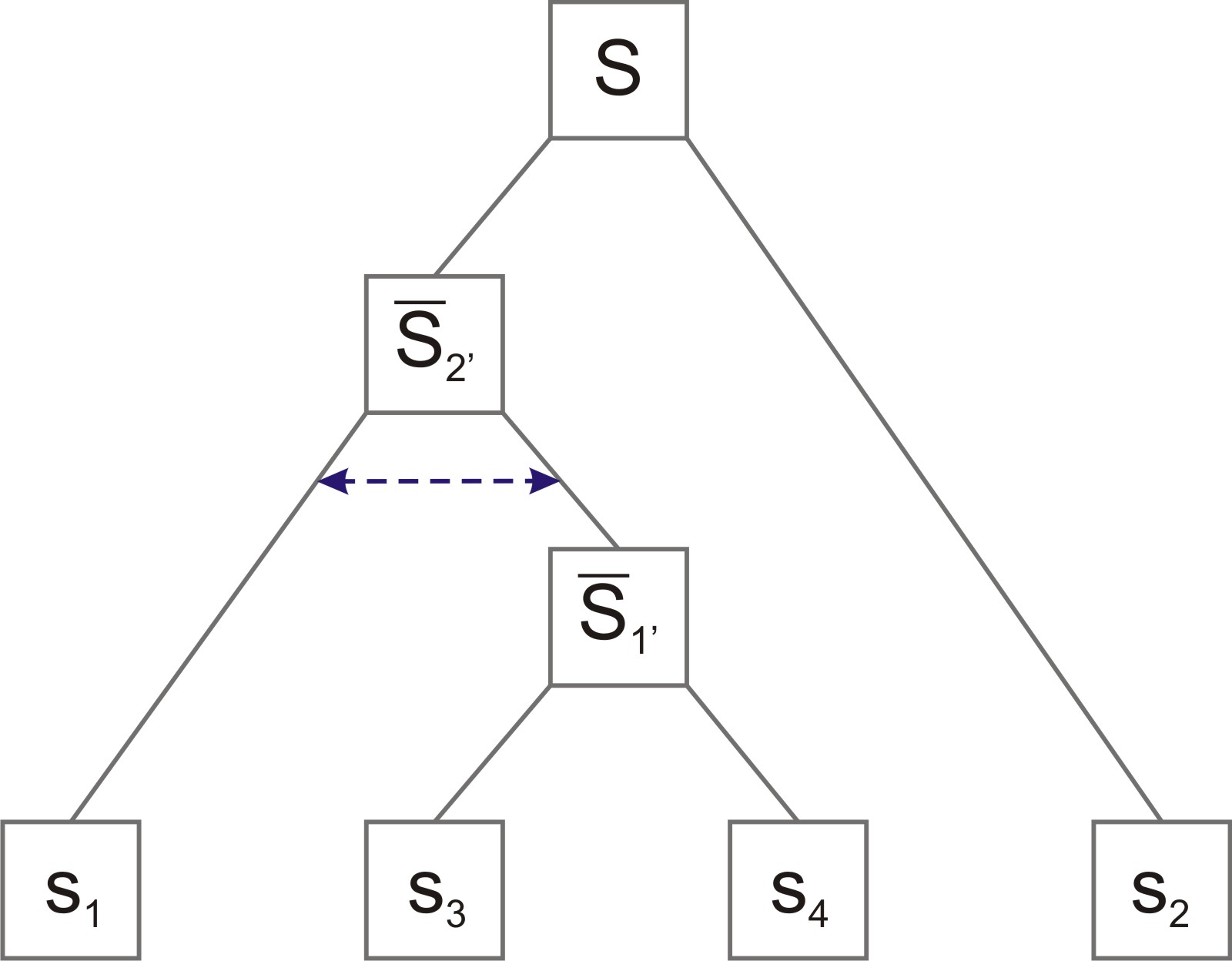} 
  }
  \caption{Sequence of operations leading to the recoupling
  coefficient $\braket{s_1 s_2 \overline{S}_{1} s_3
  \overline{S}_2 s_4 S M}{s_3 s_4 \overline{S}_{1'} s_1
  \overline{S}_{2'} s_2 S M}$.} 
  \label{fig:binary_seq} 
\end{figure}

The desired formula for the recoupling coefficient from
Eq.~\fmref{eq:Wirkung_Drehung_final} is now obtained by
performing a proper sequence of exchange and flop
operations. This sequence is in detail displayed in
Fig. \ref{fig:binary_seq}. It is not the only possible sequence
of operations on the binary tree that leads to a recoupling
formula for the discussed transition. However, in this simple
case the displayed sequence leads to an optimal formula
minimizing the number of resulting Wigner-6J symbols.

The result of the operations then reads
\begin{equation} \begin{split} \label{eq:recoupling_pi}
    & \braket{s_1 s_2 \overline{S}_{1} s_3 \overline{S}_2 s_4 S M}{s_3 s_4 \overline{S}_{1'} s_1 \overline{S}_{2'} s_2 S M}  = \\
    & \quad \sqrt{(2\overline{S}_{1'}+1)(2\overline{S}_{2}+1)(2\overline{S}_{1}+1)(2\overline{S}_{2'}+1)} \\
    & \times (-1)^{s_1+s_2+s_3+s_4+3S}  \sixj{\overline{S}_1}{s_3}{\overline{S}_2}{s_4}{S}{\overline{S}_{1'}} \sixj{s_2}{s_1}{\overline{S}_{1}}{\overline{S}_{1'}}{S}{\overline{S}_{2'}}
    \ .
\end{split} \end{equation}
This simple form was only reached because the flop operations
shown in Figs. \ref{fig:binary_1} and \ref{fig:binary_3} create
nodes that already exist in the final coupling scheme,
i.e. $\overline{S}_{1'}$ and $\overline{S}_{2'}$.

As long as such simple recoupling coefficients are considered,
the process of determining a proper sequence leading from the
initial coupling scheme to the targeted one can be easily done
by hand and does not need any automatization. Nevertheless, more
sophisticated problems result in transitions between binary
trees that cannot easily be treated. Then, it becomes necessary
to set up an algorithm that automatically creates a proper --
ideally optimal -- sequence. Regarding binary trees, known
implementations\cite{Bur:CPC70,FPV:CPC94} of such algorithms can
be seen as trial-and-error procedures.

By performing a subsequence of operations, initially containing
only one exchange or flop operation, one tries to find a tree
containing a node that is known in the targeted coupling
scheme. Whenever it is impossible to find such a tree with the
given number of operations in the subsequence, the number of
considered operations is increased, i.e. the subsequence is
extended. A successful implementation of this procedure leads to
a stepwise creation of the targeted tree in which each step is
guaranteed to be performed with the smallest (overall) number of
exchange and flop operations. However, the minimization of the
number of operations within the performed subsequences does not
assure that the resulting recoupling formula is optimal. In
general, a recoupling formula is optimal if the number of
occurring summation variables and Wigner-6J coefficients is
minimal. Since summation variables and 6J symbols are only
introduced by flop operations, generating an improved recoupling
formula directly corresponds to reducing the number of performed
flops.

\subsection{Graph theoretical solution - Yutsis graphs} \label{sec-B-2}
As it was shown in the last section, operating on binary trees
in order to generate a recoupling formula involving only phase
factors, square roots, and Wigner-6J symbols already leads to a
simple and successful procedure. However, the process of
determining an optimized sequence of operations remains
concealed. In order to improve this process and thus improving
the recoupling formula, ideas resulting from more advanced
graph-theoretical considerations can be applied. In
Refs. \onlinecite{BaK:CPC88,Lim:CPC91,FPV:CPC97,DyF:CPC2003} the
problem of generating a recoupling formula was solved with the
help of \textit{Yutsis graphs}. This procedure, providing a
technically more difficult, but at the same time theoretically
more transparent way of generating an improved recoupling
formula, shall be reviewed in this section.

The creation of Yutsis graphs is a straightforward task starting
from the background given in App. \ref{sec-B-1}. In order to
understand how these graphs evolve, an explanation of how to
construct a Yutsis graph shall be given here. Additionally, the
reduction of such a graph leading to an improved recoupling
formula will be discussed briefly. The interested reader will
find a deeper and more theoretical investigation of general
features of Yutsis graphs in the
literature.\cite{YLV:angular62,BiL:EMA81}

As already discussed in Sec. \ref{sec-A-1} an expression for a
recoupling coefficient in terms of Clebsch-Gordan coefficients
or Wigner-3J symbols can be found by decomposing the bra and the
ket states into sums of product states. In order to clarify this
procedure, the recoupling coefficient $\braket{s_1 s_2
\overline{S}_{1} s_3 \overline{S}_2 s_4 S M}{s_3 s_4
\overline{S}_{1'} s_1 \overline{S}_{2'} s_2 S M}$ shall be
discussed in detail as an example. The decomposition of the bra
state was already done in App. \ref{sec-A-5} and is given in
terms of Clebsch-Gordan coefficients in
Eq.~\fmref{eq:Uebergang_S_zu_m}. Replacing the Clebsch-Gordan
coefficients by Wigner-3J symbols yields
\begin{equation} \begin{split} \label{eq:ex_decomposition_bra}
   & \ket{s_1 s_2 \overline{S}_1 s_3 \overline{S}_2 s_4 \, S \, M} = \\
   & \sum_{\{m_i\}} \left\{ \sum_{\{\overline{M}_i\}} C(\alpha) \threej{s_1}{s_2}{\overline{S}_1}{m_1}{m_2}{-\overline{M}_1}
   \threej{\overline{S}_1}{s_3}{\overline{S}_2}{\overline{M}_1}{m_3}{-\overline{M}_2} \right. \\
   & \left. \quad \times \threej{\overline{S}_2}{s_4}{S}{\overline{M}_2}{m_4}{-M} \right\} \ket{m_1 \, m_2 \, m_3 \, m_4}
   \ ,
\end{split} \end{equation}
where $C(\alpha)$ contains the square roots as well as the phase
factors that appear when transforming Clebsch-Gordan
coefficients into Wigner symbols. The multiple sums run over all
single-spin magnetic quantum numbers $m_i$ and the magnetic
quantum numbers $\overline{M}_i$ of the intermediate spins. The
curly brackets are reminiscent of a generalized Wigner
coefficient. The same decomposition yields for the ket state
\begin{equation} \begin{split} \label{eq:ex_decomposition_ket}
   & \ket{s_3 s_4 \overline{S}_{1'} s_1 \overline{S}_{2'} s_2 \, S \, M} = \\
   & \sum_{\{m_i\}} \left\{ \sum_{\{\overline{M}_{i'}\}} C(\beta) \threej{s_3}{s_4}{\overline{S}_{1'}}{m_3}{m_4}{-\overline{M}_{1'}}
   \threej{\overline{S}_{1'}}{s_1}{\overline{S}_{2'}}{\overline{M}_{1'}}{m_1}{-\overline{M}_{2'}} \right. \\
   & \left. \quad \times \threej{\overline{S}_{2'}}{s_2}{S}{\overline{M}_{2'}}{m_2}{-M} \right\} \ket{m_1 \, m_2 \, m_3 \, m_4}
   \ .
\end{split} \end{equation}

From Eqs. \fmref{eq:ex_decomposition_bra} and \fmref{eq:ex_decomposition_ket} one immediately finds the following expression for the recoupling coefficient:
\begin{equation} \label{eq:ex_recoupling_coefficient} \begin{split}
   & \braket{s_1 s_2 \overline{S}_1 s_3 \overline{S}_2 s_4 \, S \, M}{s_3 s_4 \overline{S}_{1'} s_1 \overline{S}_{2'} s_2 \, S \, M} \\
   & = \sum_{\{m_i\},\{m'_i\}}  P^{\alpha}(\{m_i\}) \cdot P^{\beta}(\{m'_{i}\}) \\
   &\qquad \times \braket{m_1 \, m_2 \, m_3 \, m_4}{m'_1 \, m'_2 \, m'_3 \, m'_4} \\
   & = \sum_{\{m_i\},\{m'_i\}} \delta_{\{m_i\},\{m'_{i}\}} \ P^{\alpha}(\{m_i\})  \cdot P^{\beta}(\{m'_{i}\})
   \ . \end{split}
\end{equation}
The abbreviations $P^{\alpha}$ and $P^{\beta}$ stand for the
generalized Wigner coefficients that depend on the quantum
numbers of the underlying coupling schemes which the sets
$\alpha$ and $\beta$ refer to. The derived expression for the
recoupling coefficient in
Eq.~\fmref{eq:ex_recoupling_coefficient} is still involving
magnetic quantum numbers. Since a recoupling coefficient is in
general independent of any magnetic quantum number, a
simplification can be found that only involves spin quantum
numbers. Such a simplification was already found in the former
section with the help of binary trees and will now be discussed
on the basis of Yutsis graphs.

The main idea of generating a recoupling formula with the help
of Yutsis graphs is -- in a first step -- to set up a graphical
representation of generalized Wigner coefficients as they appear
in Eqs. \fmref{eq:ex_decomposition_bra} and
\fmref{eq:ex_decomposition_ket}. Afterwards, these graphs are
joined in order to build up a Yutsis graph that represents the
recoupling coefficient. A simplification of the constructed
Yutsis graph according to special operations then leads to the
desired formula that is independent of magnetic quantum numbers.

\begin{figure}[ht!]
  \centering  
  \subfigure{
    \label{fig:yutsis_triad_p}
    \includegraphics[width=21mm]{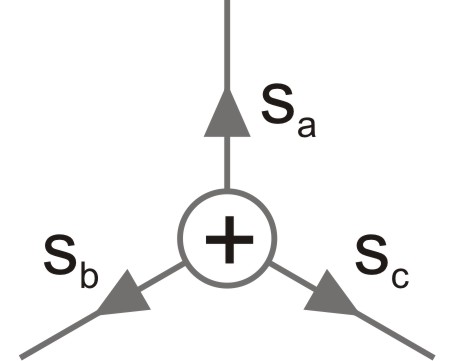} 
  }
  \hspace{10mm}
  \subfigure{
    \label{fig:yutsis_triad_m}
    \includegraphics[width=21mm]{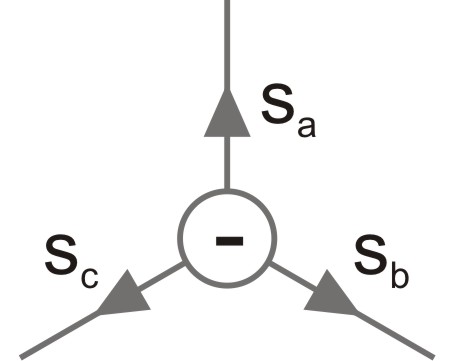} 
  }
  \caption{Two diagrams of the same Wigner-3J symbol with
  different sign of the nodes. The sign is related to the cyclic
  ordering of the lines.}
  \label{fig:yutsis_triad}
\end{figure}

The building blocks of Yutsis graphs are diagrammatic
representations of Wigner-3J symbols. Figure
\ref{fig:yutsis_triad} shows two diagrammatic representations of
the same Wigner-3J symbol
\begin{equation*}
   \threej{s_a}{s_b}{s_c}{m_a}{m_b}{m_c}
   \ .
\end{equation*}
Such a representation consists of three lines and one node. With
every spin quantum number in the Wigner symbol a line is
identified. The three lines are connected by the node. The node
is labeled with a $(+)$ or $(-)$ sign while the lines are
characterized by the direction they are pointing in. The $(-)$
sign denotes a clockwise orientation of the spin quantum numbers
within the corresponding Wigner-3J symbol
(cf. Fig. \ref{fig:yutsis_triad} r.h.s.) whereas the $(+)$ sign
indicates an anticlockwise ordering
(cf. Fig. \ref{fig:yutsis_triad} l.h.s.). The free ends of the
lines represent the projections of the spin quantum numbers,
i.e. the magnetic quantum numbers $m_a$, $m_b$, and $m_c$. If a
line leads away from the node, the corresponding magnetic
quantum number appears with a positive sign in the Wigner
symbol, whereas it appears with a negative sign if the line is
directed towards the node.

It is obvious that any operation that changes the diagrams of
the Wigner symbol in Fig. \ref{fig:yutsis_triad} will lead to a
Wigner symbol that differs from the original one. Changing the
sign of the node or simultaneously changing the directions of
all lines results in a factor that can be obtained from the
symmetry properties of the Wigner-3J symbols described in
Sec. \ref{sec-A-1}. The change of the sign corresponds to an
uneven permutation of spins within the Wigner-3J symbol
(cf. Eq.~\fmref{eq:wigner3J_symmetry}) while the change of all
directions of the lines corresponds to multiplying the lower row
of the original Wigner-3J symbol by $-1$
(cf. Eq.~\fmref{eq:wigner3J_symmetry2}). Both operations result
in a phase factor of $(-1)^{s_a+s_b+s_c}$ whereas any rotation
of the diagram has no effect on the Wigner-3J symbol since the
ordering remains unchanged.

\begin{figure}[ht!]
  \centering  
  \subfigure{
    \label{fig:yutsis_summation_open}
    \includegraphics[width=35mm]{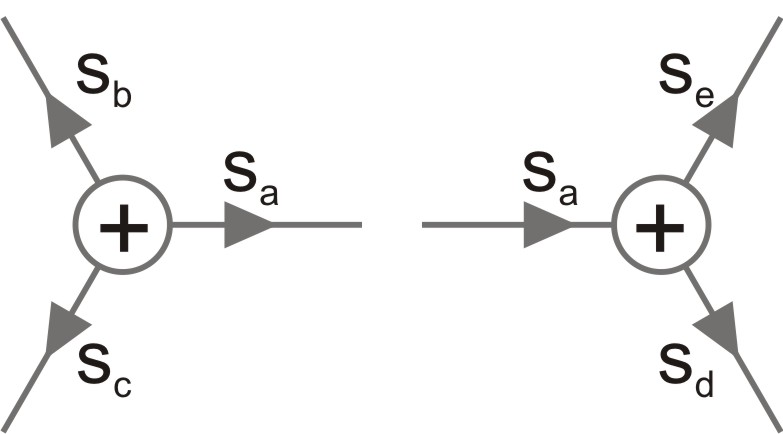} 
  } \hspace{4mm}
  \subfigure{
    \label{fig:yutsis_summation_closed}
    \includegraphics[width=27mm]{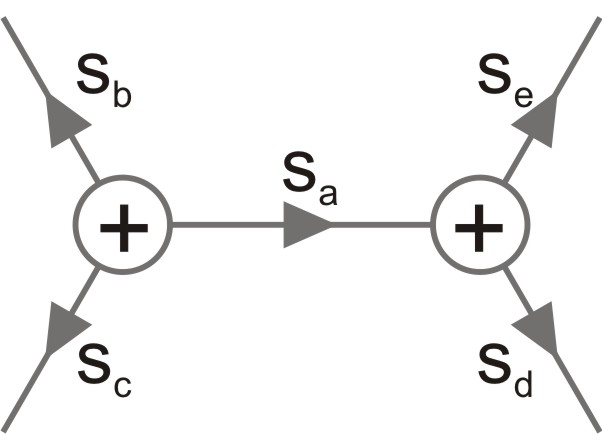} 
  }
  \caption{Contracting the open ends of two lines labeled by
  $s_a$ that leads to a summation over the magnetic quantum
  number $m_a$ of the joined lines.} 
  \label{fig:yutsis_summation}
\end{figure}

In order to construct a graph that represents a generalized
Wigner coefficient, another operation has to be introduced. As
shown in Fig. \ref{fig:yutsis_summation}, the diagrams of two
Wigner symbols can be contracted if two lines exist that are
labeled by the same quantum number and point into the same
direction. The resulting graph then represents a summation over
the corresponding magnetic quantum number given by
\begin{equation}
   \sum_{m_a} \threej{s_a}{s_b}{s_c}{m_a}{m_b}{m_c} \threej{s_a}{s_d}{s_e}{-m_a}{m_d}{m_e}
   \ .
\end{equation}

\begin{figure}[t!]
  \centering
  \subfigure[\ Initial coupling scheme]{
    \label{fig:yutsis_left}
    \includegraphics[width=43mm]{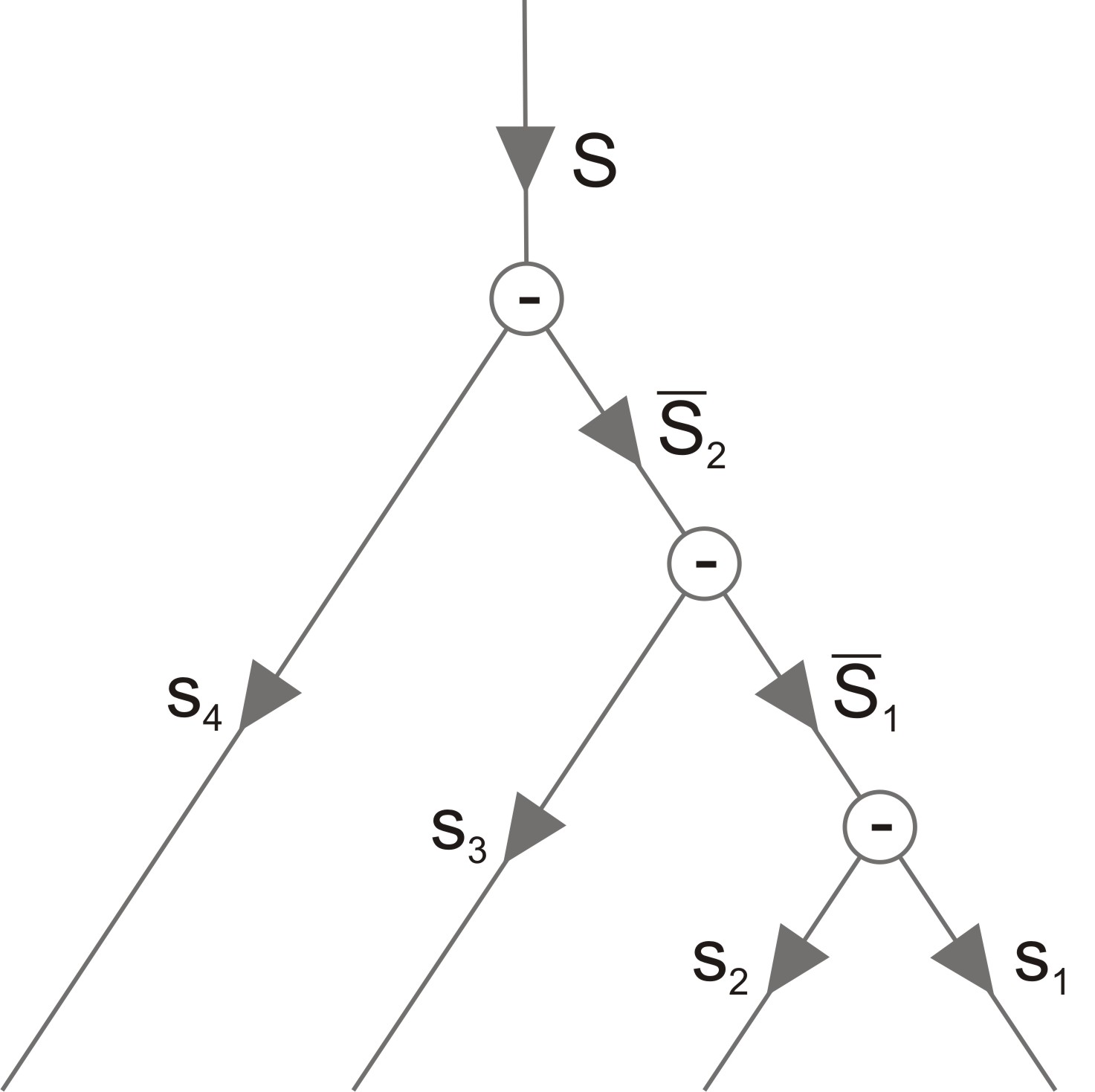} 
  }
  \subfigure[\ Targeted coupling scheme]{
    \label{fig:yutsis_right}
    \includegraphics[width=43mm]{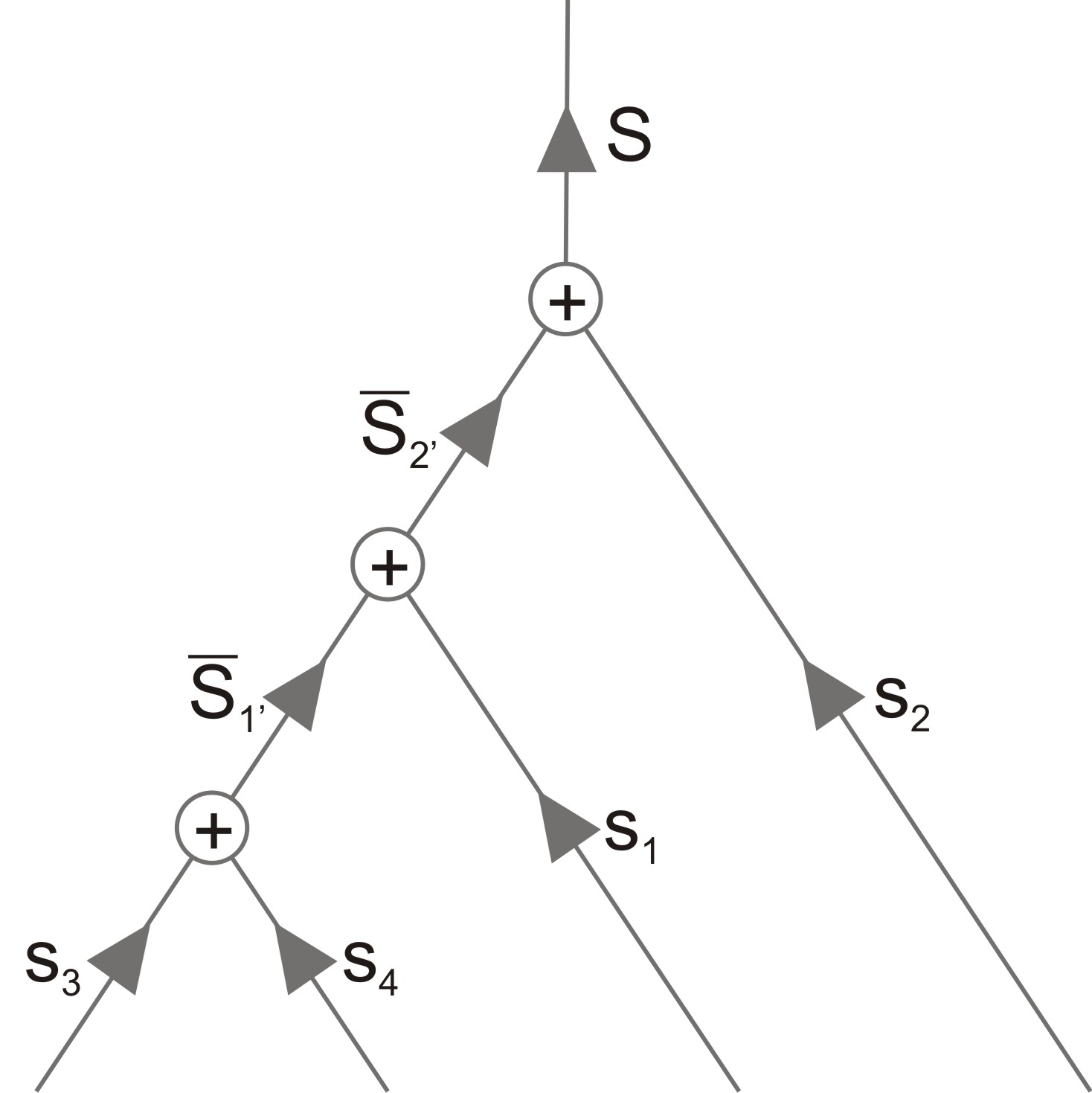} 
  }
    \subfigure[\ Yutsis graph]{
    \label{fig:yutsis_graph}
    \includegraphics[width=53mm]{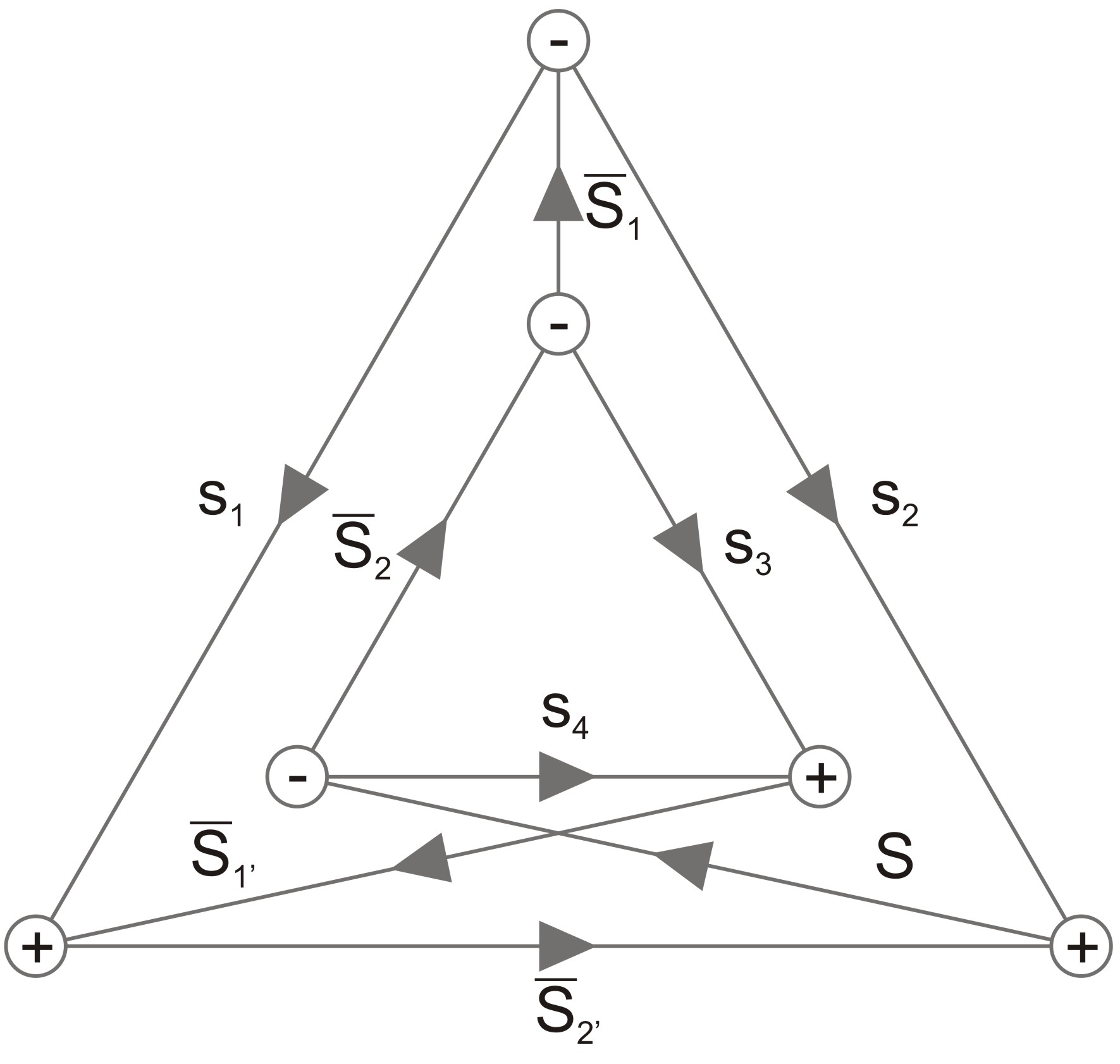} 
  }
  \caption{Graphical representation of the generalized Wigner
  coefficients for the coupling schemes contained in the
  recoupling coefficient $\braket{s_1 s_2 \overline{S}_{1} s_3
  \overline{S}_2 s_4 S M}{s_3 s_4 \overline{S}_{1'} s_1
  \overline{S}_{2'} s_2 S M}$ as well as the resulting Yutsis
  graph.} 
  \label{fig:edge-node}
\end{figure}

Figures \ref{fig:yutsis_left} and \ref{fig:yutsis_right} show
graphical representations of the generalized Wigner coefficients
as found in Eqs. \fmref{eq:ex_decomposition_bra} and
\fmref{eq:ex_decomposition_ket}. They are easily constructed
following the conventions introduced above.

The arrangement of the diagrams representing generalized Wigner
coefficients is chosen in such a way as to ease the contraction
of both graphs and was proposed in
Ref. \onlinecite{YLV:angular62}. The graph representing the left
hand side of the recoupling coefficient contains only negative
nodes with the spins being ordered clockwise around these
nodes. In the graph belonging to the right hand side of the
recoupling coefficient the spins are ordered anticlockwise
around the nodes that exclusively have a positive sign. The
directions of the lines are chosen in Fig. \ref{fig:yutsis_left}
as to match the conventional form of Wigner-3J symbols in
generalized Wigner coefficients. In Fig. \ref{fig:yutsis_right}
they are chosen in the opposite direction in order to compensate
the phase factors that result from the positive signs of the
nodes.

\begin{figure}[t!]
  \centering
  \subfigure[\ Reduction of a 2-cycle]{
    \label{fig:yutsis_2cycle}
    \includegraphics[width=72mm]{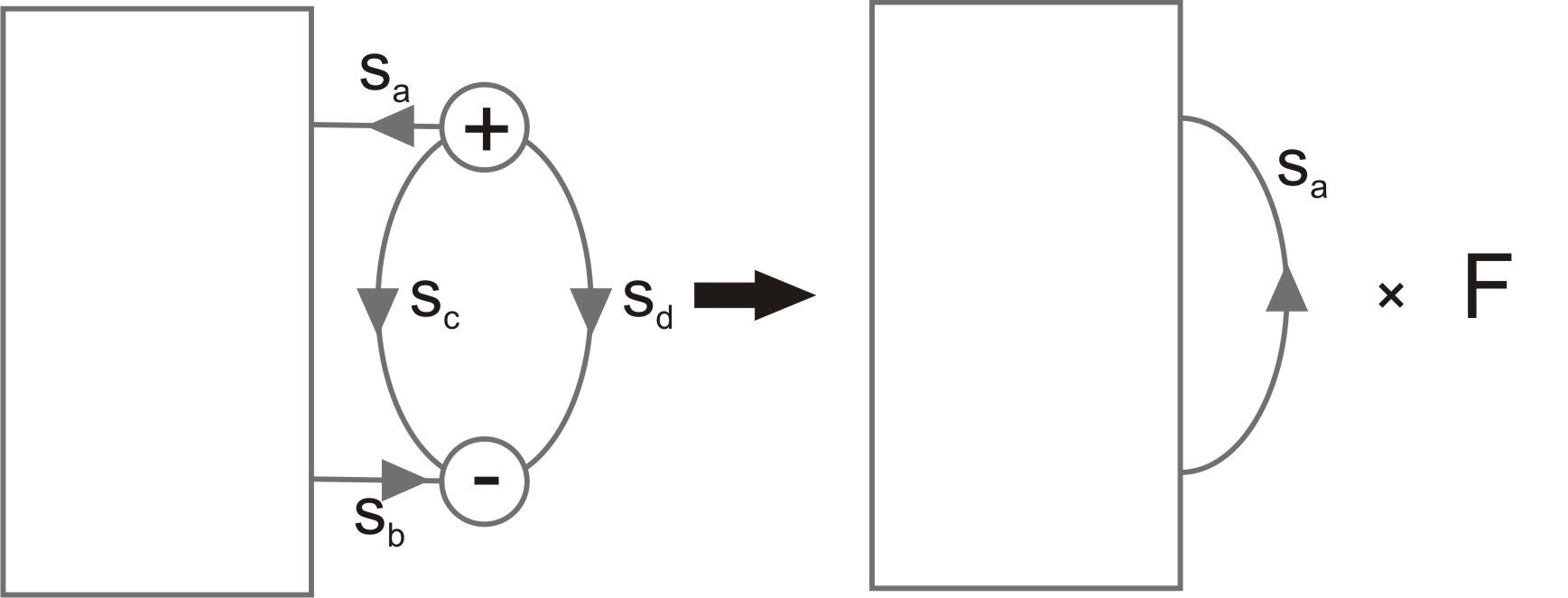} 
  }
  \subfigure[\ Reduction of a 3-cycle]{
    \label{fig:yutsis_3cycle}
    \includegraphics[width=72mm]{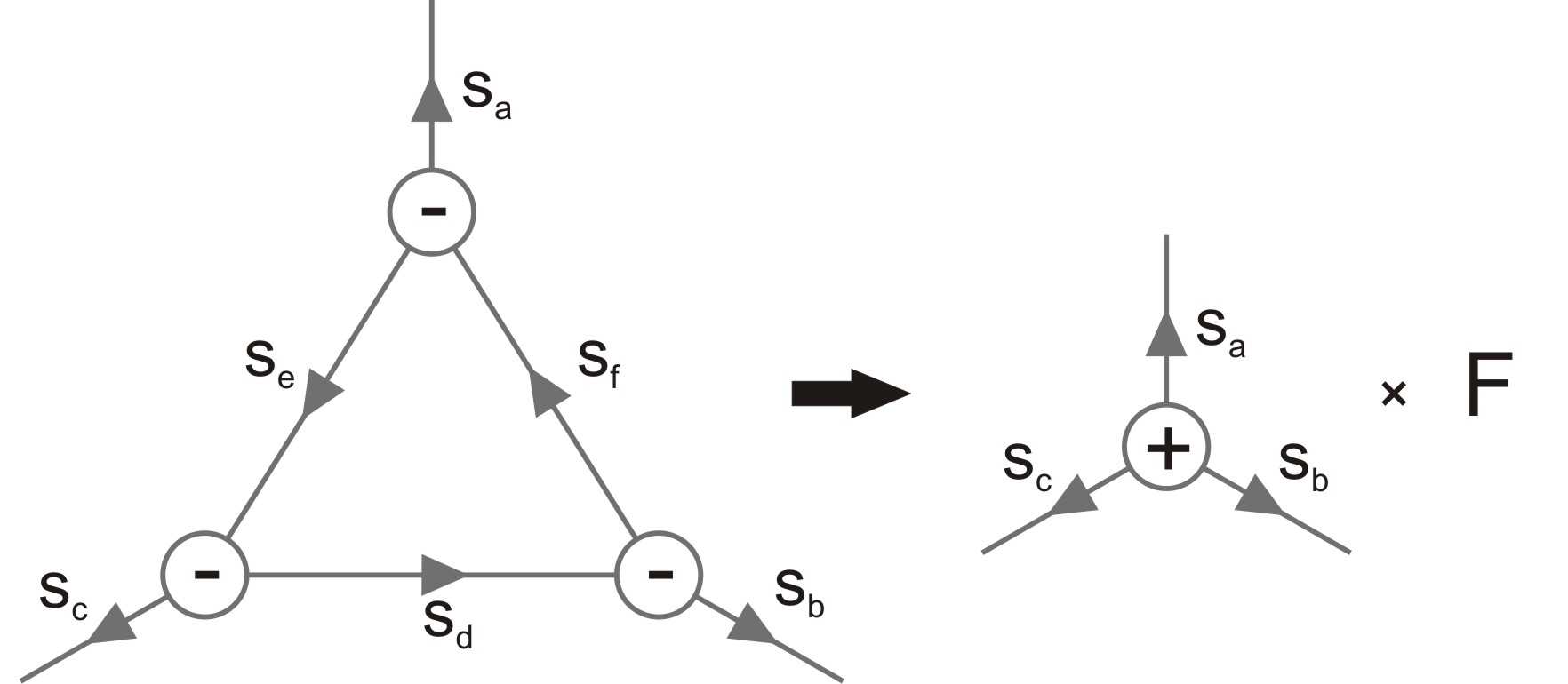}
  }
    \subfigure[\ Reduction of a 4-cycle]{
    \label{fig:yutsis_4cycle}
    \includegraphics[width=83mm]{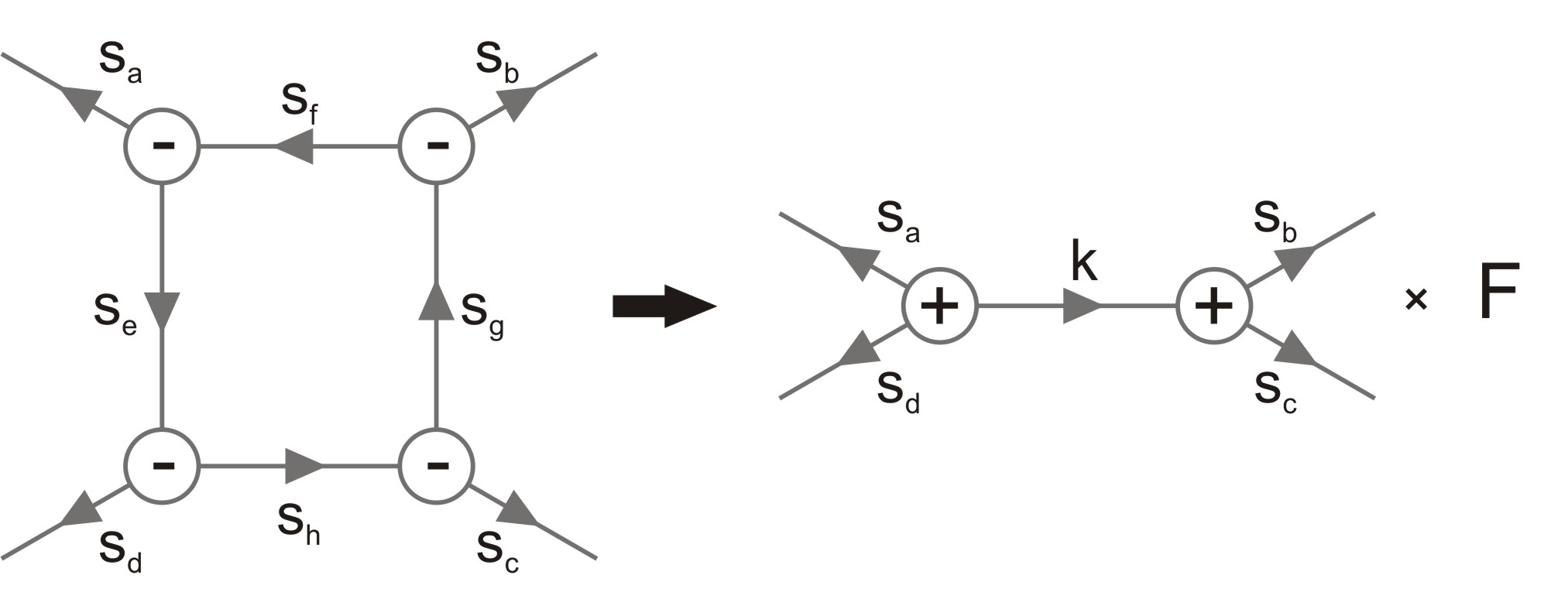}
  }
   \subfigure[\ Interchange]{
    \label{fig:yutsis_interchange}
    \includegraphics[width=77mm]{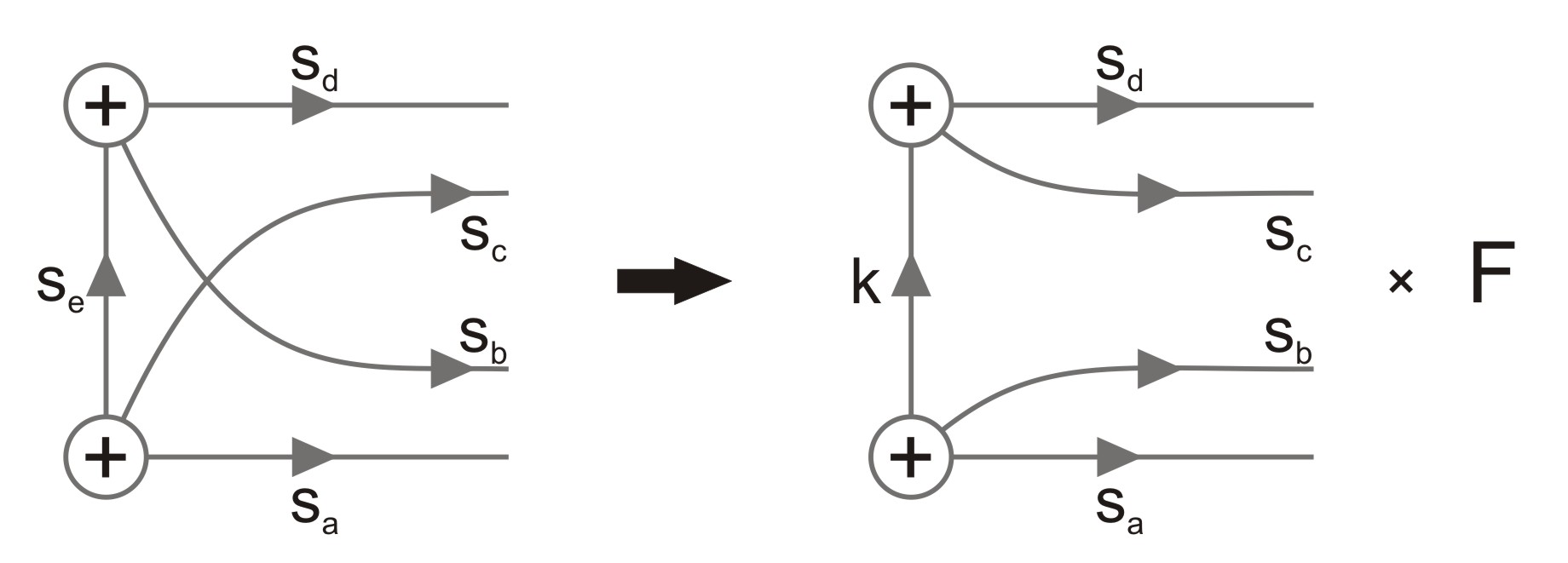}
  }
  \caption{Operations on a Yutsis graph and resulting contributions to the recoupling formula.}
  \label{fig:yutsis_operations}
\end{figure}

In Fig. \ref{fig:yutsis_graph} both diagrams of the Wigner
coefficients are contracted. The contraction corresponds to a
summation over the magnetic quantum numbers $m_i$,
$i=1,\dots,4$, as well as over $M$. The summations over the
magnetic quantum numbers of the intermediate spins
$\{\overline{M}_i\}$ and $\{\overline{M}_{i'}\}$ has already
been included in the representations of the coupling schemes in
Figs. \ref{fig:yutsis_left} and \ref{fig:yutsis_right},
respectively. The resulting Yutsis graph then represents --
apart from square roots and phase factors -- the recoupling
coefficient that is given in
Eq.~\fmref{eq:ex_recoupling_coefficient}.

In general, a recoupling coefficient is equal to a Yutsis graph,
which is constructed according to the above rules, times an
additional factor.\cite{YLV:angular62} This factor contains
phase factors and square roots that emerge from expressing
Clebsch-Gordan coefficients in terms of Wigner-3J symbols and
additional contributions from initializing the graphical
representation of the general Wigner coefficients (see
Figs. \ref{fig:yutsis_left} and \ref{fig:yutsis_right}). It can
be written as
\begin{equation} \label{eq:yutsis_factor}
   (-1)^{2\left(S + \sum_{i=1}^{N-2} \overline{S}_{i'}
   +\mathcal{S} \right)} \left[ \prod_{i=1}^{N-2}
   \left(2\overline{S}_i +1 \right) \left(2\overline{S}_{i'} +1
   \right)\right]^{1/2} 
   \ .
\end{equation}
Here $N$ is the number of single spins in the system under
consideration and the sum in the exponent is running over all
intermediate spin quantum numbers $\overline{S}_{i'}$ of the
targeted coupling scheme. $\mathcal{S}$ represents the sum of
the so-called \textit{first coupled angular momenta}, i.e. the
sum of those spins that appear in the bra-ket notation of the
recoupling coefficient in the first position of each coupling
triad. In the case of the recoupling coefficient $\braket{s_1
s_2 \overline{S}_{1} s_3 \overline{S}_2 s_4 S M}{s_3 s_4
\overline{S}_{1'} s_1 \overline{S}_{2'} s_2 S M}$, the sum of
the first coupled angular momenta is given by
\begin{equation*}
   \mathcal{S} = s_1 + \overline{S}_{1} + \overline{S}_{2} + s_3 + \overline{S}_{1'} + \overline{S}_{2'}
   \ .
\end{equation*}

\setlength{\extrarowheight}{0.2cm}
\begin{table}[t!] \centering
   \begin{tabular}{| c | c |}
      \hline
      $N$ & $F$ \\
      \hline
      \hline
      $2$ & \parbox[0pt][15mm][c]{75mm}{$(2s_a+1)^{-1} \delta_{s_a,s_b}$} \\
      \hline
      $3$ & \parbox[0pt][15mm][c]{75mm}{$\sixj{s_a}{s_b}{s_c}{s_d}{s_e}{s_f}$} \\
      \hline
      $4$ & \parbox[0pt][15mm][c]{75mm}{$\sum_k (-1)^{k+s_f-s_h} (2k+1) \sixj{s_a}{s_d}{k}{s_h}{s_f}{s_e} \sixj{s_b}{s_c}{k}{s_h}{s_f}{s_g}$} \\
      \hline
      $>4$ & \parbox[0pt][15mm][c]{75mm}{$\sum_k (-1)^{s_b+s_c+s_e+k} (2k+1) \sixj{s_a}{s_b}{k}{s_d}{s_c}{s_e}$} \\
      \hline
   \end{tabular}
   \caption{Contributions $F$ to the recoupling formula resulting from the reduction of a $N$-cycle (cf. Fig. \ref{fig:yutsis_operations}). The case $N>4$ refers to an interchange operation which is used in order to express larger cycles in terms of 2-, 3-, and 4-cycles.}
   \label{tab:yutsis_ops_F} 
\end{table}

The formula for the recoupling coefficient is now obtained by a
successive reduction of \textit{cycles} that appear in the
graph. A cycle refers to a loop that connects a certain number
of nodes. Depending on the number of connected nodes, different
operations exist that reduce the graph. Figure
\ref{fig:yutsis_operations} shows the operations leading to a
reduction of 2-, 3-, and 4-cycles. Additionally, an interchange
operation is shown that can be used in order to express cycles
which cannot be reduced immediately, i.e. cycles with more than
four nodes, in terms of 2-, 3-, and 4-cycles.\cite{FPV:CPC97}
The contributions $F$ to the recoupling formula resulting from
the shown operations are listed in
Tab. \ref{tab:yutsis_ops_F}. The values for the quantum numbers
$k$ related to the re-labeled edges in
Figs. \ref{fig:yutsis_4cycle} and \ref{fig:yutsis_interchange}
are determined by the symmetry properties of the Wigner-6J
symbols within these contributions. Whenever performing
reductions on a Yutsis graph, the directions of the edges as
well as the signs of the nodes have to be considered
carefully. Within the Yutsis graph that is supposed to be
reduced these directions and signs eventually have to be changed
in order to match the constellation of the edges and the nodes
shown in Fig. \ref{fig:yutsis_operations}. In this process the
change of the direction of an edge $s$, i.e. a contracted line,
contributes with $(-1)^{2s}$ to the phase of the recoupling
coefficient.

The contributions to the recoupling formula arising from
reducing the Yutsis graph according to the above mentioned
operations are discussed in detail in
Ref. \onlinecite{YLV:angular62} and shall not be further
explained here for the sake of brevity.

\begin{figure}[ht!]
  \centering
  \includegraphics[width=20mm]{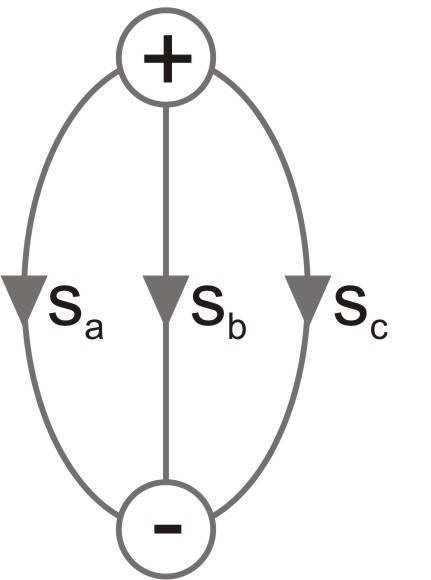}
  \caption{Graphical representation of a triangular delta.}
  \label{fig:yutsis_triangular_delta}
\end{figure}

A Yutsis graph is said to be reduced whenever a graphical
representation is obtained that corresponds to the one in
Fig. \ref{fig:yutsis_triangular_delta}. This representation is
called a \textit{triangular delta} and gives a factor $1$, if
$s_a$, $s_b$, and $s_c$ satisfy the triangular condition
(Eq.~\fmref{eq:vector_addition_rule}), and a factor $0$
otherwise.\cite{YLV:angular62}

Coming back to the example of calculating the recoupling
coefficient $\braket{s_1 s_2 \overline{S}_{1} s_3 \overline{S}_2
s_4 S M}{s_3 s_4 \overline{S}_{1'} s_1 \overline{S}_{2'} s_2 S
M}$ that is displayed in Fig. \ref{fig:yutsis_graph}, one
immediately finds two 3-cycles which can be reduced in order to
generate a recoupling formula: $s_1$-$s_2$-$\overline{S}_{2'}$
and $s_3$-$s_4$-$\overline{S}_{2}$. As a result of this
reduction, the recoupling formula contains apart from phase
factors and square roots two Wigner-6J symbols as in
Eq.~\fmref{eq:recoupling_pi}.

\begin{figure}[ht!]
  \centering
  \includegraphics[width=85mm]{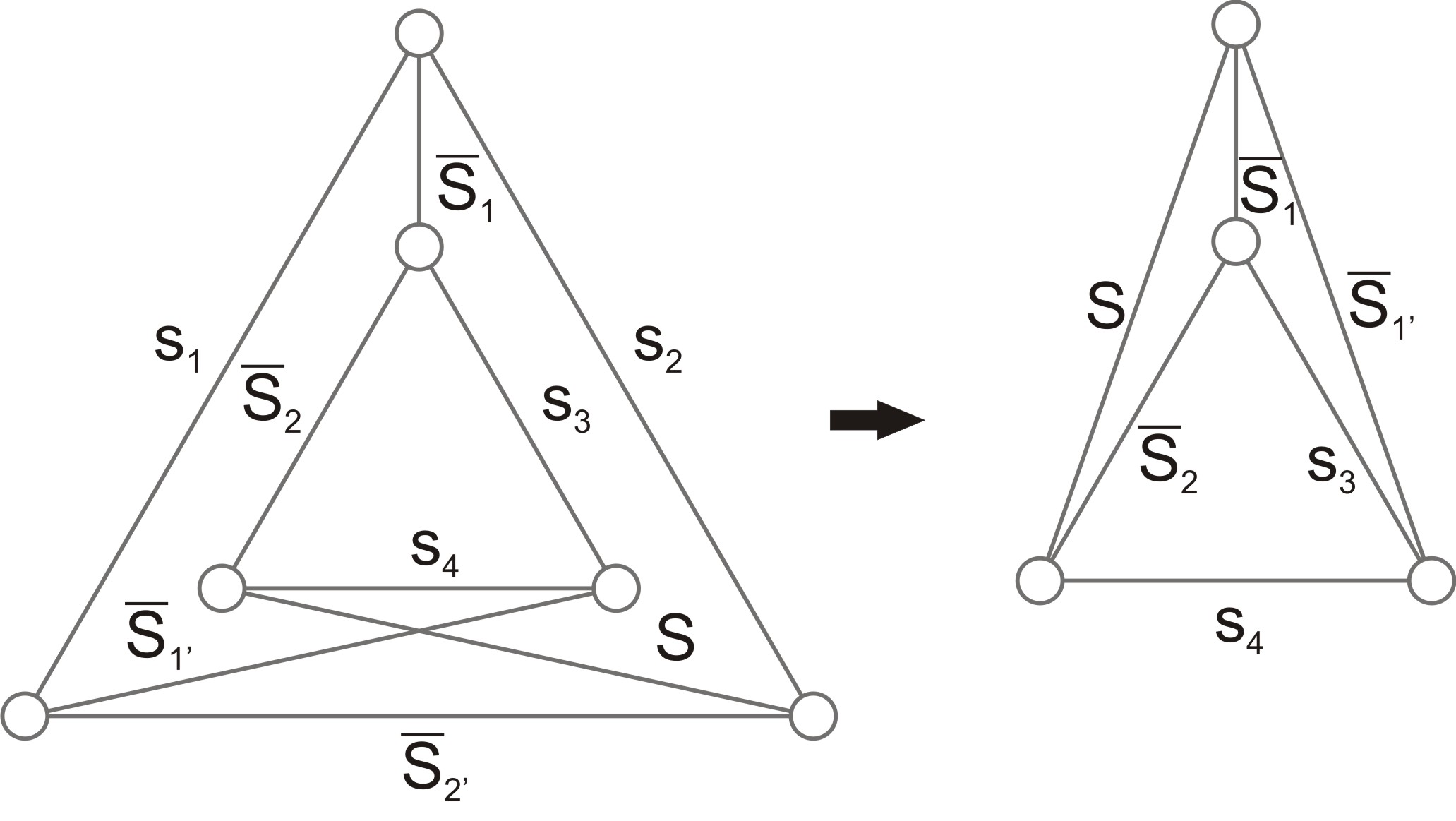} 
  \caption{Reduction of the outer triangle of the Yutsis graph from Fig. \ref{fig:yutsis_graph}.}
  \label{fig:yutsis_example}
\end{figure}

However, as already mentioned in App. \ref{sec-B-1} there is in
general more than one possibility of reducing a graph. Figure
\ref{fig:yutsis_example} shows a possible first step that
reduces the outer triangle spanned by the edges $s_1$, $s_2$,
and $\overline{S}_{2'}$. For the sake of clarity the signs of
the nodes as well as the directions of the edges are omitted
since they only contribute to the phase of the recoupling
coefficient. As one can see after reducing the outer triangle,
four different triangles appear. The reduction of each triangle
would then lead to a triangular delta and therefore to a
completed recoupling formula. These formulas look slightly
different but, can of course be transformed into each other. One
possible expression for the recoupling coefficient is given by
\begin{equation} \begin{split}
    & \braket{s_1 s_2 \overline{S}_{1} s_3 \overline{S}_2 s_4 S M}{s_3 s_4 \overline{S}_{1'} s_1 \overline{S}_{2'} s_2 S M}  = \\
    & \quad (-1)^{-s_1-s_2-s_3-s_4+\overline{S}_{1}-\overline{S}_{1'}+2\overline{S}_{2'}} \\
    & \quad \times \sqrt{(2\overline{S}_{1'}+1)(2\overline{S}_{2'}+1)(2\overline{S}_{1}+1)(2\overline{S}_{2}+1)} \\
    & \quad \times  \sixj{S}{\overline{S}_1}{\overline{S}_{1'}}{s_1}{\overline{S}_{2'}}{s_2} \sixj{s_4}{S}{\overline{S}_{2}}{\overline{S}_{1}}{s_3}{\overline{S}_{1'}}
    \ .
\end{split} \end{equation}

So far, this section has dealt with the construction of a Yutsis
graph and those operations that reduce this graph to a
triangular delta. In principle, one could generate recoupling
formulas and calculate general recoupling coefficients with this
information. In the discussed example of
Fig. \ref{fig:yutsis_graph} only triangles appear leading to an
easy reduction that contributes two Wigner-6J symbols to the
recoupling formula. However, in larger systems with high
symmetry often more complicated recoupling coefficients have to
be calculated. In order to minimize the computational effort it
is desirable to generate a formula that contains as few
Wigner-6J symbols and summation indices as possible. Again, they
result from triangles, squares, and cycles of higher order.

The most intuitive way of generating an improved recoupling
formula is to reduce the smallest cycles first. This idea was
implemented in Refs. \onlinecite{FPV:CPC97,BaK:CPC88,Lim:CPC91}
and already yields considerably improved formulas in comparison
to the use of a trial-and-error technique. However, these
formulas can be further improved by using a more sophisticated
strategy of reducing cycles.\cite{DyF:CPC2003}

Summarizing this section, one can say that with the help of
graph-theoretical methods the effect of general point-group
operations on vector-coupling states can be determined. Once
this effect is known, the eigenstates of the system under
consideration can be labeled with respect to irreducible
representations of the point group. This characterization not
only reduces the dimensions of the Hamilton matrices saving
hardware resources, but it also provides deeper insight into the
physics of the system arising from its geometry.

%


\end{document}